\newcommand{\teff}{T$_{\mathrm{eff}}$}

\newcommand{\logg}{log $g$}

\newcommand{\kms}{km~s$^{-1}$}
\newcommand{\vsini}{V~$\sin{i}$}
\newcommand{\vmicro}{$\rm{v_{micro}}$}

\newcommand{\project}[1]{\textsl{#1}}
\newcommand{\gaia}{\project{Gaia}}

\documentclass[a4paper,fuseAMS,leqn,usenatbib]{mnras}

\usepackage{graphicx}
\usepackage{amssymb}
\usepackage{amsmath}
\usepackage{times}
\usepackage{subfigure}
\usepackage[T1]{fontenc}
\usepackage{ae,aecompl}
\usepackage{textcomp}

\usepackage{hyperref}

\title[COMBS III]{The COMBS Survey - III. The Chemodynamical Origins of Metal-Poor Bulge Stars  \thanks{Based on observations collected at the European Southern Observatory under ESO programme: 089.B-069 }}

 \author[Lucey, et al. 2021]{Madeline~Lucey$^{1}$\thanks{E-mail:m\_lucey@utexas.edu}, Keith~Hawkins$^{1}$, Melissa~Ness$^{2,3}$,
 Tyler~Nelson$^{1}$,
 \newauthor Victor~P.~Debattista$^4$, 
 Alice~Luna$^{1,5}$,
 Thomas~Bensby$^{6}$, 
 Kenneth~C.~Freeman$^{7}$,  
 \newauthor
 and Chiaki~Kobayashi$^{8}$ 
\\
$^{1}$Department of Astronomy, The University of Texas at Austin, 2515 Speedway Boulevard, Austin, TX 78712, USA \\
$^{2}$Center for Computational Astrophysics, Flatiron Institute,
162 5th Ave., New York, NY 10010, USA\\
$^{3}$Department of Astronomy, Columbia University, 550 W 120th St., New York, NY, 10027, USA \\
$^4$Jeremiah Horrocks Institute, University of Central Lancashire, Preston, PR1 2HE, UK \\
$^{5}$Department of Astronomy \& Astrophysics, University of Chicago, Chicago, IL 60637, USA \\
$^{6}$Lund Observatory, Department of Astronomy and Theoretical Physics, Box 43, SE-221\,00 Lund, Sweden \\
$^{7}$Research School of Astronomy and Astrophysics, The Australian National University, Canberra, ACT 2611, Australia \\
$^{8}$Centre for Astrophysics Research, Department of Physics, Astronomy and Mathematics, University of Hertfordshire, Hatfield AL10 9AB, UK \\
}

\date{Accepted . Received ; in original form }

\pubyear{2021}

\begin{document}
\label{firstpage}
\pagerange{\pageref{firstpage}--\pageref{lastpage}}
\maketitle

\begin{abstract}
The characteristics of the stellar populations in the Galactic Bulge inform and constrain the Milky Way's formation and evolution. The metal-poor population is particularly important in light of cosmological simulations, which predict that some of the oldest stars in the Galaxy now reside in its center. The metal-poor bulge appears to consist of multiple stellar populations that require dynamical analyses to disentangle. In this work, we undertake a detailed chemodynamical study of the metal-poor stars in the inner Galaxy.  Using R$\sim$ 20,000 VLT/GIRAFFE spectra of 319 metal-poor (-2.55 dex$\leq$[Fe/H]$\leq$0.83 dex, with $\overline{\rm{[Fe/H]}}$=-0.84 dex) stars, we perform stellar parameter analysis and report 12 elemental abundances (C, Na, Mg, Al, Si, Ca, Sc, Ti, Cr, Mn, Zn, Ba, and Ce) with precisions of $\approx$0.10 dex. Based on kinematic and spatial properties, we categorise the stars into four groups, associated with the following Galactic structures: the inner bulge, the outer bulge, the halo, and the disk. We find evidence that the inner and outer bulge population is more chemically complex (i.e., higher chemical dimensionality and less correlated abundances) than the halo population. This result suggests that the older bulge population was enriched by a larger diversity of nucleosynthetic events. We also find one inner bulge star with a [Ca/Mg] ratio consistent with theoretical pair-instability supernova yields and two stars that have chemistry consistent with globular cluster stars.

\end{abstract}


\section{Introduction}

\label{sec:Introduction}
The goal of Galactic archaeology is to understand the Milky Way's (MW) formation and evolution through the chemodynamical properties of its stars. Using observations \citep{Ortolani1995,Kuijken2002,Zoccali2003,Clarkson2011,Brown2010,Valenti2013,Calamida2014,Howes2014} and simulations \citep{Tumlinson2010,Kobayashi2011a,Starkenburg2017a,El-Badry2018b}, the bulge of the MW has been shown to contain many of the oldest stars in our Galaxy. Studies of the chemodynamics of these old stars can reveal new insights into the formation and early chemical evolution of the MW.

The bulge is a complex Galactic component, with many overlapping stellar populations. Spectroscopic studies of the stars in the bulge have revealed a metallicity distribution function (MDF) with multiple components. Specifically, the Abundances and Radial velocity Galactic Origins Survey \citep[ARGOS;][]{Freeman2013} found that the MDF of the bulge has five components \citep{Ness2013a}. The two most metal-rich components, which are associated with the bulge, peak at [Fe/H] = +0.12 dex and -0.25 dex. The other three components, which peak at [Fe/H] = -0.70 dex, -1.18 dex and -1.70 dex, they associate with the thin disk, thick disk and halo components of the MW, respectively. However, it is important to note that the metal-rich components dominate with only 5\% of bulge stars having [Fe/H] < -1 dex \citep{Ness2016}. Although many studies have found similar results \citep[e.g.,][]{Zoccali2008,Johnson2013a,Rojas-Arriagada2014,Zoccali2017,Rojas-Arriagada2017,Duong2019},  \citet{Johnson2020} argue that the multi-modal MDF is only valid for the outer bulge and that inside a Galactic latitude of ($b$) $\sim$ 6$^{\circ}$ the MDF is consistent with a closed box model (a single peak with a long metal-poor tail). However, \citet{Bensby2013,Bensby2017} found strikingly similar results to \citet{Ness2013a} using bulge micro-lensed dwarf stars within -6$^{\circ}$<$b$<-2$^{\circ}$.

The discovery of metallicity-dependent structure and kinematics in the bulge provides further evidence for multiple stellar populations \citep{Ness2013a,Ness2013b}. Today, it is generally accepted that the majority of the mass in the bulge participates in a boxy/peanut-shaped (B/P) bulge \citep{Howard2009,Shen2010,Ness2013b,Debattista2017}. A B/P bulge is a rotation-supported structure, which is the result of secular disk and bar evolution \citep{Combes1981,Combes1990,Raha1991,Merritt1994,Quillen2002,Bureau2005,Debattista2006,Quillen2014,Sellwood2020}. However, it is also suggested that the MW may host a less-massive metal-poor classical bulge component \citep{Babusiaux2010,Hill2011,Zoccali2014}, which is a spheroidal, pressure-supported structure formed by hierarchical accretion \citep{Kauffmann1993,Kobayashi2011a,Guedes2013}.  Evidence for a metal-poor classical bulge has been found in studies of the kinematics of bulge stars as a function of metallicity. Specifically, metal-poor stars in the bulge rotate slower and have a higher velocity dispersion than the metal-rich stars \citep{Ness2013b,Kunder2016,Arentsen2020}. However, \citet{Debattista2017} demonstrated that these observations may be the result of an overlapping halo population rather than a classical bulge. In fact, \citet{Kunder2020} found that 25\% of the RR Lyrae stars currently in the bulge are actually halo interlopers. Similarly, \citet{Lucey2021} found that about 50\% of their sample of metal-poor giants are halo interlopers and that the fraction of interlopers increases with decreasing metallicity. When they removed the halo interlopers from the sample, \citet{Lucey2021} found that the velocity dispersion decreased and there was no evidence for a classical bulge component in the kinematics.

With the advent of metallicity-sensitive photometric surveys such as the Skymapper \citep{Casagrande2019} and Pristine \citep{Starkenburg2017b} surveys, there is great potential to target and study the metal-poor stars in the Galactic bulge. These metal-poor stars are especially exciting because previous work on old stars have focused on the Galactic halo, where the majority of stars are metal-poor \citep[e.g.,][]{Frebel2006,Norris2007,Christlieb2008,Keller2014}. Simulations now indicate that targeting metal-poor stars in the bulge is most conducive to the discovery of ancient stars. For example, simulations predict that if Population III stars exist in our Galaxy, they are most likely to be found in the bulge \citep{White2000,Brook2007,Diemand2008}. Furthermore, simulations predict that stars of a given metallicity are more likely to be older if they are found closer to the Galactic center \citep{Salvadori2010,Tumlinson2010,Kobayashi2011a}. Specifically, metal-poor bulge stars are ancient in that they formed before $z$ > 5 and are older than 12 Gyr \citep{Kobayashi2011a}.

The chemistry of ancient stars is of special interest, given that they are thought to be primarily enriched by Population III stars. Therefore, their chemistry can provide insight into the properties of Population III stars and the early universe in which they formed. Several studies have found that a significant fraction of Population III stars would explode as pair-instability supernovae (PISNe) given that simulations of metal-free star formation yield a top-heavy initial mass function \citep[IMF;][]{Tumlinson2006,Heger2010,Bromm2013}. Results of simulated yields from PISNe predict that a star which is 90\% enriched by a PISNe would have [Fe/H]~$\approx-2.5$~dex \citep{Karlsson2008} and would contain barely any elements heavier than Fe \citep{Karlsson2008,Kobayashi2011b,Takahashi2018}. Recently, \citet{Takahashi2018} found that the two most discriminatory abundance ratios that indicate enrichment from PISNe are [Na/Mg] $\approx-1.5$~dex and [Ca/Mg]$\approx$0.5-1.3 dex. Excluding PISNe (i.e., if the IMF is truncated at $< 140M_\odot$), ancient stars are expected to have higher levels of $\alpha$-element enhancement than typical MW stars due to the top-heavy IMF of Population III stars and the mass-dependent yields of Type II supernovae \citep{Tumlinson2010,Heger2010,Bromm2013}. Another important chemical signature of ancient stars is lower copper (Cu), manganese (Mn), sodium (Na), and aluminum (Al) abundances with respect to typical MW stars given the metallicity dependence of these yields in Type II supernovae \citep{Kobayashi2011a}.

Recently, there have been many spectroscopic surveys targeting the metal-poor stars in the bulge \citep[e.g.,][]{Howes2014,Howes2015,Howes2016,Duong2019,Duong2019b,Lucey2019,Arentsen2020b}. The first installment of the Chemical Origins of Metal-poor Bulge Stars (hereafter COMBS I) studied the detailed chemistry of 26 metal-poor bulge stars \citep{Lucey2019}. One of the major results from this work was the discovery of higher levels of calcium enhancement in the bulge compared to Galactic halo stars of similar metallicity. Furthermore, COMBS I found lower scatter in many elemental abundances for very metal-poor bulge stars compared to halo stars. The HERMES Bulge Survey \citep[HERBS;][]{Duong2019} and \citet{Fulbright2007} found similar results with respect to higher levels of Ca enhancement and lower scatter for their sample of metal-poor stars. Further differences between metal-poor bulge stars and halo stars include the rate of carbon (C) and neutron process enhancements. C-Enhanced Metal-Poor (CEMP) stars occur at a rate of 15-20\% among halo stars with [Fe/H]<-2 dex \citep{Yong2013}. However, in the bulge, the rate of CEMP stars is estimated at $\sim$6\% for the same metallicity range \citep{Arentsen2021}. Furthermore, neutron-capture element-enhanced stars are rarely observed in bulge spectroscopic surveys \citep{Johnson2012,Koch2019,Lucey2019,Duong2019b}.

It is important to note, however, that $\sim$25-50\% of metal-poor stars in the bulge are actually halo interlopers \citep{Kunder2020,Lucey2021}. Therefore, it is unclear if these chemistry results simply apply to the Galactic halo in the inner Galaxy, to the Galactic bulge, or both. Consequently, dynamical analysis is essential to study these populations separately. Given results from simulations \citep{Tumlinson2010}, metal-poor stars on tightly bound orbits are expected to have formed as early as $z$ $\sim$ 20 while stars on loosely bound orbits only form as early as $z$ $\sim$ 10-13. This is because stars on loosely bound orbits, which are accreted more recently, originate from small dark matter halos which form later than the most massive main progenitors \citep{Tumlinson2010}. This is consistent with recent simulation results demonstrating that the majority of stars within 2 kpc of the Galactic center formed in the most massive main progenitor of the MW \citep{Santistevan2020}. Therefore, we expect stars confined to the inner bulge region are more ancient than loosely bound halo stars. However, it is essential to combine chemical and dynamical information to test this prediction and compare these populations in detail. 

In this work, we aim to determine the origins of the metal-poor stars in the Galactic bulge through chemodynamial analysis. Specifically, we will test predictions from simulations that the metal-poor bulge stars are ancient and search for signatures of PISNe. To accomplish this, we present the stellar parameters and elemental abundances for a sample of 319 stars selected to be metal-poor bulge stars using SkyMapper photometry. We combine this analysis with dynamical results from the second installment of the COMBS survey \citep[][hereafter COMBS II]{Lucey2021} for a full chemodynamical picture. In Section \ref{sec:data} we present the VLT/GIRAFFE observations and data reduction method. The stellar parameter and elemental abundance analysis are described in Sections \ref{sec:param} and \ref{sec:abund}, respectively. We perform a comparison between our analysis, the ARGOS survey and the HERBS survey in Section \ref{sec:comp}. We present our MDF and elemental abundance results in Sections \ref{sec:met} and \ref{sec:results}. We separate our population into four dynamical groups and compare their chemistry in Section \ref{sec:dyn_grp}. We discuss chemical signatures of pair-instability supernovae in Section \ref{sec:pisne} and possible globular cluster origins for our stars in Section \ref{sec:glob}. Last, we present our final conclusions in Section \ref{sec:conclu}.

\section{Data} \label{sec:data}
Given the high levels of extinction and primarily metal-rich population, obtaining large spectroscopic samples of metal-poor stars in the Galactic bulge has historically been difficult. With the advent of metallicity-sensitive photometric surveys, like the SkyMapper \citep{Wolf2018} and Pristine \citep{Starkenburg2017b} surveys, it is now possible to target and observe these rare stars in large numbers. In this work, we use SkyMapper photometry and ARGOS spectra \citep{Freeman2013} to select metal-poor giants for spectroscopic follow-up. For further information on the target selection, we refer the reader to Section 2 of COMBS I. 

The observations presented in this work are from the FLAMES spectrograph \citep{Pasquini2002} on the European
Southern Observatory's (ESO) Very Large Telescope (VLT). The FLAMES instrument is fiber-fed with fibers going to both the UVES and GIRAFFE spectrographs. Therefore, observations with both spectrographs can be simultaneously obtained. For the COMBS survey, we observed 555 stars with the GIRAFFE spectrograph along with 40 stars with the UVES spectrograph. For the UVES spectra, we used the RED580 setup which has a resolution (R=$\lambda/\Delta \lambda \approx$ 47,000) and wavelength coverage 4726-6835 \AA. The stellar parameters and elemental abundances of the UVES spectra have already been published in COMBS I. In this work, we present the stellar parameter and chemical abundance analysis of the GIRAFFE spectra.

\subsection{Medium Resolution GIRAFFE Spectra}

For the GIRAFFE spectra, we use the HR06 and HR21 setups. The HR06 setup has resolution R$\approx$24,300 and wavelength coverage 4538-4759 \AA, while the HR21 setup has resolution R$\approx$18,000 and wavelength coverage 8484-9001 \AA. The HR21 spectra contain the Calcium II near-infrared triplet (CaT), which is useful for determining accurate radial velocities. The HR06 spectra contain many metal lines including iron (Fe) lines for constraining the metallicity and even a barium (Ba) line (4554 \AA) in order to measure the s-process abundance. It also contains a number of $\rm{C_2}$ Swan band features with band heads at approximately 4715 \AA, 4722 \AA, 4737 \AA, and 4745 \AA. Therefore, we can also determine if a star is a C-enhanced metal-poor (CEMP) star with or without s-process enhancement (CEMP-s or CEMP-no).

As these spectra were used to perform kinematic analysis in COMBS II, the full description of the reduction process can be found in Section 2.2 of that paper. In short, we use the EsoReflex\footnote{\url{ https://www.eso.org/sci/software/esoreflex/}} workflow to perform the bias and flat-field subtraction, along with fiber-to-fiber corrections, cosmic ray cleaning, wavelength calibration, and extraction. We then use IRAF to perform sky subtraction. Last, we use iSpec \citep{Blanco-Cuaresma2014} to radial velocity (RV) correct, coadd and normalize the spectra. During the radial velocity determination, we find two possible spectroscopic binary stars (labeled as 6406.0 and 6400.2 in the ESO Phase 3 Data Products archive\footnote{\url{http://archive.eso.org/wdb/wdb/adp/phase3_spectral/form}}) which both have two significant peaks (peak probability > 0.5) in the cross-correlation function. As unresolved spectroscopic binaries can lead to systematic biases in stellar parameters \citep[e.g.][]{El-Badry2018a}, we do not perform stellar parameter analysis on these stars. 

We estimate the signal-to-noise ratio (SNR) using the flux uncertainty estimates from the EsoReflex pipeline which are propagated through the reduction process.  We do not use any individual spectra with SNR < 10 $\rm{pixel^{-1}}$. Out of 555, there are 545 stars with HR21 spectra with SNR > 10 $\rm{pixel^{-1}}$ and only 389 stars with both HR06 and HR21 spectra having SNR > 10 $\rm{pixel^{-1}}$. It is expected that the HR06 spectra have lower SNR on average compared to the HR21 spectra since they are bluer and therefore more impacted by the high levels of extinction towards the Galactic center. In this work, we analyze only stars that have both HR06 and HR21 spectra for consistency. Therefore, after removing the two possible binary stars, there are a total of 387 stars for which we perform stellar parameter analysis.

\section{Stellar Parameter Analysis}
\label{sec:param}
\begin{figure}
    \centering
    \includegraphics[width=\columnwidth]{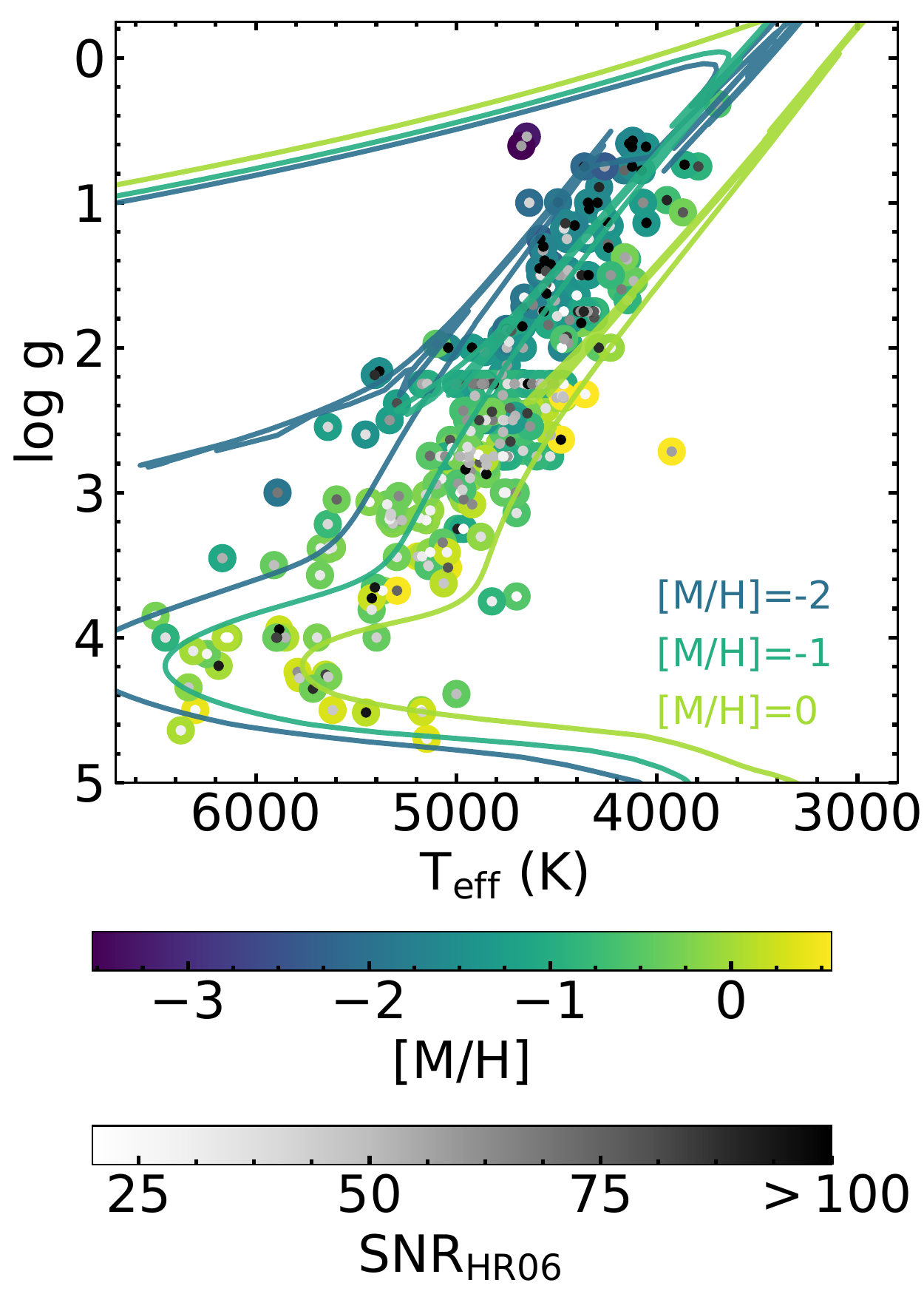}
    \caption{ Kiel diagram for our sample of 319 stars for which we report stellar parameters and elemental abundances. The points are colored by SNR in the center and metallicity in the outer ring. We also plot 10 Gyr MIST isochrones with [M/H]=0, -1 and -2 dex in green, light blue and dark blue lines, respectively. These lines match the metallicity color scale. Our data are well represented by the models, with the exception of outliers which typically have low SNR.}
    \label{fig:kiel}
\end{figure} 

Given the wavelength coverage and resolution of our spectra, there are not enough clean Fe I and Fe II lines to perform the standard Fe-excitation-ionization balance technique to determine the stellar parameters. Therefore, in this work we use a full-spectrum $\chi^2$ fitting technique to determine the effective temperature (\teff), surface gravity (\logg), metallicity ([M/H]), and rotational velocity (\vsini). 

The model spectra, which we use to compare to the observed spectra, are synthesized using Spectroscopy Made Easy (SME) v574 \citep{Valenti1996,Piskunov2016}. To synthesize spectra, we utilize the 1D, local thermodynamic equilibrium (LTE) MARCS model atmosphere grid \citep{Gustafsson2008} and the fifth version of the Gaia-ESO atomic line list which includes hyperfine structure \citep{Heiter2020}. In addition, we use solar abundances from \citet{Grevesse2007}. We incorporate non-LTE (NLTE) line formation for a number of elements using grids of departure coefficients. We use all grids available with SME v574 which includes lithium \citep[Li;][]{Lind2009}, oxygen \citep[O;][]{Amarsi2016}, Na \citep[]{Lind2011}, magnesium \citep[Mg;][]{Osorio2015}, Al \citep[][]{Nordlander2017}, silicon \citep[Si;][]{Amarsi2017}, calcium \citep[Ca;][]{Mashonkina2008}, titanium \citep[Ti;][]{Sitnova2020}, Fe \citep{Amarsi2016} and Ba \citep{Mashonkina2008}. 

As we targeted stars in the bulge, which is over 5 kpc away from the Sun, we expect most of our stars to be giants with \logg\ < 3 dex given that only giants would be sufficiently luminous to be observed at the bulge. However, the results from COMBS II indicate that our target selection has been contaminated by a number of nearby disk stars. Therefore, we require a synthetic grid with a wide range of possible parameters, including dwarf, giant, metal-rich, and metal-poor stars. Our grid covers the following range:

\begin{itemize}
    \item 2500 K $\leq$ \teff $\leq$ 6500 K, steps = 250 K
    \item -0.5 dex $\leq$ \logg $\leq$ 5 dex, steps = 0.25 dex
    \item -5 dex $\leq$ [M/H] $\leq$ 0.75 dex, steps = 0.25 dex
\end{itemize}
We scale the microturbulence (\vmicro) with \teff\ using the relationship calibrated from the Gaia-ESO survey \citep{Smiljanic2014}:
\begin{equation}
\begin{split}
    \text{\vmicro} = &1.1 + 1.0\times 10^{-4}\times(\text{\teff}-5500)\\ &+4.0\times 10^{-7}\times(\text{\teff}-5500)^2
\end{split}
\end{equation}
We also scale the global [$\alpha$/Fe] with [M/H] as follows:

\begin{equation}
     [\alpha/\text{Fe}] =  \begin{cases} 0, & \text{if } [\text{M/H}] \geq0  \\  -0.4\times[\text{M/H}], & \text{if } -1\leq[\text{M/H}]\leq 0 \\  0.4, & \text{if } [\text{M/H}] <-1,\end{cases}
    \label{eqn:alpha}
\end{equation}
in order to match the model atmospheres as well as empirical MW chemical evolution. 

Following \citet{Carroll1933a,Carroll1933b}, we add a convolution term to account for rotational  (\vsini) and instrumental broadening. We allow this term to vary between 0 \kms\ $\leq$ \vsini\ $\leq$ 30 \kms. However, since we have two unique parts of our spectra (HR06 and HR21) which have different wavelength resolutions (R$\approx$24,300 and R$\approx$18,000, respectively) the convolution term must be different for each part. Therefore, we multiply the convolution term by 1.35 (the ratio of the resolutions) before applying it to the HR21 spectra. We attempt to fit the convolution terms for the HR21 and HR06 spectra separately, but the degeneracy between the effect of \logg\ and convolution on the CaT is too strong. Therefore, we must use what we know about the convolution from the HR06 spectra to constrain the HR21 convolution. To interpolate between grid points, we use a piece wise linear interpolator. 

In order to avoid getting stuck in a local minimum when performing the $\chi^2$ fit, we ensure that we start with an accurate guess for the stellar parameters. We do this by performing a quick cross-correlation with a grid of model spectra that is similar, but smaller than our grid for the $\chi^2$ fit. This smaller grid covers the following range:

\begin{itemize}
    \item 3500 K $\leq$ \teff $\leq$ 6500 K, steps= 250 K
    \item 0.5 dex $\leq$ \logg $\leq$ 4 dex, steps=0.5 dex
    \item -5 dex $\leq$ [M/H] $\leq$ 0.5 dex, steps=0.5 dex
\end{itemize}

There are many observational and modeling effects that may cause our model spectra to differ from the observed spectra in ways that can negatively impact the fit. For example, the cores of strong lines, like the CaT, are known to be strongly impacted by NLTE, even when using departure coefficients for population levels. Therefore, we mask pixels that are not well-matched by the model spectra in order to minimize their impact on the spectral fitting. To do this, we compare our model spectra to \gaia\ Benchmark stars \citep[GBS;][]{Blanco-Cuaresma2014}. As these stars are observed in the Gaia-ESO survey \citep{Gilmore2012}, they have GIRAFFE HR21 spectra. However, they do not have HR06 spectra. Instead, we download reduced HARPS spectra \citep{Mayor2003} from the ESO archive\footnote{\url{https://archive.eso.org/scienceportal/}} and degrade the resolution and wavelength coverage to match that of HR06 spectra. We then compare the observed spectra to synthesized spectra of the corresponding parameters derived in \citet{Jofre2014,Heiter2015}. We mask any pixels that differ from the observed spectra by $>$ 0.1 in normalized flux. As the ability of the synthesis to accurately reproduce each pixel of the observed spectra is a function of the stellar parameters, we make the masks using four different benchmark stars depending on the stellar parameters. Specifically, we use the initial guess parameters to chose between four different spectra: (1) for metal-poor giants (\logg\ $<$2.5 dex and [M/H] $\leq$ -1.5 dex) we use HD 122563, (2) for metal-rich giants (\logg\ $<$ 2.5 dex and [M/H] $>$ -1.5 dex) we use Arcturus, (3) for metal-poor sub-giants/dwarfs (\logg\ $\geq$ 2.5 and [M/H] $\leq$ -1.5 dex) we use HD 140283, and (4) for metal-rich sub-giants/dwarfs (\logg $\geq$ 2.5 dex and [M/H] $>$ -1.5 dex) we use $\epsilon$ For.

In addition to the masking, we also use the difference between the observed benchmark spectra and the corresponding model spectra as an uncertainty term in our fit ($\sigma_{synth}$). Therefore, we essentially underweight pixels in the $\chi^2$ fit that are not well reproduced by the model spectra. We add this term in quadrature with the flux uncertainties. We then use this combined uncertainty in the $\chi^2$ fit.

Thus, the $\chi^2$ equation which we minimize is:
\begin{equation}
    \chi^2= \sum \frac{(observed-model)^2}{(\sigma_{flux}^2+\sigma_{synth}^2)}
\end{equation}
where $observed$ is the observed flux, $model$ is the synthesis flux, $\sigma_{flux}$ is the flux uncertainties and $\sigma_{synth}$ is the synthesis uncertainty as described above. We use the Nelder-Mead algorithm to find the global minimum. 

Of the 387 stars for which we attempt stellar parameter analysis, we find a number of stars that we are unable to fit. Upon visual inspection, it is clear that one of these stars (899.0) is a CEMP-s star from the overwhelming $\rm{C_2}$ Swan band features and strong Ba line absorption at 4554 \AA. However, we do not report results for this star in this work, as it requires separate analysis and will be thoroughly studied in a future installment of the COMBS survey. We also find 2 stars (1386.0 and 1659.0) that may show C enhancement and are unable to be fit by our pipeline. Although we will attempt to analyze them in future work with 899.0, these stars are not obviously CEMP stars. In addition, we find 7 stars that continually give solutions at the edge of our grid, with \teff=6500 K. We exclude these stars given that solutions at the edge of the grid are not trustworthy.

Upon visual inspection, we choose to only perform elemental abundance analysis for spectra with SNR > 20 $\rm{pixel^{-1}}$. Of the 377 stars with SNR > 10 $\rm{pixel^{-1}}$ for which we have stellar parameter solutions, 344 have SNR > 20 $\rm{pixel
^{-1}}$. Furthermore, 319 of these stars have a match in \gaia\ DR2 within 1 arcsecond and a \gaia\ DR2 renormalized unit weight error (ruwe) <1.4 \citep{Lindegren2018b}. Therefore, only these 319 stars have measured dynamics from COMBS II. For the rest of this work, we focus on these 319 stars since combining the dynamical analysis with the measured chemistry is essential to the goal of this work. 

We present a Kiel diagram of these 319 stars in Figure \ref{fig:kiel}. The center of the points is colored by the SNR of the HR06 spectra. We also create rings around the points that are colored by the metallicity. Along with our data, we also show 10 Gyr MIST isochrones with various metallicities \citep{Dotter2016,Choi2016,Paxton2011,Paxton2013,Paxton2015}. Our data are well represented by these models, which is consistent with MW bulge age estimates \citep{Zoccali2003}.

\subsection{Stellar Parameter Uncertainties}
\begin{figure}
    \centering
    \includegraphics[width=\columnwidth]{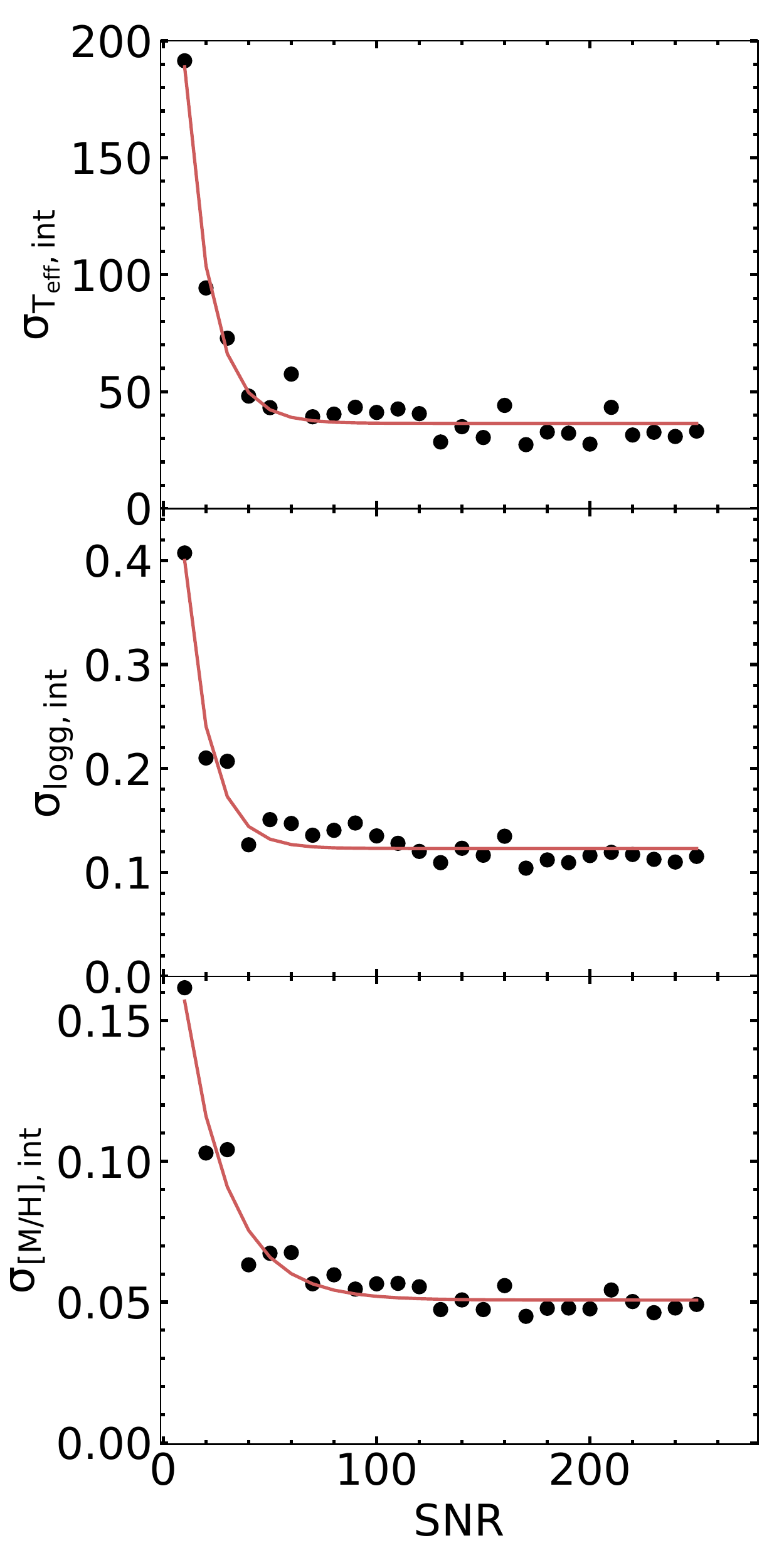}
    \caption{The estimates of the internal uncertainty for \teff, \logg, and [M/H] as a function of SNR. The black points represent the standard deviations of the differences between the derived and synthesized parameters for 100 random synthetic spectra as a function of the SNR. The red lines are the best fit exponentially decreasing functions which are then used to determine the internal uncertainty estimates for our observed data. }
    \label{fig:int_errors}
\end{figure}

\begin{figure}
    \centering
    \includegraphics[width=\columnwidth]{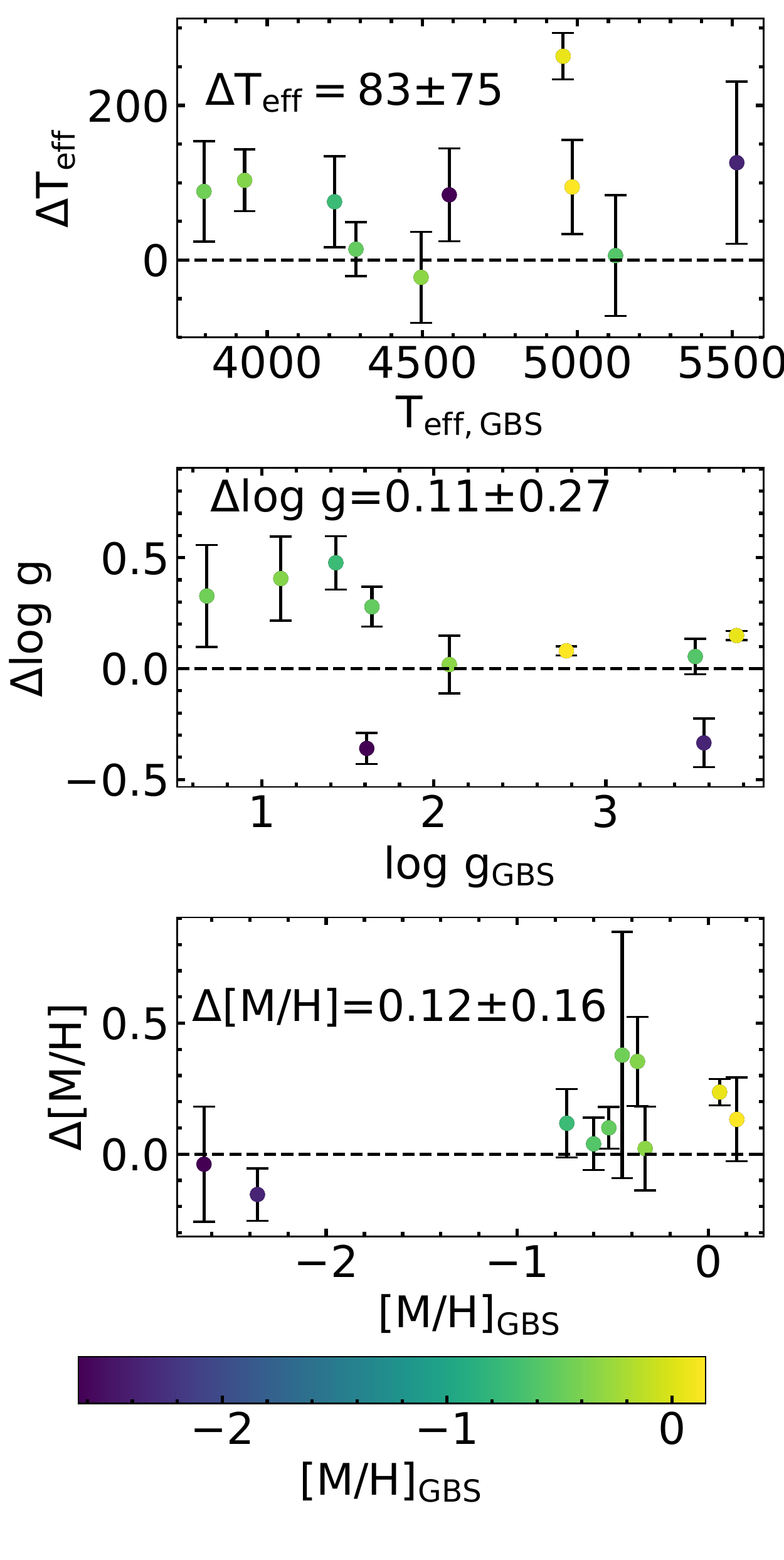}
    \caption{ The differences between our derived values and the reference values from \citet{Heiter2015} and \citet{Jofre2014} for 10 \gaia\ Benchmark stars (GBS). The differences are (this work -- GBS). The points are colored by the reference metallicity. The error bars shown are for the reference values. We also provided the mean and standard deviation of the differences for each parameter in the black text. The standard deviation of the differences is used as our external uncertainty estimate for the given parameter.  }
    \label{fig:gbs_blue}
\end{figure} 

In order to accurately evaluate the uncertainties on the stellar parameters, we must take into account the internal uncertainties, caused by noise in the data and biases in the fitting procedure, as well as the external uncertainties, caused by imperfections in the model spectra. To account for the internal uncertainties, we aim to evaluate the precision of our fitting procedure as a function of SNR. To do this, we run our fitting procedure on synthetic spectra with known stellar parameters and various SNRs. To create these spectra, we use the same synthesis method as was used to create the model spectra grid and we randomly select 100 sets of parameters where 3500 K $\leq$ \teff $\leq$ 5500 K, 0.5 dex $\leq$ \logg\ $\leq$ 4 dex and -5 dex $\leq$ [M/H] $\leq$ 0.5 dex. After synthesizing these 100 spectra with random parameters, we add synthetic Gaussian noise according to the desired SNR. As we aim to evaluate the precision of our method across the entire SNR range of our observed sample, we add noise in order to create spectra with 10 $\rm{pixel^{-1}}$ $\leq$ SNR $\leq$ 250 $\rm{pixel^{-1}}$ in steps of 10 $\rm{pixel^{-1}}$. We do this for each of our 100 synthetic spectra with random parameters, resulting in a total of 2,500 spectra with varying parameters and SNRs with which we can evaluate our precision. 

We put each of the 2,500 synthetic spectra through our parameter analysis pipeline and compare the derived parameters to the true values. For every 100 spectra with the same SNR, we take the standard deviation of the differences between the derived and true values. We use this value as our estimate for the internal precision at that SNR. Therefore, we have internal precision estimates for 25 different SNR values.

We show the calculated internal precision for a range of SNRs in Figure \ref{fig:int_errors}. We fit exponentially decreasing functions to estimate the precision, or internal uncertainty, as a function of SNR. We find the internal uncertainties are best described as:
\begin{equation}
    \sigma_{T_{eff},int} = 345\text{ K }e^{-0.082 \times SNR} + 36\text{ K} 
\end{equation}
\begin{equation}
    \sigma_{log g,int} = 0.653 e^{-0.086 \times SNR} +0.123
\end{equation}
\begin{equation}
    \sigma_{[M/H],int} = 0.173 e^{-0.049 \times SNR} +0.051
\end{equation}
Therefore, we can use these equations to evaluate the \teff, \logg\ and [M/H] internal uncertainties for each of our stars. Specifically, we calculate the internal uncertainties using the SNR estimates for the HR06 spectra which are always lower than the SNR estimates for the HR21 spectra. Given that the SNR was the same for both HR06 and HR21 in our synthetic analysis, we may be slightly overestimating our uncertainties since the HR21 spectra will have higher SNR in our observations.

To evaluate the external uncertainties, we use a sample of 10 \gaia\ Benchmark giant and subgiant stars. These stars are common calibration stars that are frequently used to evaluate the accuracy and precision of stellar parameter pipelines \citep[e.g.,][]{Smiljanic2014,Buder2018,Duong2019}. They are especially useful to compare to spectroscopically-derived parameters since their reference \teff\ and \logg\ values are determined independently from their spectra. Specifically, the bolometric flux and angular diameter are used to determine the \teff. The \logg\ is then determined using the angular diameter and mass estimate. 

In Figure \ref{fig:gbs_blue}, we show the comparison of our results to the reference values for 10 GBS. We color each point by metallicity in order to track the impact of metallicity on the \teff\ and \logg\ determination. The differences on the y-axis are (this work -- GBS). For \teff, we find a mean bias of 83 K with a standard deviation of 75 K. For \logg, we find a bias of 0.11 dex with a standard deviation of 0.27 dex. Lastly, for [M/H], we find a bias of 0.12 dex with a standard deviation of 0.16 dex. However, it is important to note that we are comparing our global metallicity value to their metallicity derived from only Fe lines, which may introduce some bias as a function of [M/H]. Overall, these results are comparable to the HERBS survey which has a similar sample and analysis method as this work \citep[see Figure A1 in][]{Duong2019}. 

We use the derived standard deviations of the differences for \teff, \logg, and [M/H] as our external uncertainty estimates. Our overall uncertainty estimate is calculated by adding the internal and external uncertainty estimates in quadrature. The external uncertainty is larger than the internal uncertainty for \teff, \logg\ and [M/H] at high SNR (SNR $\gtrsim$ 100 $\rm{pixel^{-1}}$). Therefore, the external uncertainty dominates our stellar parameter uncertainties for stars with SNR $\gtrsim$ 100 $\rm{pixel^{-1}}$ and the internal uncertainty only becomes important at SNR $\lesssim$ 50 $\rm{pixel^{-1}}$.

\section{Elemental Abundance Analysis}
\label{sec:abund}
Once the stellar parameters are determined, we perform a line-by-line $\chi^2$ fit to determine the individual elemental abundances. For each line, we compute synthetic spectra using the same method as in the stellar parameter analysis, including all of the same NLTE departure coefficient grids. Specifically, we compute five different spectra with [X/Fe] = (-0.6, -0.3, 0.0, 0.3, 0.6) dex. If the derived solution is [X/Fe]=0.6 dex or -0.6 dex, we repeat the analysis but add or subtract 1 dex from the synthesized [X/H] values. We use the derivatives of the spectrum with respect to the elemental abundance to determine the pixel selection. Explicitly, going out from the line core, we include all pixels until the derivative changes sign or becomes < 0.01 $\rm{dex^{-1}}$. However, we also force the minimum line window to be 0.2 \AA\ wide and the maximum line window to be 10 \AA\ wide. This method is similar to what is applied in other spectroscopic codes \citep[e.g. the BACCHUS code;][]{Masseron2016, Hawkins2015}.

As the strength of absorption features is strongly dependent on the metallicity, we find that it is necessary to use a metallicity-dependent line selection to avoid weak, blended, or saturated lines across our entire metallicity range. Specifically, we have a very metal-poor ([M/H]$\leq$-2.0 dex), metal-poor (-2.0 dex < [M/H] $\leq$ -0.5 dex ) and metal-rich ([M/H] > -0.5 dex) line selection. However, we include many of the same lines between the selections to ensure continuity.

Although we report the abundance derived from each individual line, we use the mean of the lines as our final [X/H] value. We report elemental abundances for C, Na, Mg, Al, Si, Ca, Ti, chromium (Cr), Mn, Fe, zinc (Zn), Ba,  and cerium (Ce). Of those, the only elements for which we do \textit{not} use NLTE departure coefficient grids are C, Cr, Mn, Zn, and Ce. We note that the NLTE effects of Cr and Mn are important when we constrain the enrichment source from the abundance pattern, in particular for low-$\alpha$ stars \citep{Kobayashi2014}.

For each atomic line, we determine an associated uncertainty for the derived abundance based on the $\chi^2$ fit.  The uncertainty is the distance in abundance space from the minimum $\chi^2$ to where the reduced $\chi^2$ equals the minimum $\chi^2$ plus one \citep[e.g., FERRE\footnote{Available from \url{http://hebe.as.utexas.edu/ferre}} code;][]{Allende-Prieto2004,AllendePrieto2006,AllendePrieto2008,Allende-Prieto2016}. After visual inspection of 50 stars with varying SNR, we find that an individual line abundance uncertainty $\gtrsim$ 0.25 dex tends to indicate an untrustworthy fit and requires further visual inspection to determine if the line fit should be discarded. We also inspect stars whose line-by-line scatter in the abundance is $\gtrsim$ 0.25 dex. For our final abundance uncertainties, we propagate the individual line-by-line abundance uncertainties through the mean. The result is the individual line-by-line uncertainties added in quadrature and then divided by the number of lines used.

\section{Comparison with ARGOS and HERBS Surveys}
\label{sec:comp}
\begin{figure}
    \centering
    \includegraphics[width=\columnwidth]{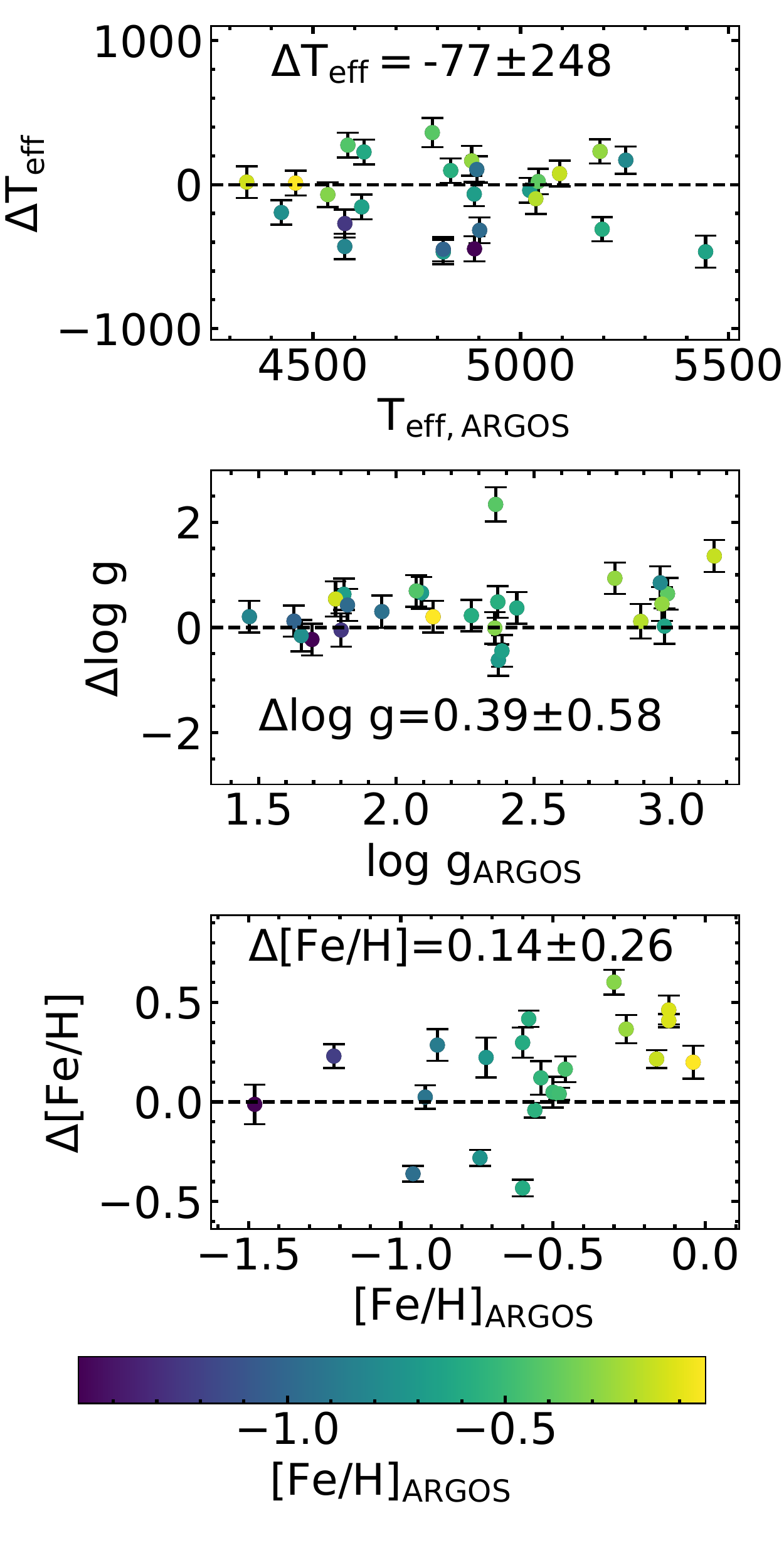}
    \caption{ Comparison of the derived stellar parameters compared to results from the ARGOS survey \citep{Freeman2013} for 26 stars in common. The differences shown are (this work - ARGOS). The points are colored by the [Fe/H] from the ARGOS survey to ensure there are no trends in accuracy and precision of the stellar parameters with [Fe/H]. The error bars are the uncertainties for our derived parameters. The text gives the mean and standard deviations of the differences for each stellar parameter.  }
    \label{fig:argos}
\end{figure}

In order to test the accuracy and precision of our stellar parameters, we compare them to other large Galactic bulge surveys. Specifically, we compare to the ARGOS survey which uses R$\sim$11,000 spectra of $\sim$28,000 stars \citep{Freeman2013}. This survey measured the RV, \teff, \logg, [Fe/H], and [$\alpha$/Fe] ratio of their program stars. Our work has 26 stars in common with the ARGOS survey. In addition, we also compare to the HERBS survey which uses R$\sim$28,000 spectra of 832 stars \citep{Duong2019,Duong2019b}. However, we only observed 3 stars in common with the HERBS survey, which is not enough for a thorough comparison. Fortunately, the HERBS survey performs a detailed comparison with the ARGOS survey. Therefore, we can compare to the HERBS survey through a comparison with the ARGOS survey.

In Figure \ref{fig:argos}, we show the comparison between our derived stellar parameters and the values from the ARGOS survey. The differences shown are (this work -- ARGOS). The points are colored by the ARGOS-derived metallicity. The error bars are the uncertainties on our derived parameters. In the bottom panel, we compare the ARGOS metallicity to our [Fe/H] value derived from Fe lines, rather than the global [M/H] derived during the stellar parameter analysis. However, we have also performed the comparison using the global [M/H] and found the results to be similar to [Fe/H]. We find that the mean difference in \teff\ is -77 K with a standard deviation of 248 K. The mean difference in \logg\ is 0.39 dex with a standard deviation of 0.58 dex, while the mean difference in [Fe/H] is 0.14 dex with a standard deviation of 0.26 dex.

When comparing to the ARGOS survey, the HERBS survey reports the median, 1$\sigma$ and standard deviation (after excluding 3$\sigma$ outliers) of the differences between derived stellar parameters \citep{Duong2019}. They find a median difference in \teff\ of -64 K, which is consistent with our value of -77 K. However our 1$\sigma$ value (246 K), which is also very similar to our standard deviation before (248 K) and after excluding 3$\sigma$ outliers (248 K), is significantly larger than the value reported by the HERBS survey (117 K). We expect that this difference is largely due to the different metallicity distribution of our sample. As the ARGOS survey derives the \teff\ using the photometric colors, it is reasonable to assume that their \teff\ precision would be metallicity-dependent, given that metallicity also impacts the photometric colors. Specifically, it is possible that the ARGOS survey may have worse \teff\ precision for metal-poor stars. In fact, \citet{Freeman2013} notes that using different empirical \teff\ - colour calibrations lead to differences in \teff\ estimates up to 200 K for metal-poor stars \citep{Bessell1998,Alonso1999}. Given that our survey is significantly more metal-poor than the HERBS survey, we would therefore expect the ARGOS precision to be worse for our sample than the HERBS sample. Furthermore, we note that for the 3 stars we have in common with the HERBS survey we find the standard deviation for the differences in \teff\ between our values and the HERBS values is 168 K. In addition, it is interesting to note that when comparing APOGEE DR16 stellar parameters \citep{Ahumada2020} to ARGOS, \citet{Wylie2021} find the differences in \teff\ have a standard deviation of 321 K, which is significantly larger than our value of 248 K. 

For \logg, we find that our results are very consistent with the HERBS survey. Specifically, our median difference is 0.39 dex while the HERBS survey reports a median difference of 0.29 dex. The 1$\sigma$ difference for our work is 0.30 dex while the HERBS survey finds a 1$\sigma$ of 0.29 dex. Last, the standard deviation we find after removing 3$\sigma$ outliers is 0.34 dex, while the HERBS survey reports 0.38 dex. These results indicate that our stellar parameter analysis is consistent with the results from the HERBS survey.

Last, for [Fe/H], we find a median difference of 0.20 dex between our [Fe/H] and the values from ARGOS, while the HERBS survey reports a value of 0.04 dex. From Figure \ref{fig:argos}, it is clear that our large bias is mostly due to our [Fe/H] being significantly larger than the ARGOS values for stars with [Fe/H] $>$ -0.5 dex in ARGOS. We note that the median offset between our [Fe/H] results and the HERBS survey for the 3 stars in common is 0.03 dex. It is also important to note that these 3 stars have -0.7 dex <$\rm{[Fe/H]_{HERBS}}$ <-0.3 dex, which is the same range where we are most inconsistent with ARGOS. We find that our spread in [Fe/H] differences with ARGOS is similar to the differences reported in the HERBS survey. Specifically, we find a 1$\sigma$ of 0.17 dex while HERBS reports a 1$\sigma$ of 0.14 dex. After removing 3$\sigma$ outliers, we find a standard deviation of 0.17 dex. While using the same method, HERBS finds a standard deviation of 0.16 dex. Therefore, we find our stellar parameter results to be generally consistent with the HERBS survey.

\section{Metallicity Distribution Function}
\label{sec:met}

\begin{figure}
    \centering
    \includegraphics[width=\columnwidth]{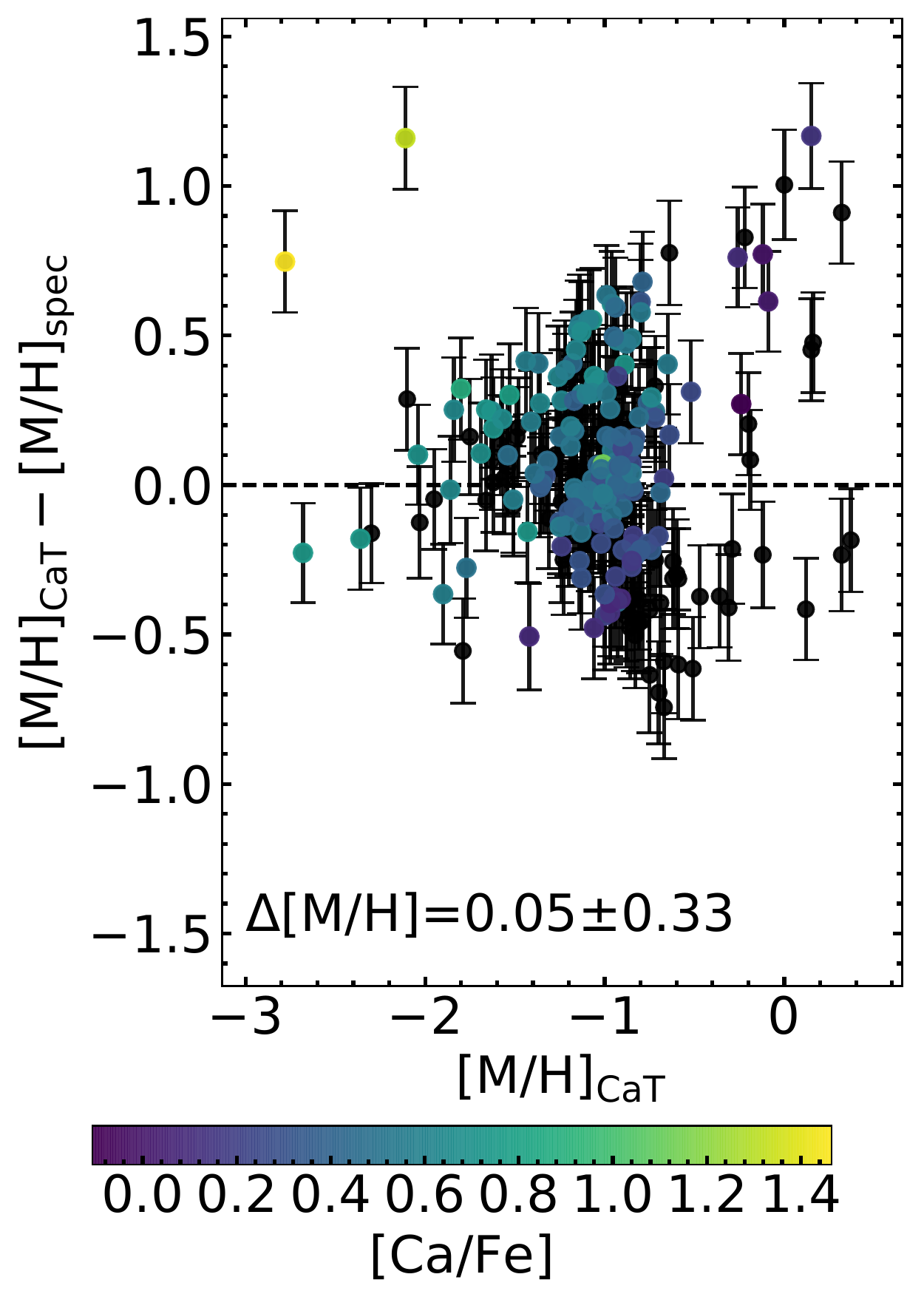}
    \caption{ Comparison between metallicity estimates from the CaT presented in COMBS II to the [M/H] results presented in this work. The points are colored by the [Ca/Fe] abundance when available. The error bars shown are the uncertainty estimates on [M/H] from this work. The black text shows the bias, or mean difference, (0.05 dex) and the standard deviation of the differences (0.33 dex).}
    \label{fig:cat}
\end{figure}

\begin{figure}
    \centering
    \includegraphics[width=\columnwidth]{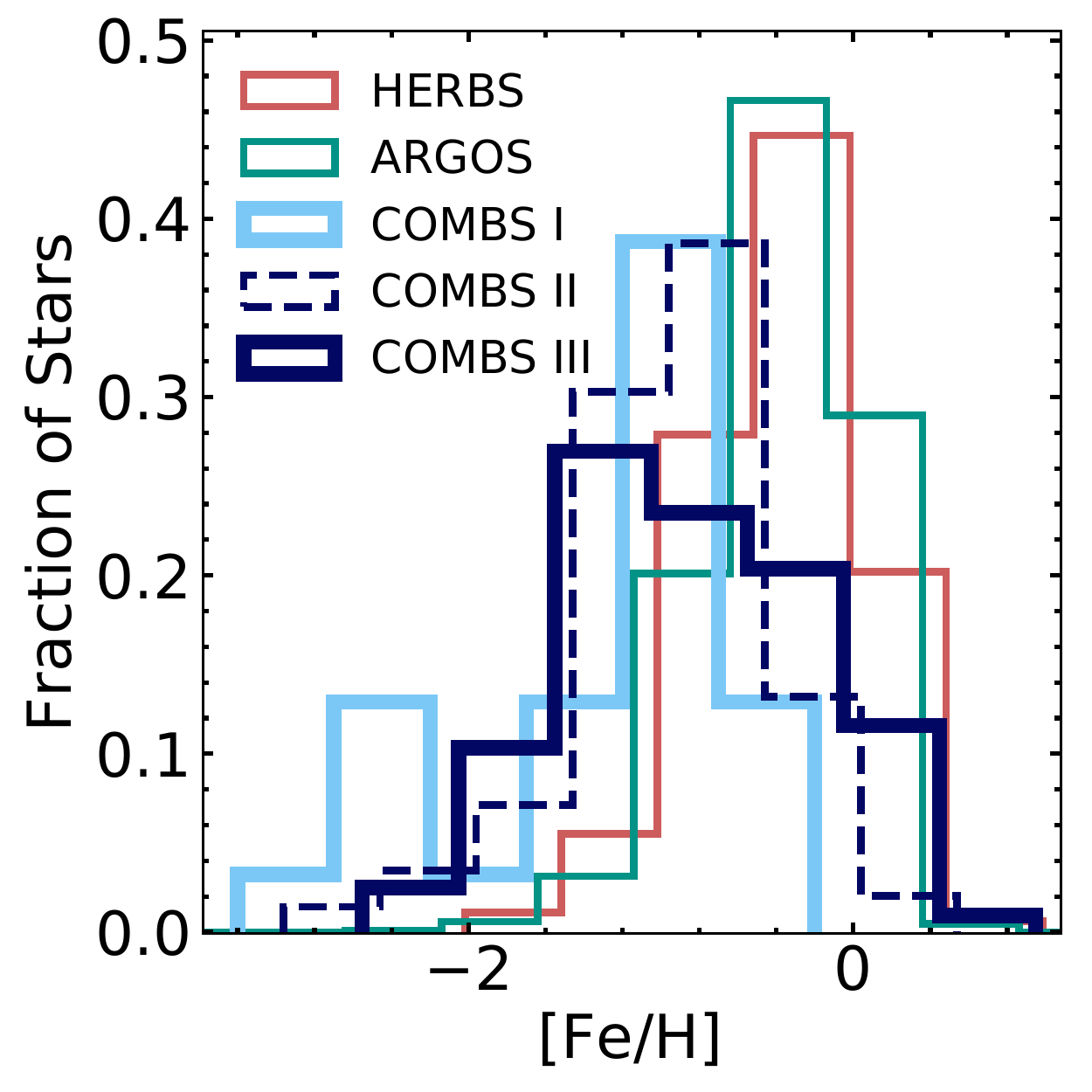}
    \caption{ The metallicity distribution function (MDF) for our results (dark blue solid line), compared to results from COMBS II (dark blue dashed line), COMBS I (light blue solid line), ARGOS \citep[green solid line;][]{Freeman2013,Ness2013a}, and HERBS \citep[red solid line;][]{Duong2019}. Our MDF is more metal-poor than surveys that did not target metal-poor stars \citep[ARGOS and][]{Bensby2017}. Therefore, our selection of metal-poor stars with SkyMapper photometry was successful. }
    \label{fig:mdf}
\end{figure}

The MDF of the Galactic bulge is well-studied through photometric and spectroscopic surveys and is primarily composed of a metal-rich population with [Fe/H] > -1 dex \citep{Zoccali2008,Ness2013a,Johnson2013a,Zoccali2017,Bensby2013,Rojas-Arriagada2014,Bensby2017,Rojas-Arriagada2017,GarciaPerez2019,Duong2019,Rojas-Arriagada2020,Johnson2020}. In this work, we have used SkyMapper photometry to target the metal-poor tail of the Galactic bulge MDF. Therefore, we expect our sample to have an MDF that is on the metal-poor end with [Fe/H] < -1 dex.

In COMBS II, metallicity estimates were determined from the CaT using the same spectra presented in this work.  In Figure \ref{fig:cat}, we show a comparison between the results presented in COMBS II and the [M/H] results determined in the stellar parameter analysis of this work.  For this figure, we only show results for stars with \logg\ $\leq$ 3 dex as our CaT method was designed to be applied to giant stars similar to previous work on metallicity estimates from the CaT \citep{Armandroff1988,Olszewski1991,Armandroff1991,Cole2004,Battaglia2008,Starkenburg2010,Li2017}. The error bars shown are those derived for the [M/H] value in this work. We color the points by the [Ca/Fe] abundance. The metallicity estimates from COMBS II are generally consistent with the [M/H] results from this work, with only a 0.05 dex bias. the standard deviation of the differences is 0.33 dex which is only slightly larger than the uncertainty on the metallicity estimates from the CaT (0.22 dex) added in quadrature with the mean [M/H] uncertainty in this work (0.17 dex).

We present the MDF of our sample in Figure \ref{fig:mdf} using the derived [Fe/H] abundances (dark blue solid line). We also show the results from COMBS II (dark blue dashed line), COMBS I (light blue solid line), the ARGOS survey \citep[green solid line;][]{Freeman2013,Ness2013a} and the HERBS survey \citep[red solid line;][]{Duong2019}. Our MDF peaks at [Fe/H] $\approx$ -1 dex, while the results for the surveys which did not target metal-poor stars (ARGOS and HERBS) peak at [Fe/H] $\gtrsim$ -0.5 dex. Therefore, our use of SkyMapper photometry to select metal-poor stars was successful. However, we have relatively fewer stars with [Fe/H] <-2 dex compared to COMBS I. This is expected, given that the most promising metal-poor targets were prioritized for the high-resolution UVES spectra which were presented in COMBS I. Compared to COMBS II, we see a stronger metal-rich tail which broadens the MDF. This was likely missed in COMBS II because the [Ca/Fe] ratio decreases at [Fe/H] > -1 dex. This causes a smaller increase in [Ca/H] for a given increase in [Fe/H]. Therefore, [Fe/H] values estimated from Ca lines would be underestimated in this [Fe/H] range.

\section{Elemental Abundance Results}
The chemical abundances of stars provide unique insight into the formation and evolution of stellar populations. However, in order to interpret the abundances, we need to contextualize our results in terms of other stellar populations and nucleosynthetic pathways. In this section, we present our abundance results and discuss the formation mechanisms for each element. We also compare our results with other MW populations and literature samples.

\label{sec:results}
\begin{figure*}
    \centering
    \includegraphics[width=\linewidth]{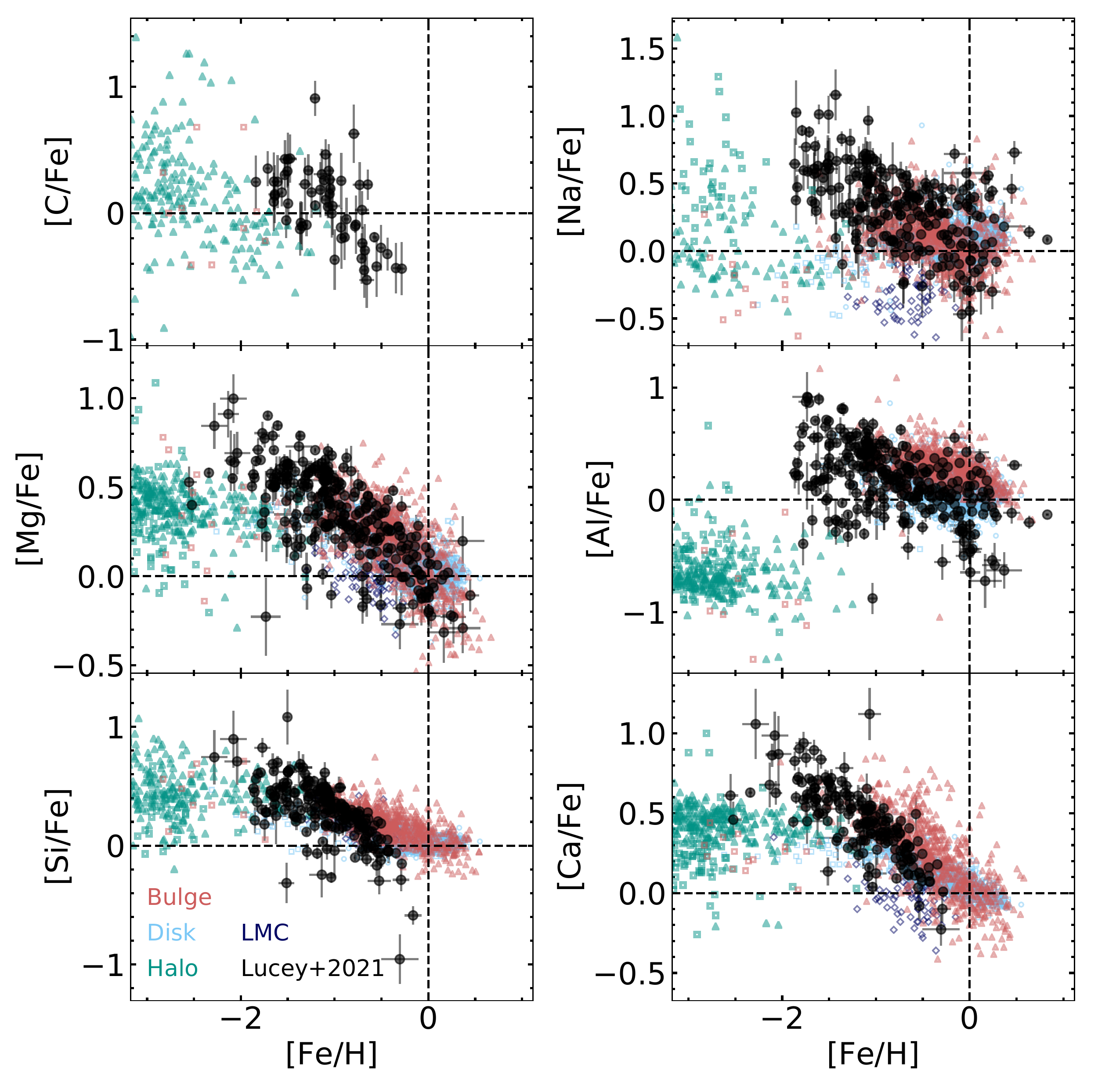}
    \caption{Light and $\alpha$-element abundances for all of the stars in our sample (black circles) compared to other Milky Way samples from the literature.  Specifically, we show other Milky Way bulge samples in red, including results from the HERBS survey \citep[red open triangles;][]{Duong2019,Duong2019b} and results from the EMBLA survey \citep[red open squares;][]{Howes2016}. Also shown are abundances for the halo (green), the Large Magellanic Cloud (LMC; dark blue), and the disk (light blue). The halo abundances are from \citet[][green triangles]{Roederer2014}  and \citet[][green open squares]{Yong2013}. The LMC abundances are from \citet[][dark blue open diamonds]{VanderSwaelmen2013}. The disk abundances are from \citet[][light blue open squares]{Bensby2014}, \citet[][light blue open circles]{Adibekyan2012}, and \citet[][light blue open diamonds]{Battistini2015}.}
    \label{fig:alpha}
\end{figure*} 
\subsection{C}
C is primarily produced in massive stars (> 10$\rm{M_{\odot}}$) and low-mass asymptotic giant branch (AGB) stars. The [C/Fe] yield is especially increased in low-mass stars where the Fe yield is essentially zero \citep{Kobayashi2011c}. In addition, high levels of C-enhancement ([C/Fe] >1 dex) among metal-poor stars is thought to come from Population III supernovae, specifically faint supernovae \citep[e.g.,][]{Nomoto2013}.

In this work, we measure elemental C abundances from the atomic line at 8727 \AA. However, in stars with [Fe/H] <-2 dex, we find that this line is too weak to measure an accurate abundance from. Furthermore, as stars move up the red giant branch (RGB), they experience the second dredge-up which depletes the photospheric C abundance. To account for this depletion, we apply a correction factor to our derived C abundances. These correction factors come from \citet{Placco2014} and are a function of the \logg, [Fe/H] and uncorrected [C/Fe] ratio. We show the corrected abundances in Figure \ref{fig:alpha}. We note that the shown literature abundances from \citet{Roederer2014} and \citet{Howes2016} have not been corrected. 

At [Fe/H] $\gtrsim$ -1 dex, the [C/Fe] ratio decreases with increasing [Fe/H]. This is consistent with chemical evolution models and the onset of Type Ia Supernovae (SNe Ia) which overproduce Fe with respect to C. At [Fe/H] $\lesssim$ -1 dex,  generally [C/Fe] >0 dex. In order to reach this level of C enhancement, it is likely that inhomogenous mixing needs to be taken into account which allows AGB stars to contribute C yields at [Fe/H] $\lesssim$ -1.5 dex \citep{Kobayashi2014,Vincenzo2018a}.

\subsection{$\alpha$-elements}
The $\alpha$-elements are generally divided into two categories based on their formation site. Specifically, the hydrostatic $\alpha$-elements (Mg) primarily form in the hydrostatic burning phase of massive stars, while the explosive $\alpha$-elements (Ca and Si) are primarily produced through explosive nucleosynthesis of core-collapse, or Type II, supernovae \citep[SNe II;][]{Woosley1995,Woosley2002}. Specifically, Mg is produced from C and neon (Ne) burning, while Si and Ca are primarily synthesized from explosive O burning. Thus the yields of explosive $\alpha$-elements depend on the explosion energy \citep{Kobayashi2006}. Although they have different formation sites, the hydrostatic and explosive elements tend to trace each other as they are usually mixed during supernova explosions and dispersed into the interstellar medium (ISM). At low metallicities, before the onset of SNe Ia, the $\alpha$-element abundances are generally indicative of the initial mass function (IMF) of the enriching stellar population, given that their yields in SNe II are mass-dependent. On the other hand, SNe Ia overproduce Fe with respect to the $\alpha$-elements and cause the [$\alpha$/Fe] ratio to decrease. Therefore, the [Fe/H] value at which the [$\alpha$/Fe] ratio begins to decrease specifies the amount of Fe built up by SNe II before the onset of SNe Ia. Furthermore, the behavior of the [$\alpha$/Fe] ratio as a function of [Fe/H] is indicative of the star formation timescale, where a short star formation timescale leads to a large build-up of Fe in the ISM before the onset of SNe Ia. 
\subsubsection{Mg}
Mg abundances at [Fe/H] $\gtrsim$ -1 dex are slightly higher in the bulge than in the disk \citep{McWilliam1994,Rich2000,McWilliam2004,Fulbright2007,Johnson2014,Gonzalez2015,Bensby2017,Duong2019}. This is consistent with a shorter star formation timescale causing a larger build-up of Fe and Mg from SNe II before the contribution from SNe Ia begins. As shown in Figure \ref{fig:alpha}, our results are consistent with the literature at [Fe/H] $\geq$ -1 dex. 

At [Fe/H] $\lesssim$ -1 dex, our abundance measurements generally continue to show high levels of Mg enhancement. Specifically, when compared to the EMBLA survey \citep{Howes2015}, our Mg abundances are generally higher. Furthermore, the Mg abundances reported by the EMBLA survey appear to be more consistent with a Galactic halo population, while our Mg abundances are higher than results from the halo \citep{Roederer2014,Yong2013}. However, it is difficult to draw strong conclusions here because there may be systematic offsets between surveys that impact the comparative results. 
\subsubsection{Ca and Si}

Measurements of Ca and Si abundances in the bulge at [Fe/H] $\gtrsim$ -1 dex are generally higher than what is found in the disk  \citep{McWilliam1994,Rich2000,McWilliam2004,Fulbright2007,Johnson2014,Bensby2017,Duong2019}. However, our abundances at [Fe/H] $\gtrsim$ -1 dex are slightly lower than literature values for the bulge. This is likely a systematic effect possible from our photometric targeting method, or offsets between surveys resulting from differences in analysis methods.

We measure high levels of Ca and Si enhancement at [Fe/H] $\lesssim$ -1 dex. Similar to Mg, we find that our Ca and Si abundances are generally higher than what has been observed in the bulge by the EMBLA survey \citep{Howes2015} and in the halo \citep{Yong2013,Roederer2014}. Our Ca abundances are especially high. This is interesting given that many Population III stars are thought to explode as PISNe which are theorized to have high Ca yields with [Ca/Fe] as high as 2 dex. However, high [Ca/Fe] itself does not suggest PISNe since faint SNe give high [(Mg,Si,Ca)/Fe] as well. We discuss further signatures of PISNe, including the discriminatory [Ca/Mg] ratio, in Section \ref{sec:pisne}.

\subsection{Odd-Z elements}
The odd-Z elements are light elements that have an odd atomic number and therefore could not be produced by successive addition of $\alpha$ particles. In this work, we measure Na and Al. The yields of Na and Al from SNe II are metallicity-dependent, with higher yields from more metal-rich stars \citep{Kobayashi2006}. This leads to an increase in the [(Na,Al)/Fe] ratio with increasing [Fe/H]. However, Fe is overproduced relative to Na and Al in SNe Ia which causes the [(Na,Al)/Fe] ratio to decrease as these types of explosions become relevant. 

\subsubsection{Na}
At [Fe/H] $\gtrsim$ -1 dex, the bulge and disk show similar trends in [Na/Fe] \citep{Bensby2017,Duong2019b}. Consistent with our observations, the [Na/Fe] ratio decreases with metallicity indicating contributions from SNe Ia, similar to the behaviour of $\alpha$ elements. At [Fe/H] $\lesssim$ -1 dex, we generally measure [Na/Fe] >0 dex, while the results from the EMBLA survey generally have [Na/Fe] <0 dex \citep{Howes2015}. The results from \citet{Yong2013} in the halo show high levels of [Na/Fe]. However, when they take NLTE into account, their results approach [Na/Fe] $\approx$ 0 dex, similar to \citet{Roederer2014}. Our results already take NLTE into account and use the same NLTE corrections as \citet{Roederer2014} and \citet{Howes2015}. Therefore, it is unlikely that our higher [Na/Fe] abundances, with respect to halo observations, are merely a NLTE effect. However, it is possible that differences in analysis methods (e.g., line lists, model atmospheres, etc.) causes systematic offsets between ours and other survey's abundances. 

Assuming systematic offsets do not entirely account for the higher [Na/Fe] ratio we measure in the bulge compared to the Galactic halo population, we can infer some of the differences in their chemical evolution histories.  Given the metallicity dependence of Na yields from SNe II, where more metal-rich stars have higher yields, our stars must have been enriched by a more metal-rich population than stars of similar metallicity in the Galactic halo. Therefore, our results indicate a short star formation timescale, and rapid enrichment consistent with chemical evolution models for the bulge \citep[e.g.,][]{Kobayashi2011a}. However, it is important to note that the Na lines used (4668.6 \AA\ and 4751.8 \AA) are too weak to measure low Na abundances for metal-poor stars. Therefore, it is possible that our lack of stars with [Na/Fe] <0 dex at low metallicity is a measurement effect. 

\subsubsection{Al}
Al abundances in the bulge are typically higher than in the disk at [Fe/H]$\gtrsim$-1 dex \citep{Bensby2017,Duong2019b}. However, our [Al/Fe] abundances are generally consistent with the disk at [Fe/H] $\gtrsim$ -1 dex, although they show large scatter. At [Fe/H] $\lesssim$ -1 dex, our reported [Al/Fe] abundances continue to show a large scatter. However, they are generally higher than abundances from the EMBLA survey \citep{Howes2015} and the Galactic halo \citep{Roederer2014,Yong2013}. Unlike our abundances,  \citet{Howes2015,Yong2013,Roederer2014} do not perform NLTE line corrections, which may account for some of the offset at low metallicity. Similar to the Na abundances, the Al abundances are consistent with a short star formation timescale and rapid chemical evolution. 

The high scatter in the Al abundances may indicate inhomogeneous mixing or multiple populations. We note that the standard deviation of our [Al/Fe] abundances is $\sim$0.51 dex while the mean uncertainty is $\sim$ 0.07 dex. Therefore, it is unlikely that the observed scatter is merely due to uncertainties in the abundances. Interestingly, large Al enhancement is a signature of second-generation globular cluster stars. This signature has been identified in a couple of stars in this work and will be further discussed in Section \ref{sec:glob}.

\subsection{Fe-peak elements}

\begin{figure*}
    \centering
    \includegraphics[width=\linewidth]{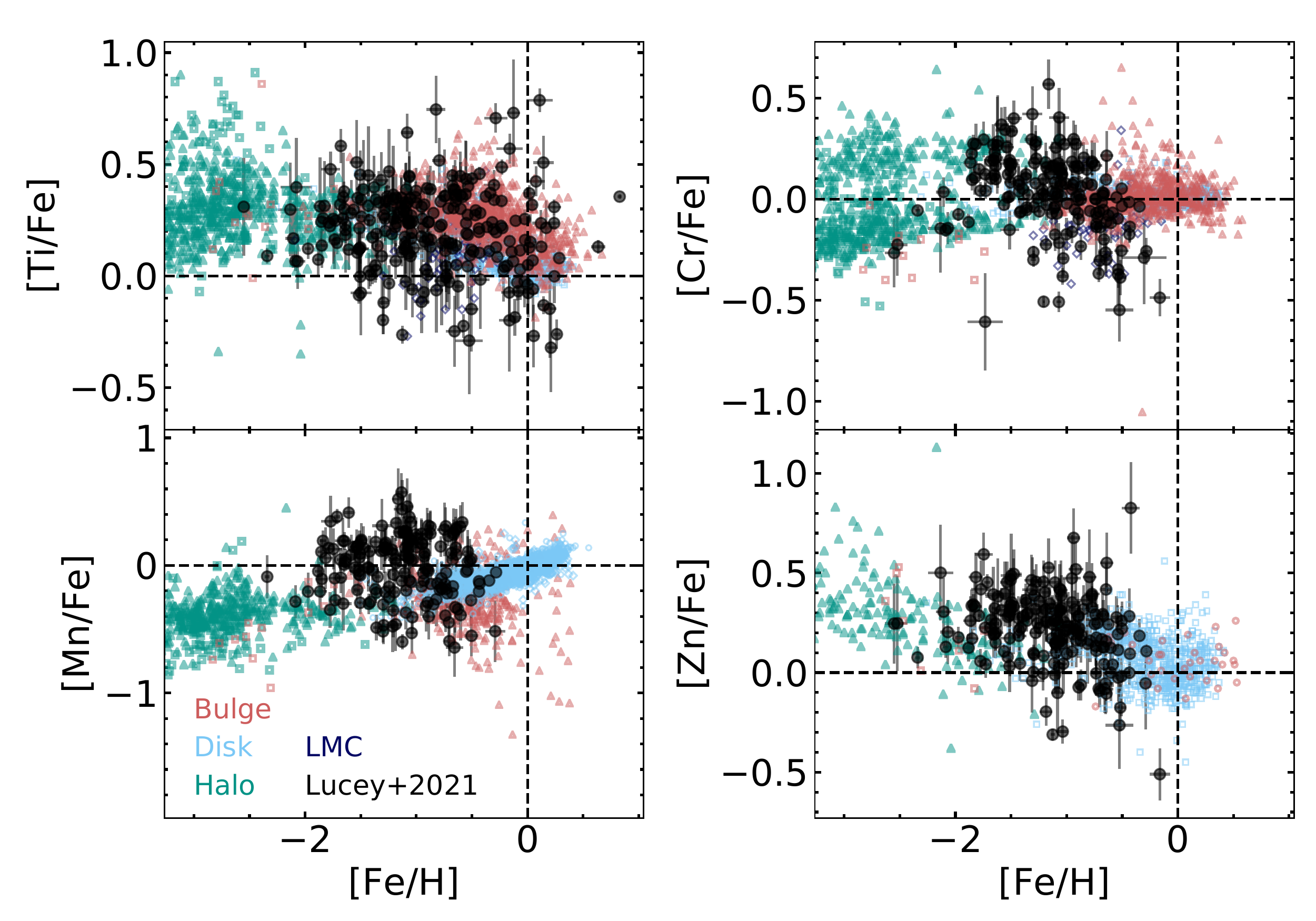}
    \caption{Abundance ratios as a function of metallicity for the Fe-peak elements (Ti, Cr, Mn, and Zn). Symbols are the same as in Figure \ref{fig:alpha}. However, we also include Zn abundances from \citet[][red open circles]{Bensby2017}. }
    \label{fig:fe}
\end{figure*} 

The Fe-peak elements, although formed in a variety of ways, generally trace the Fe abundance with only small variations \citep{Iwamoto1999,Kobayashi2006,Nomoto2013}. However, these slight variations can be extremely informative for supernova physics and chemical evolution models \citep{Kobayashi2011a}. Of the Fe-peak elements, we measure Ti, Cr, Mn, and Zn. We show these results in Figure \ref{fig:fe} compared to other MW populations from the literature.

\subsubsection{Ti}
Frequently considered an $\alpha$-element, Ti is similarly overproduced in SNe II and underproduced in SNe Ia with respect to Fe. Therefore, it is expected to behave similarly to the $\alpha$-elements. In the bulge, the [Ti/Fe] ratio is generally higher than the disk at [Fe/H] $\gtrsim$ -1 dex \citep{Bensby2017,Duong2019}. Our abundances, however, show a large scatter even at high metallicity, with some stars matching the low Ti abundances observed in the Large Magellanic Cloud \citep[LMC;][]{VanderSwaelmen2013}. This is difficult to draw strong conclusions from given that our analysis uses NLTE while the other bulge surveys do not \citep{Bensby2017,Duong2019}. Nonetheless, this result suggests that our sample is a mixed stellar population with a variety of origins.

\subsubsection{Cr}
Cr abundances in the bulge and disk closely follow the Fe abundance \citep{Bensby2014,Bensby2017,Duong2019b}. This is mostly true for our sample at [Fe/H] $\gtrsim$ -1 dex, although we observe large scatter and an overabundance of stars with [Cr/Fe] <0 dex. Interestingly, we observe a number of stars with [Cr/Fe] abundance ratios similar to the LMC \citep{VanderSwaelmen2013}. At [Fe/H] $\lesssim$ -1 dex, the [Cr/Fe] ratio decreases with decreasing metallicity similar to results from the EMBLA survey \citep{Howes2015}. It is interesting to note that low [Cr/Fe] is inconsistent with chemical enrichment from PISNe.

\subsubsection{Mn}
Mn has a metallicity-dependent yield in SNe II. In general, it is thought to be underproduced with respect to Fe in SNe II and overproduced in SNe Ia. Therefore, at [Fe/H] $\gtrsim$ -1 dex in the MW  [Mn/Fe] increases. This trend is observed in the disk, as shown in Figure \ref{fig:fe}. However, this is not observed in our sample or the sample from the HERBS survey \citep{Duong2019b}. Both of these bulge samples show high scatter that is generally centered at [Mn/Fe] $\approx$ 0 dex, although the number of stars with [Mn/Fe] < 0 dex increases with increasing [Fe/H]. This result is interesting and likely indicates inhomogeneous mixing of the ISM or that the bulge is made up of multiple stellar populations with different chemical evolution histories. Of the samples shown in Figure \ref{fig:fe}, the only work to perform NLTE line corrections is \citet{Battistini2015}, shown in light blue open diamonds. 

\subsubsection{Zn}
Zn is produced in core-collapse supernovae with high explosion energy (i.e., hypernovae) and its yields depend strongly on supernova physics. At [Fe/H] $\gtrsim$ -1 dex the [Zn/Fe] ratio decreases with increasing metallicity in our sample as well as literature samples for the disk and bulge \citep{Bensby2014,Bensby2017}. This is consistent with yields from SNe Ia. At [Fe/H] $\lesssim$ -1 dex, our observed [Zn/Fe] ratios are consistent with the EMBLA survey \citep{Howes2015} and the Galactic halo \citep{Roederer2014}. This is also consistent with Galactic chemical evolution models where the large spread in [Zn/Fe] is a result of the metallicity and mass-dependent yields from hypernovae \citep{Kobayashi2011a,Kobayashi2020}.

\subsection{Neutron-Capture Elements}

\begin{figure}
    \centering
    \includegraphics[width=\columnwidth]{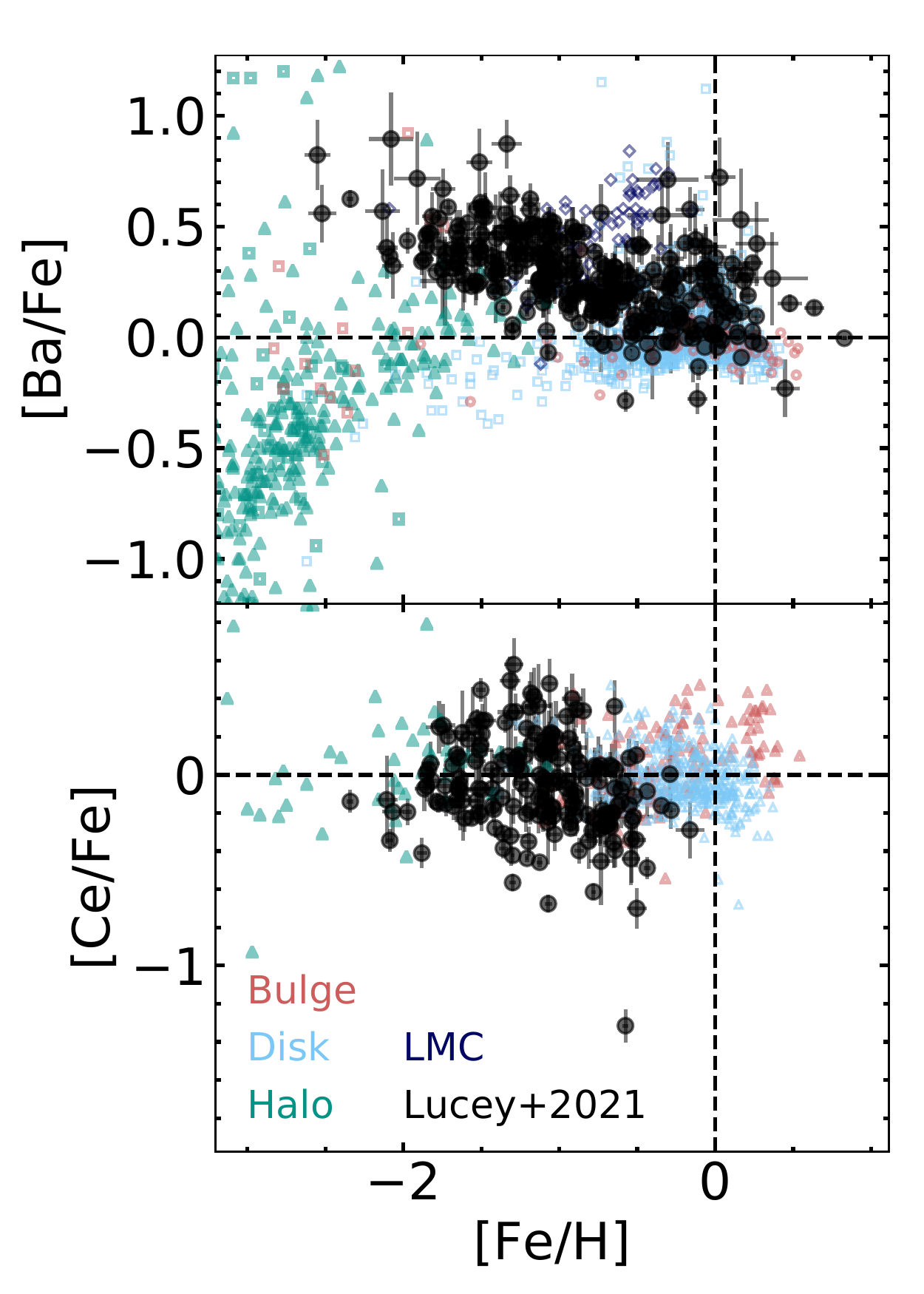}
    \caption{Abundance ratios as a function of metallicity for the neutron-capture elements Ba and Ce. Symbols are the same as in Figure \ref{fig:alpha} and \ref{fig:fe}, with the addition of Ce abundances from \citet{Battistini2016} in red triangles. }
    \label{fig:neutron}
\end{figure} 

Neutron-capture elements are produced through the successive capture of neutrons either through a rapid (r) process or a slow (s) process. In this work, we measure Ba and Ce abundances which are thought to be primarily produced through s-processes, specifically in AGB stars. However, they can both be produced in r-process sites as well \citep{Kobayashi2020}. We show the results for these elements in Figure \ref{fig:neutron}.

\subsubsection{Ba}
Generally, stars in the MW with [Fe/H] $\gtrsim$ -1 dex, show [Ba/Fe] ratios that are roughly solar \citep{Bensby2014,Bensby2017}. In the Galactic halo for stars with [Fe/H] $\lesssim$ -1 dex a large scatter in [Ba/Fe] is observed \citep{Yong2013,Roederer2014}. Nonetheless, the general trend in the halo is [Ba/Fe] decreasing with decreasing metallicity. However, r-process events, like electron-capture (EC) supernovae \citep{Truran1981,Cowan1991}, magneto-rotationally driven (MRD) supernovae \citep{Winteler2012,Nishimura2015}, or neutron star mergers \citep{Rosswog1999}, for example, can enhance the [Ba/Fe] ratio to values > 1 dex \citep{Cescutti2014}. 

The EMBLA survey found that most of their metal-poor bulge stars show a decreasing trend of [Ba/Fe] with decreasing metallicity, and therefore did not find evidence for an r-process event \citep{Howes2015}. However, our sample shows the opposite trend with the [Ba/Fe] ratio increasing at lower metallicities. We note that our survey performs NLTE abundance corrections for Ba while the EMBLA survey does not \citep{Howes2016}. In order to have [Ba/Fe] $>$ 0 and increasing at lower metallicities, it is likely that an r-process event enriched the gas from which these stars formed.

Given the predictions from cosmological simulations that the metal-poor stars in the bulge are ancient \citep{Tumlinson2010,Kobayashi2011a,Starkenburg2017a,El-Badry2018b}, the r-process event which enriched these stars must have occurred on a short timescale. As neutron star mergers are thought to occur on timescales $\gtrsim$ 4 Gyr, it is unlikely that our sample received its r-process material from one of these events. MRD SNe have the shortest timescale, with 1-10\% of stars with 10 $\rm{M_{\odot}}$ $\leq$ M $\leq$ 80 $\rm{M_{\odot}}$ exploding as MRD SNe \citep{Woosley2006,Winteler2012,Cescutti2014}. EC SNe, on the other hand, are thought to occur for all stars with 8 $\rm{M_{\odot}}$ $\leq$ M $\leq$ 10 $\rm{M_{\odot}}$ \citep{Cescutti2013}. \citet{Cescutti2018} demonstrated that the MRD SNe scenario occurs on a fast enough timescale to enhance [Ba/Fe] ratios in metal-poor bulge stars, while the EC SNe scenario does not. 

\subsubsection{Ce}

In the MW, at all metallicities, Ce tracks the Fe abundance, with the [Ce/Fe]$\approx$ 0 dex \citep{Battistini2016,Roederer2014,Duong2019b}. However, at low metallicities in the Galactic halo, there is large scatter in the [Ce/Fe] ratio, similar to [Ba/Fe]. Unlike [Ba/Fe], our stars do not show r-process enhancement in the [Ce/Fe] ratio. Given that the r-/s- process ratio for Ba and Ce are very similar \citep{Simmerer2004}, it is expected that they would be equally enhanced in r-process events and display similar trends with [Fe/H]. However, unlike Ba, we do not perform NLTE abundance corrections for Ce. Therefore, it is possible that NLTE effects may be obscuring a trend in [Ce/Fe] with [Fe/H]. Future work to measure further NLTE neutron-capture abundances is essential for constraining the chemical enrichment history of the metal-poor bulge.

\section{Dynamically Separating the Mixed Stellar Populations} \label{sec:dyn_grp}

Results from COMBS II demonstrate that metal-poor bulge stars ([Fe/H] $<$ -1 dex) are comprised of multiple stellar populations that can be separated dynamically. Specifically, COMBS II separated these stars into a population that stays confined to within 3.5 kpc of the Galactic center throughout their orbits and those that do not. In this work, we go one step further and divide the unconfined population into multiple dynamically defined groups. 

\subsection{Selection Method}

In total, we separate our observed stars into four groups. The groups are defined using the probability of confinement (P(conf.); see Section 4.2 in COMBS II for more details on how this is determined), the apocenter ($r_{apo}$), and the maximum distance from the Galactic plane that the stars reach during their orbit ($z_{max}$). The orbital properties are calculated in COMBS II using GALPY \citep{Bovy2015}. Specifically, we use a Dehnen bar potential \citep{Dehnen2000} generalized to 3D \citep{Monari2016} with parameters designed to match the long, slow bar model put forth by \citet{Portail2017}. The selection of the dynamical groups is described in Table \ref{tab:table1}.
\begin{table*}
\caption{Properties of the Dynamical Groups.}
\label{tab:table1}
\begin{tabular}{ccccccccc}
\hline\hline
Associated Structure & P(conf.) & $r_{apo}$ & $z_{max}$ & Number of Stars & Min. [Fe/H] & Max. [Fe/H] & Mean [Fe/H] \\ 
& & (kpc) & (kpc) & & & &\\
 \hline
 Inner Bulge & >0.5 & & &136  & -2.52 &  0.37 &-0.92 \\
 Outer Bulge & $\leq$0.5 & $\leq$5 &$\leq$ 2.5 & 84  & -2.55 &  0.28 & -0.89 \\
 Halo & $\leq$0.5& & >2.5 & 32 & -1.97 & 0.01 &-1.09 \\
 Disk & $\leq$0.5 & >5 &$\leq$ 2.5 &67 &-2.11 & 0.83  &-0.50  \\

\hline
\end{tabular}
\
\end{table*}

\begin{figure}
    \centering
    \includegraphics[width=\columnwidth]{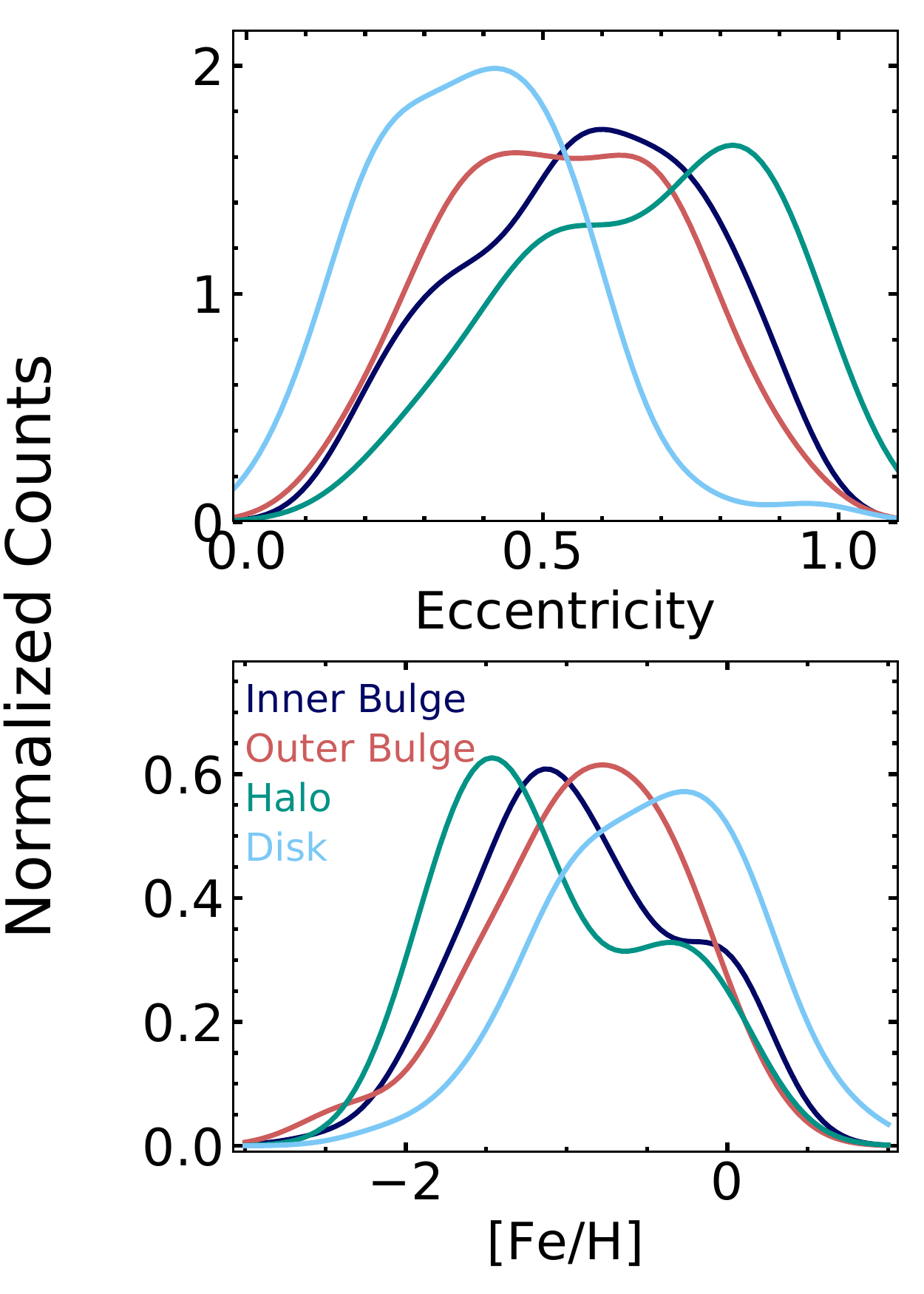}
    \caption{ Properties of the populations we associate with various Galactic structures. In the top plot, we show the eccentricity distributions, while the bottom plot shows the metallicity distribution functions. The inner (136 stars; dark blue) and outer bulge (84 stars; red) have eccentricity distributions consistent with expectations for the Milky Way's  bulge. The halo population (32 stars; green) is consistent with highly eccentric halo stars that pass through the Galactic center region. The disk population (67 stars; light blue) has an eccentricity distribution consistent with the Milky Way disk. The metallicity distribution functions are consistent with expectations given our photometric selection method. Namely, the halo is the most metal-poor component and the disk is the most metal-rich.}
    \label{fig:dyn_grp}
\end{figure} 

We label the groups based on the Galactic structures to which the majority of the stars belong. However, as with most methods of tagging stars to Galactic structures, there is likely contamination since the structures overlap spatially and kinematically \citep{Carrillo2020}. Our inner bulge population is based on an apocenter cut of <3.5 kpc. However, it is now thought that the bar likely extends out to 5 kpc \citep{Wegg2015}. Therefore, we also define an outer bulge population which is likely part of the bulge but does not stay confined to within 3.5 kpc of the Galactic center. To separate the outer bulge and halo population we use a $z_{max}$ cut of 2.5 kpc. This cut is based on the $z_{max}$ distribution of the inner bulge stars. Given the X-shape of the MW bulge, it is possible that the outer bulge is more flared and reaches larger heights above and below the Galactic plane than in the inner regions. This would cause contamination of the halo population by stars belonging to the bulge. However, after visual inspection of many of the orbits of stars in the halo group, the overwhelming majority have Galactic halo-like orbits and clearly do not belong to the bulge. To separate the outer bulge from the disk population, we use an $r_{apo}$ cut of 5 kpc based on the proposed length of the bar from \citet{Wegg2015}. It is important to note that there is certainly a disk population within 5 kpc of the Galactic center which is included in our outer bulge population. However, we are most concerned with simply removing the solar vicinity disk contamination from our sample, rather than selecting bar/bulge stars. Primarily, we aim to compare the metal-poor stars in the inner-most region of the Galaxy to the metal-poor stars in the surrounding regions. Therefore, we mostly focus on the inner bulge, outer bulge and halo populations for the rest of this work.

In Figure \ref{fig:dyn_grp}, we show the  properties of our four groups to confirm that they match expectations for the associated structures. For each group we apply a Gaussian kernel density estimator (KDE) to the eccentricity and metallicity distributions. The eccentricity distributions of the inner and outer bulge populations are very similar, consistent with the stars being different parts of the same structure. Furthermore, our halo population has highly eccentric orbits with a median eccentricity of $\sim$0.67. Lastly, the disk is the least eccentric population, which also matches expectations.

We show the MDFs of the dynamically defined groups in the bottom panel of Figure \ref{fig:dyn_grp}. These distributions are determined using KDEs. It is important to note that we do not expect these MDFs to represent the associated structures, given our photometric selection method. Nonetheless, our results generally match expectations for the given structures and selection method. Specifically, we see that our target selection was generally successful and our inner bulge population peaks at [Fe/H]$\approx$-1 dex. However, there is some contamination by metal-rich bulge stars, as the inner bulge distribution also contains a metal-rich peak at [Fe/H]$\approx$0 dex. The outer bulge distribution is very similar to the inner bulge distribution, although the peak's metallicity is slightly higher. In addition, the outer bulge distribution does not have a second metal-rich peak. Although, this is likely a selection effect.

The halo population has the most metal-poor peak, consistent with results from COMBS II, which found that the fraction of halo interlopers increases with decreasing metallicity. However, it is interesting to note the second metal-rich peak. We confirm these stars have high eccentricity ($e$>$\sim$0.5) and their orbits match expectations for halo stars. However, more precise positional and kinematic data is required to confirm the existence of this metal-rich halo population in the inner Galaxy. The most metal-rich population in our sample is the disk, but it has a large metal-poor tail. This is expected for the disk population, as it is known to have a metal-weak component \citep{Beers2014,Carollo2019}.

\subsection{Distinct Chemical Distributions of Dynamically-Defined Groups}
\begin{figure*}
    \centering
    \includegraphics[width=\linewidth]{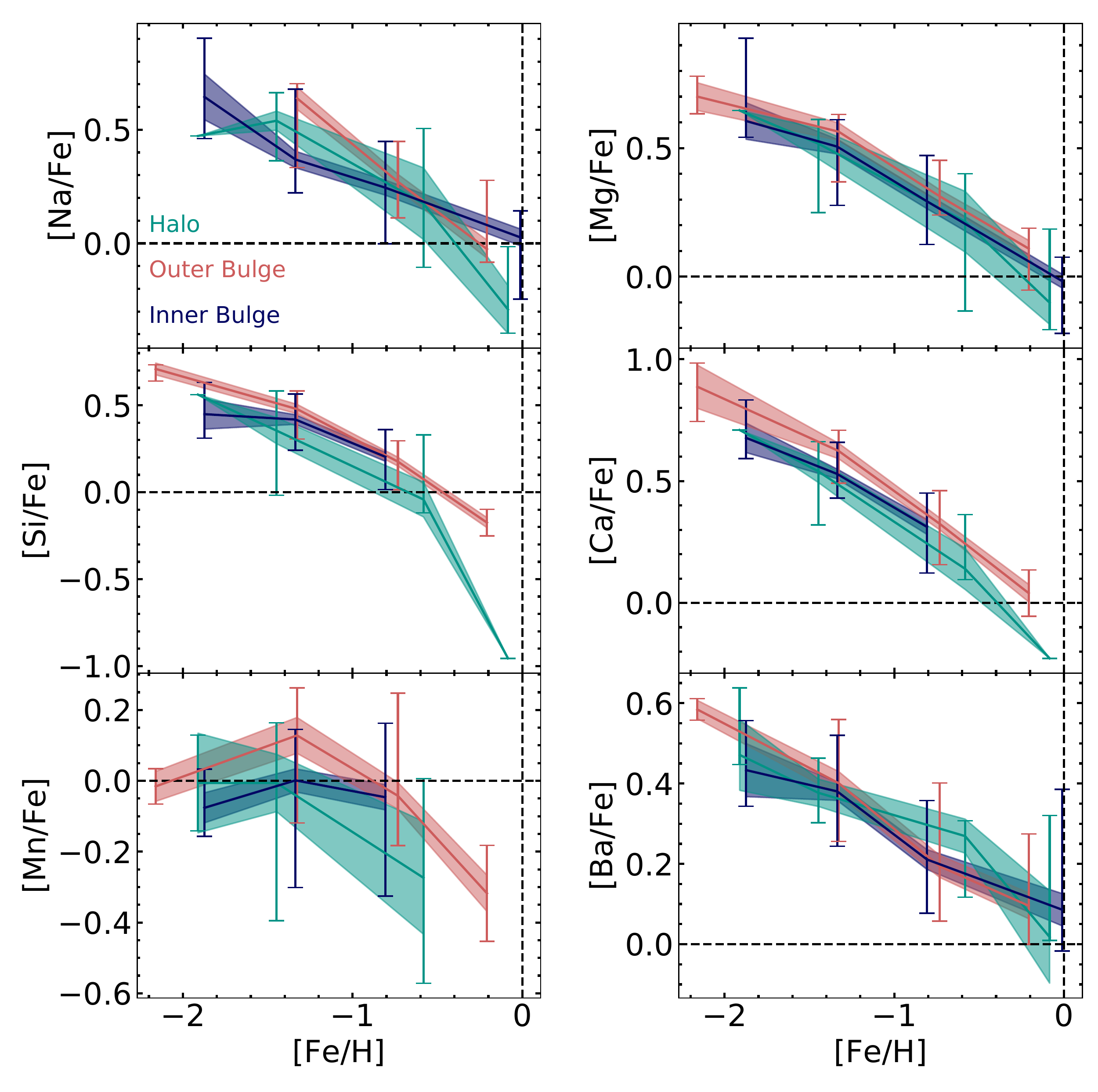}
    \caption{ The distributions of a number of key elements as a function of metallicity for the halo (red), outer bulge (green) and inner bulge (dark blue) samples. Specifically, we show the $\alpha$-elements (Mg, Si and Ca), along with one odd-Z (Na), Fe-peak (Mn) and neutron-capture element (Ba). The lines shown are the median values of the distributions and the error bars correspond to the scatter. We also show the uncertainty on the medians as the shaded regions.}
    \label{fig:alpha_dyn}
\end{figure*}

In addition to dynamical and metallicity differences, the halo, outer bulge, and  inner bulge all show differences in their abundance trends. In Figure \ref{fig:alpha_dyn}, we show the abundance trends for the halo, outer bulge, and inner bulge populations as a function of metallicity for a number of key elements. Specifically, we show the $\alpha$-elements (Mg, Si, and Ca), one odd-Z element (Na), one Fe-peak element (Mn), and one neutron-capture element (Ba). For each population, the lines shown are the median values, and the error bars correspond to the asymmetric 1$\sigma$ spread. We also show the uncertainty on the median as $\sigma/(N-1)^{0.5}$ where $N$ is the number of stars.

The comparison of $\alpha$-element trends between the halo, outer bulge, and inner bulge populations shows a consistent story. At low metallicities ([Fe/H] $\lesssim$ -1 dex), the three populations show similar plateau values. However, the outer bulge consistently has the highest median, followed by the halo and then the inner bulge population. It is interesting to note that the difference between the median $\alpha$-abundance trends at the lowest [Fe/H] is smallest for Mg ($\sim$0.1 dex) which is a hydrostatic $\alpha$-element, while the explosive $\alpha$-elements, Si and Ca, have larger differences ($\sim$0.2-0.3 dex). At this metallicity, the inner bulge population's median [$\alpha$/Fe] is lower than the halo and outer bulge values because of a few stars with especially low [$\alpha$/Fe] ratios. It is important to note the inner bulge population contains stars with [$\alpha$/Fe] values as high as the most $\alpha$-enhanced outer bulge stars, while the halo population does not. 

The halo population's [$\alpha$/Fe] ratio decreases sharply, with the halo becoming the least $\alpha$-enhanced population at -2 dex $\lesssim$ [Fe/H] $\lesssim$ -1 dex. This change might be due to the onset of contributions from  SNe Ia. However, the decreasing trend of [Mn/Fe] cannot be explained by SN Ia enrichment. This trend continues to higher metallicities, with the halo population generally having lower [$\alpha$/Fe] values, indicating a longer star formation duration. On the other hand, the outer and inner bulge populations have very similar [$\alpha$/Fe] distributions at high metallicities indicating similar star formation histories. 

To test the statistical significance of the differences in the distributions, we perform 2D Kolmogorov-Smirnov tests \citep{Peacock1983,Fasano1987}. We perform the test 1000 times sampling the abundances from a Gaussian distribution centered on their measured value with a width corresponding to the uncertainty. We then report the mean p-value of those 1000 test as our final confidence level. The [Ca/Fe] and [Mg/Fe] distributions as a function of [Fe/H] for the outer bulge are different from the halo distributions to a $>$90\% confidence level. However, the inner bulge and halo distributions are not significantly different in [Ca/Fe] or [Mg/Fe], as they both have large scatter. On the other hand, the differences between [Si/Fe] as a function of [Fe/H] distributions for the halo compared to both the inner and outer bulge populations are statistically significant to $>$ 90\% confidence. The inner and outer bulge distributions are not significantly different for any $\alpha$-elements.

The only elements for which the differences between the outer and inner bulge populations are statistically significant is Mn and Na, which both have metallicity-dependent yields in SNe II. At low metallicities ([Fe/H] $\lesssim$ -1 dex), the inner bulge population has lower values in [Mn/Fe] and [Na/Fe] than the outer bulge population. Therefore, the inner bulge stars were generally enriched by a more metal-poor population than the outer bulge stars. This is consistent with results from simulations indicating that more tightly bound stars are older than less tightly bound stars of similar metallicity \citep{Tumlinson2010,El-Badry2018b}.

The difference between [Mn/Fe] and [Na/Fe] as a function of [Fe/H] distributions for the inner bulge and halo populations is not statistically significant. The [Mn/Fe] and [Na/Fe] distributions for the halo population have large scatter with generally lower values than the outer and inner bulge populations. For [Mn/Fe], the halo population starts to decrease at [Fe/H] $\sim$-1.5 dex, consistent with the onset of SNe Ia and a longer star formation timescale than the inner and outer bulge populations. The difference between [Mn/Fe] and [Na/Fe] as a function of [Fe/H] distributions for the halo compared to the outer bulge population is statistically significant to the $>$90\% level.

The [Ba/Fe] distributions of the three groups are surprisingly similar to the [$\alpha$/Fe] distributions at low metallicity. Specifically, we see the same trend in that the three populations all have similar values at the lowest metallicities, but the outer bulge population has the highest level of enhancement, followed by the halo population and then the inner bulge population. This may indicate that the origin of Ba in the low metallicity stars of these populations is similar to the origin of the $\alpha$-elements. However, the similarity to the [$\alpha$/Fe] distribution ceases at higher metallicities where the [Ba/Fe] ratio for the halo is not significantly lower than for the inner and outer bulge populations. In general, the distribution in [Ba/Fe] is much more scattered in the inner and outer bulge populations than in the halo. The differences between the halo distribution of [Ba/Fe] as a function of [Fe/H] and the outer bulge distribution are statistically significant to the $>$90\% confidence level. On the other hand, the outer and inner bulge populations show strikingly similar distributions in [Ba/Fe] as a function of [Fe/H]. This is similar to the Ca, and Mg abundances, providing further evidence that the halo population has a significantly different chemical evolution history than the outer bulge population.

\subsection{Chemical Complexity of Inner and Outer Bulge Compared to Halo Population}

\begin{figure*}
    \centering
    \includegraphics[width=\linewidth]{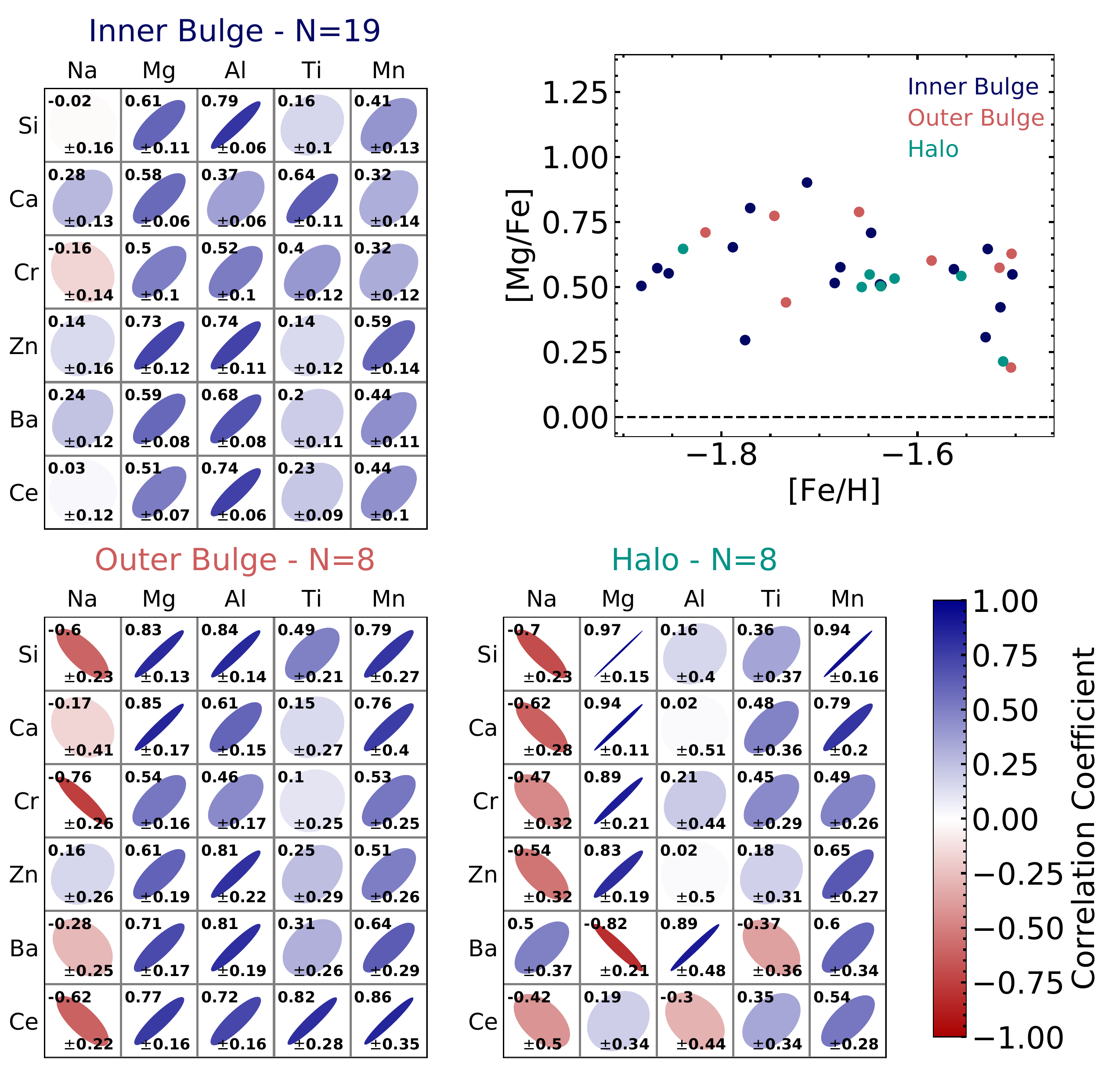}
    \caption{ The Pearson correlation coefficients between a number of key elements for the halo (bottom right), outer bulge (bottom left) and inner bulge (top left) populations. Specifically we compare (Na, Mg, Al, Ti and Mn) to (Si, Ca, Cr, Zn, Ba and Ce). The correlations are calculated using [X/Fe] for stars with -2 dex $\lesssim$ [Fe/H] $\lesssim$ -1.5 dex and SNR> 40 $\rm{pixel^{-1}}$. The top right plot shows the [Mg/Fe] abundances as a function of [Fe/H] for inner bulge (dark blue), outer bulge (green) and halo (red) stars, for reference. To aid in the visualization, we show an oval for each correlation coefficient whose ellipticity, rotation, and color corresponds to the strength and direction of the correlation. We also print the correlation coefficient value in the top left of each correlation box and the corresponding uncertainty on the correlation coefficient in the bottom right. We include the number of stars used in calculating the correlations for the different groups in the title above each correlation plot.  }
    \label{fig:corr}
\end{figure*}

\begin{table}
\caption{Chemical Complexity of the Dynamical Groups.}
\label{tab:table2}
\begin{tabular}{cccc}
\hline\hline
Associated & Mean Absolute  & Variance  & Relative  \\ 
Structure & Correlation & Explained by&  Chemical \\ 
& Strength &  4 Components &  Complexity \\ 
 \hline
 Inner Bulge & 0.36$\pm$0.12 & 96.6$\pm$0.4\% & Highest \\
 Outer Bulge & 0.55$\pm$0.25  & 95.0$\pm$1.0\%  &  \\
 Halo & 0.59$\pm$0.30  & 98.6$\pm$0.5\%  & Lowest
 \\

\hline
\end{tabular}
\flushleft{We define stellar populations that have higher chemical dimensionality and less correlated abundances as being more \textit{chemical complex} than populations with lower dimensionality and more highly correlated abundances. }
\end{table}

In addition to the individual elemental distributions, we also study the correlation between elements and the chemical dimensionality of the inner bulge, outer bulge and halo populations. Through this analysis, we shed light on the diversity of nucleosynthetic events that enriched each population. 

In Figure \ref{fig:corr}, we show the Pearson correlation coefficients for pairs of a number of key elements in the inner bulge, outer bulge, and halo populations. Specifically, we calculate the correlation coefficient for the [X/Fe] values, comparing (Na, Mg, Al, Ti, and Mn) to (Si, Ca, Cr, Zn, Ba, and Ce). The correlation coefficients are calculated using stars with SNR> 40 $\rm{pixel^{-1}}$ and -2 dex $<$[Fe/H] $<$ -1.5 dex, in order to isolate yields from core-collapse supernovae and limit the impact of metallicity on the correlations. In the top left, we show the results for the inner bulge population which uses 19 stars, while the outer bulge is shown in the bottom left, using 8 stars. We also show results for the halo population on the bottom right, using 8 stars. The top right plot shows [Mg/Fe] as a function of [Fe/H] for the stars used in the correlation plots, as a reference. This figure demonstrates that the stars used for the halo (red), outer bulge (green), and inner bulge (dark blue) populations span similar metallicity ranges. In the three correlation plots, each small box contains an ellipse whose eccentricity and color corresponds to the strength of the correlation. In addition, we print the correlation coefficient in the top right corner, with the corresponding uncertainty on the coefficient in the bottom right corner. 

The uncertainties are calculated using a bootstrap method in order to propagate the impact of the abundance uncertainties and the limited number of stars. To account for the abundance uncertainties, we recalculate the correlation coefficient 1000 times with new abundance values each time. These values are randomly selected from Gaussian distributions that are centered on the measured abundance value with a width equivalent to the uncertainty. We then define the correlation coefficient's uncertainty due to abundance uncertainties as the median of the differences between the original correlations and the recalculated values. Similarly, to account for the limited number of stars, we recalculate the coefficients N times dropping out 1 star from the sample each time. Again, we use the median of the differences between the original correlations and the recalculated values as the uncertainty due to the limited number of stars. We then add the uncertainties due to the number of stars and the uncertainties due to the abundance uncertainties in quadrature for our total uncertainty values. We note that the uncertainties due to the abundance uncertainties are dominant with the uncertainties due to the number of stars being on the order of $\sim$0-0.01. 

Overall, the inner bulge population shows the weakest correlations, followed by the outer bulge population and then the halo. As they are all $\alpha$-elements, it is expected for Mg, Si, and Ca to be tightly correlated. This is observed in the halo population, but the correlations are weaker in the outer and inner bulge populations. This is especially interesting given that PISNe yields have [Ca/Mg] and [Si/Mg] abundance ratios that are mass-dependent. Therefore, yields from PISNe of varying masses would cause the correlation between Mg, Si, and Ca to weaken and become noisier as seen in the inner and outer bulge populations. 

Furthermore, the correlation of Ca and Si with Mn in the halo is strikingly strong. This is surprising given that Mn is thought to have metallicity-dependent yields in SNe II while Si and Ca do not. This may indicate that the halo stars were enriched by a population with a narrow metallicity range. This correlation becomes sequentially weaker as we move to more tightly bound stars in the outer and inner bulge populations. In general, the inner bulge population only shows weak correlations with Mn. In addition, the negative correlation of Na with Si, Ca, and Zn are significant in the halo, but almost completely disappear in the outer bulge population and are non-existent in the inner bulge population. Similar results are found for the positive Na to Ba correlation.

Excluding Al, which is discussed later, the mean of the absolute correlations of the abundance pairs shown in Figure \ref{fig:corr} is 0.59$\pm$0.30 for the halo population, while the means for the outer and inner bulge populations are 0.55$\pm$0.25 and 0.36$\pm$0.12, respectively. The uncertainties on the mean are determined by recalculating the mean using the correlation strengths with the individual correlation uncertainties added/subtracted.  Therefore, we find evidence that the elemental abundances in the halo population are generally more correlated than in similar metallicity stars in the inner bulge populations. This result may indicate a less diverse chemical enrichment history in the halo population as compared to the inner bulge.

Furthermore, the abundance pairs which do not follow the above trend can provide interesting insight into the possible differences between the chemical enrichment histories of these populations. For example, the Al abundances show the opposite trend in that the inner and outer bulge populations have strong correlations while the Al abundances for the halo population are generally not correlated with any elements, except for Ba . This is especially difficult to interpret given that Na and Al are thought to be produced in similar ways, but the inner bulge population does not show strong correlations for Na with any elements. This result solicits further investigation into possible nucleosynthetic sites, beyond SNe II, for Al in the inner bulge population. 

Another striking difference between the outer bulge, inner bulge, and halo populations is the strength of the positive Mg, Ba, and Ce correlations in the outer and inner bulge populations, while the halo population shows negative or weak correlations. This is further evidence for the similar origin of Ba and $\alpha$-elements at low metallicities in the inner and outer bulge populations. Specifically, this result further supports MRD SNe as the origin for Ba in these ancient stars \citep{Kobayashi2020}. MRD SNe produce high levels of Ba and Ce as well as $\alpha$-elements \citep{Yong2013}. However, further work analyzing and comparing MRD SNe theoretical yields to the observed abundances are required.

Ti is another interesting case that does not match the general trend of strong correlations in the halo and weak correlations in the inner bulge population. Specifically, Ti generally does not have significantly strong correlations with any element except for with Ca in the inner bulge population. We note that the outer bulge shows a somewhat strong correlation between Ti and Ce, however, the uncertainty is large at 0.28. It is possible that the Ca and Ti correlation in the inner bulge population is insignificant, but it may also indicate an interesting origin for some of the Ca in this population. 

In addition to the correlation analysis, we perform a chemical dimensionality analysis to further explore the differences in the inner bulge, outer bulge and halo populations. Specifically, we perform Principal Component Analysis (PCA) on the elemental abundances for each population. Essentially, PCA sequentially finds orthogonal components which explain the most variance in the given data. For the analysis, we include all stars in each population with SNR> 40 $\rm{pixel^{-1}}$, [Fe/H]$<$-1 dex, and a complete set of elemental abundances for Na, Mg, Al, Si, Ca, Ti, Cr, Mn, Zn, Ba and Ce. Similar to the correlation analysis, we choose to only focus on metal-poor stars ([Fe/H] <-1 dex) in this analysis to limit the impact of SNe Type Ia. However, since the PCA analysis requires a complete set of abundances, we are left with only 7 stars from the halo population, 11 stars from the outer bulge population and 20 stars from the inner bulge population, even though the metallicity range used is larger than for the correlation analysis. To account for the uncertainties in the abundances, we perform the PCA analysis 1000 times with new abundances sampled from a normal distribution centered on the measured abundance with a width corresponding to the abundance uncertainty.

To explore the comparative dimensionality of the elemental abundances, we investigate the percentage of variance explained by each component derived from the PCA. Consistently, we find that for the same number of components, a higher percentage of the variance in the halo population is explained compared to the inner and outer bulge populations. For example, 92.2$\pm$1.7\% of the variance is explained by 2 components in the halo population while only 85.7$\pm$2.6\% and 91.3$\pm$1.0\% is exaplined in the outer and inner bulge populations, respectively. Furthermore, when using 4 components, 98.6$\pm$0.5\% of the variance in the halo is explained, while only 95.0$\pm$1.0\% and 96.6$\pm$0.4\% of the variance is explained in the outer and inner bulge populations, respectively. Therefore, we find evidence that the halo population has lower chemical dimensionality than the inner and outer bulge populations.

To describe the combination of our correlation and dimensionality analysis, we define a new term: \textit{chemical complexity}. In total, we find that the elemental abundances in the halo population are highly correlated with a mean correlation of 0.59$\pm$0.30, while the mean for the outer and inner bulge populations are 0.55$\pm$0.25 and 0.36$\pm$0.12, respectively. Furthermore, we found that the halo population has lower chemical dimensionality than the inner and outer bulge populations. Specifically, when using 4 components, only 96.6$\pm$0.4\% and 95.0$\pm$1.0\% of the variance  in the elemental abundances is explained for the inner and outer bulge populations while the same is true for 98.6$\pm$0.5\% of the elemental abundance variance in the halo population. \textit{ Therefore, we describe the highly-correlated, lower dimensional halo population as less chemically complex compared to the inner and outer bulge populations whose elemental abundances are less correlated and have higher dimensionality.} This measure of relative chemical complexity is indicative of the diversity of chemical enrichment events. Therefore, we suggest that the inner and outer bulge populations have a higher diversity of enrichment events compared to the halo population. However, it is important to note that these results may also be impacted by the rate of mixing in the ISM at the different formation times. Specifically, higher chemical complexity could also indicate a less well-mixed ISM.

In total, we discover a number of key results from our comparison between the abundances of the inner bulge, outer bulge, and halo populations. First, we find that the inner and outer bulge populations have shorter star formation timescales and more rapid chemical evolution than the halo population. In addition, our results solicit further investigation into the nucleosynthetic origins of Ba and Al in metal-poor inner bulge stars. Furthermore, we find that the abundances are consistent with the inner bulge being the oldest population, compared to the outer bulge and halo populations. We also find that at low-metallicity, the inner bulge is the most chemically complex population, followed by the outer bulge and then halo population. Combined, these results suggest that older bulge populations are more chemically complex. This may be due to a combination of diversity of chemical enrichment events (e.g., PISNe, EC SNe, MRD SNe, and other SNe predictions for Population III stars), as well as inhomogeneous mixing of the ISM. 

\section{Pair-Instability supernovae signatures} 
\label{sec:pisne}
\begin{figure}
    \centering
    \includegraphics[width=\columnwidth]{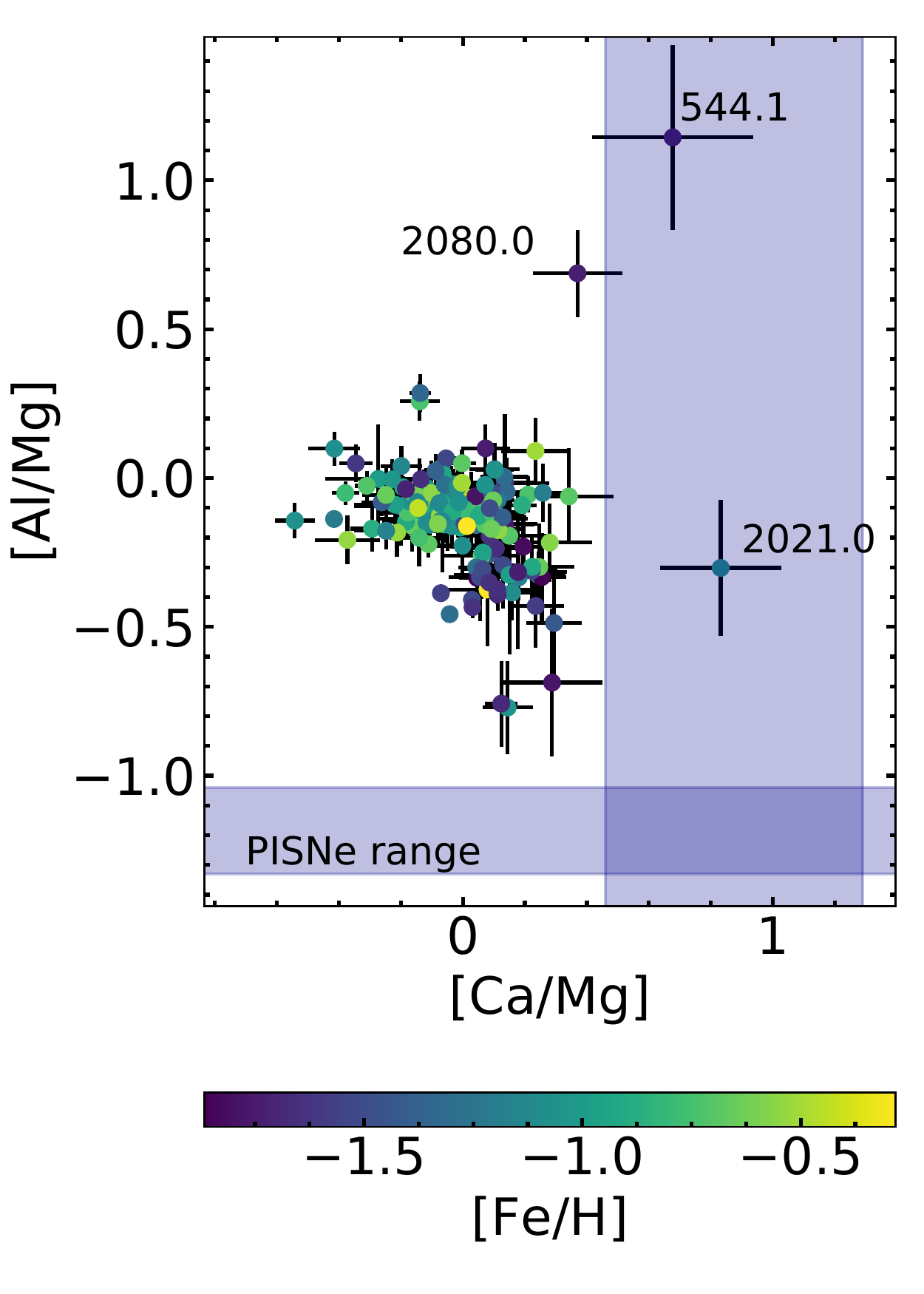}
    \caption{ The [Al/Mg] ratios as a function of [Ca/Mg] for our sample stars, colored by [Fe/H]. We shade the regions corresponding to simulated PISNe yields \citep{Takahashi2018}. We have two stars (544.1 and 2021.0) with [Ca/Mg] ratios consistent with PISNe predictions, but their [Al/Mg] ratios are significantly higher than predictions. }
    \label{fig:pisne}
\end{figure} 

\begin{figure}
    \centering
    \includegraphics[width=\columnwidth]{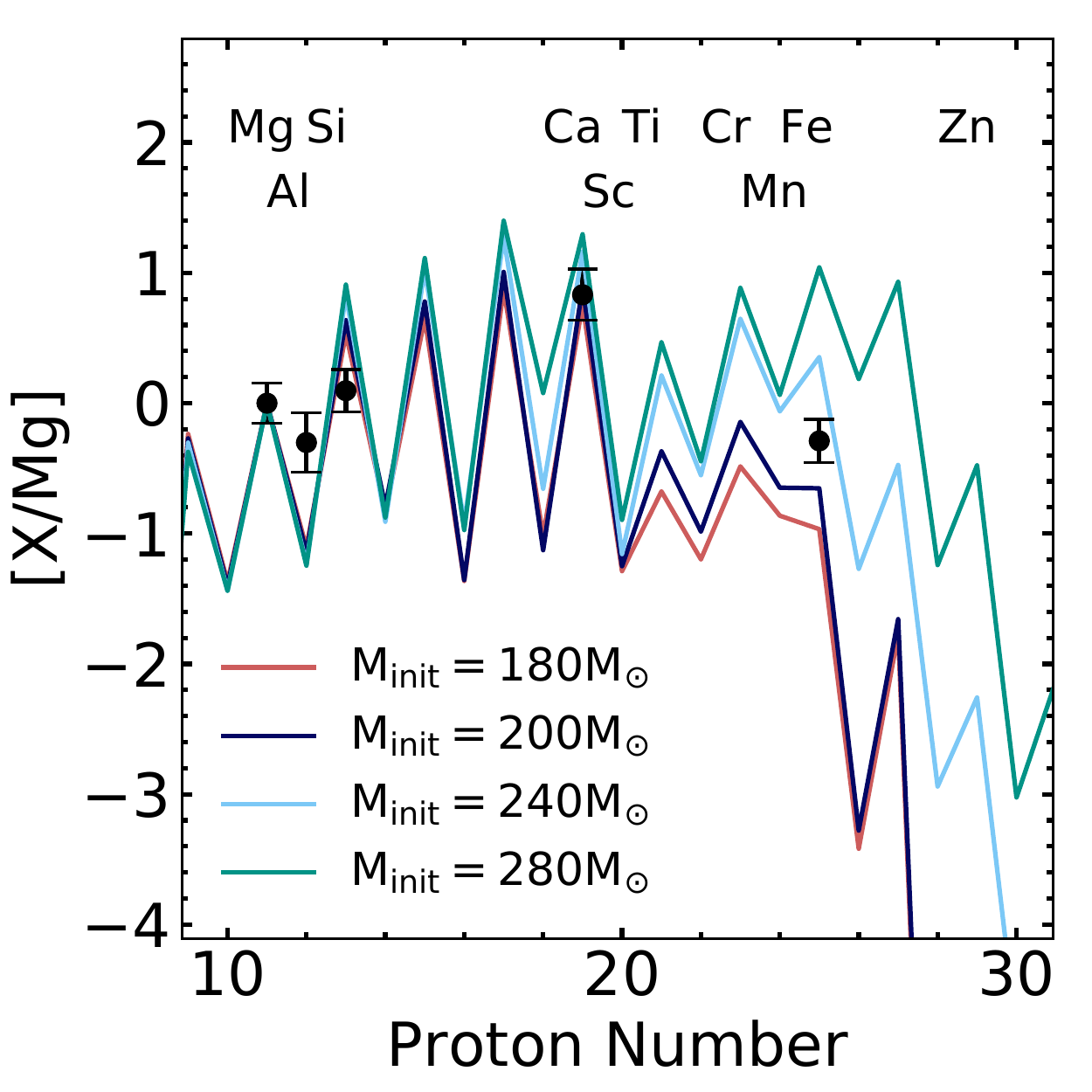}
    \caption{ The [X/Mg] ratios for star 2021.0 compared to non-rotating model PISNe yields of various initial masses from \citet{Takahashi2018}. We are only able to measure Mg, Al, Si, Ca, and Fe for this star as it has SNR=21 $\rm{pixel^{-1}}$. The [Ca/Mg] and [Fe/Mg] ratios match PISNe signatures, but the [Al/Mg] and [Si/Mg] ratios do not.  }
    \label{fig:pisne_2}
\end{figure}

PISNe are highly energetic thermonuclear explosions that occur after the hydrodynamical collapse caused by electron-positron pair production in massive (> 25 $\rm{M_{\odot}}$) CO cores \citep{Barkar1967,Rakavy1967}. It is predicted that $\sim$25\% of the first stars would explode as PISNe \citep{Hirano2015}. Therefore, it is expected that $\sim$ 1/400 stars with [Ca/H] < -2 dex would be enriched by a PISN \citep{Takahashi2018}. However, chemical signatures of PISNe in studies of metal-poor stars have been elusive. A number of candidates have been put forward, but none perfectly match the predicted abundance trends from simulations \citep{Takahashi2018}. Whether the lack of PISNe chemical signature detections is an observational effect or the result of incorrect simulated rates and yields is yet to be determined. 

In Figure \ref{fig:pisne}, we plot our sample's abundances with respect to the predicted PISNe abundance trends from \citet{Takahashi2018}. Specifically, we plot abundance ratios that are thought to be especially discriminatory in PISNe yields: [Al/Mg] and [Ca/Mg]. The range of predicted PISNe abundance yield ratios is shown in the blue-shaded regions. The measured abundances of our sample are shown in points colored by their metallicity, with corresponding uncertainties as black error bars. We discover two stars that have [Ca/Mg] ratios consistent with predictions for PISNe yields. However, neither of these stars has a consistent [Al/Mg] ratio. Out of the two stars (544.1 and 2021.0) with high enough [Ca/Mg] ratios, we focus on 2021.0 which has an [Al/Mg] ratio closest to PISNe yield predictions.

Star 2021.0 has [Fe/H]=-1.07 dex and [Ca/H] = 0.05 dex. We show all measured abundances for this star in Figure \ref{fig:pisne_2}. Specifically, we show [Mg/Mg], [Al/Mg], [Si/Mg], [Ca/Mg] and [Fe/Mg]. We also attempted to measure the other elements presented in this work, but the HR06 spectrum has only SNR=21 $\rm{pixel^{-1}}$, making many of the elements difficult to measure reliably. In Figure \ref{fig:pisne_2}, we also plot predicted PISNe yields for non-rotating models with various initial masses from \citet{Takahashi2018}. Our [Ca/Mg] and [Fe/Mg] ratios match predictions for PISNe, but our [Al/Mg] and [Si/(Mg,Ca,Fe)] ratios do not. In fact, our measured [Si/(Mg,Fe)] ratio is rather consistent with normal core-collapse supernovae. Furthermore, the metallicity of 2021.0 is higher than expectations for a Population II star which was enriched solely by a single PISNe \citep{Karlsson2008}. Therefore, it is possible that this star was enriched by a PISNe along with an SNe II. However, further observations are needed to measure more elemental abundances in this star to confirm the PISNe signature. It is also important to note that that star 2021.0 is part of the inner bulge population with a P(conf.)=0.93. Furthermore, star 2021.0 is tightly bound with a pericenter of 0.49 kpc, $r_{apo}$= 2.31 kpc and $z_{max}$=1.15 kpc.  Therefore, it is likely that this star formed in the first few Gyrs of star formation in the Universe. However, asteroseismology is required to further constrain its age.

\section{Globular Cluster Origin}
\label{sec:glob}
\begin{figure}
    \centering
    \includegraphics[width=\columnwidth]{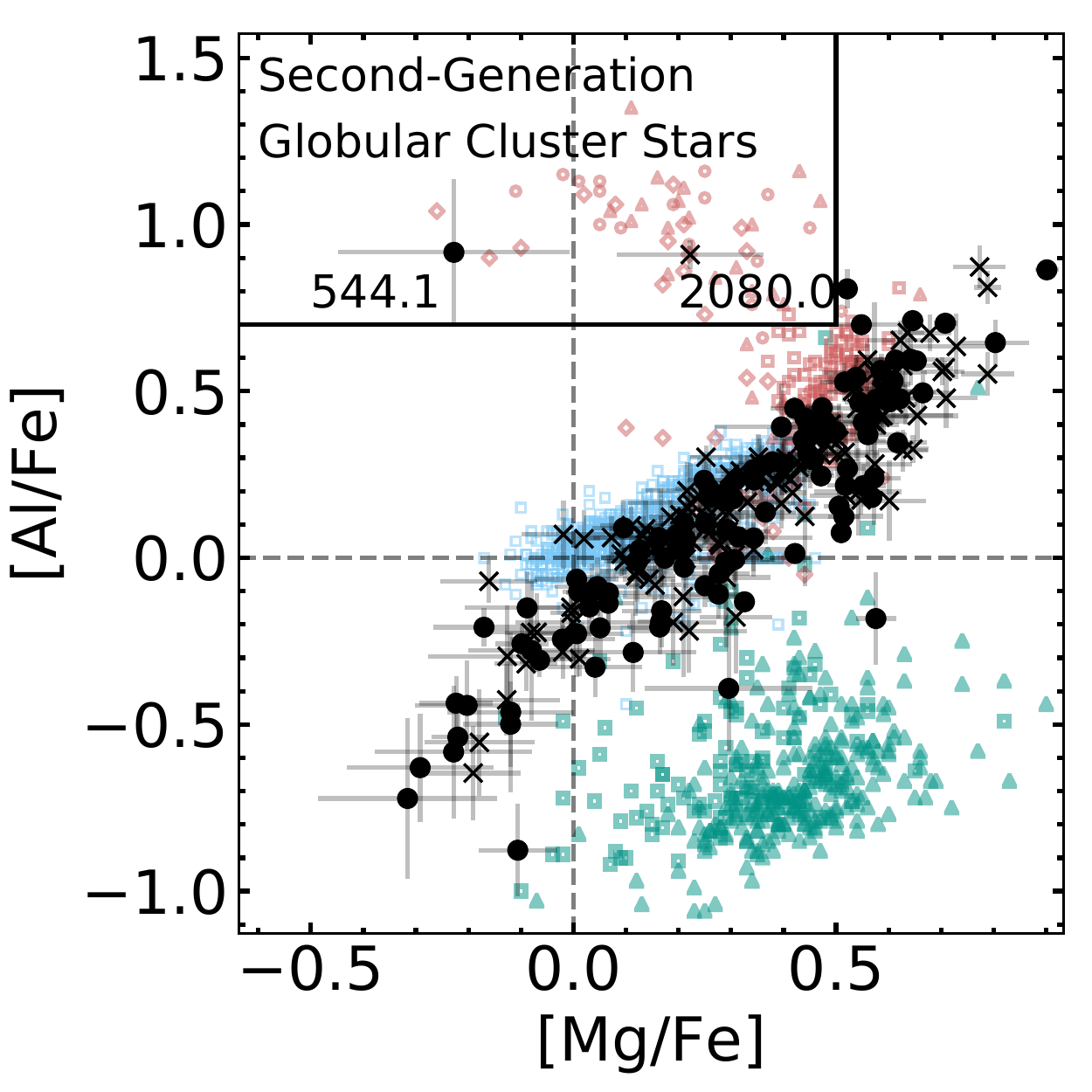}
    \caption{The [Al/Fe] abundance ratios as a function of [Mg/Fe] for our sample. We show confined bulge stars (P(conf.)>0.5) as black points and unconfined (P(conf.)$\leq$0.5) stars as black crosses. We also show halo \citep[][green open triangles and squares, respectively]{Roederer2014,Yong2013}, disk \citep[][light blue open squares]{Bensby2014}, and globular cluster literature samples for comparison. Specifically, we show NGC~4833 (red open circles), NGC~7089 (red open triangles), and NGC~2808 (red open diamonds) from \citet{Pancino2017}, along with NGC~6121 (red open squares) from \citet{Marino2008}. We have two stars (544.1 and 2080.0) that have chemistry consistent with second-generation globular cluster stars. }
    \label{fig:glob}
\end{figure}

Recent work suggests that the metal-poor bulge may be at least partially built up by dissipated globular clusters \citep{Kruijssen2015,Shapiro2010,Bournaud2016}. To date, a significant number of stars in the bulge with chemistry consistent with globular clusters have been detected \citep{Schiavon2017,Fernandez2017,Lucey2019}. Specifically, the chemical signatures encountered include nitrogen enhancement and the Al-Mg and Na-O anti-correlations which are signatures of second-generation globular cluster stars \citep{Gratton2004}. However, the rate at which these stars occur among metal-poor bulge stars and whether they are confined stars as opposed to interloping halo stars is yet to be determined. 

In this work, we find two stars (544.1 and 2080.0) that have enhanced Al with respect to their Mg abundances. In Figure \ref{fig:glob}, we show the [Al/Fe] abundances as a function [Mg/Fe] for our sample. The confined bulge population is shown as black points while the unconfined stars are shown as black crosses. We also show a number of surveys from the literature for comparison. Specifically, we show halo samples from \citet[][green open triangles]{Roederer2014} and \citet[][green open squares]{Yong2013}, along with a disk sample from \citet[][light blue open squares]{Bensby2014}. In addition, we show abundances from the globular clusters NGC~4833 (red open circles), NGC~7089 (red open triangles), and NGC~2808 (red open diamonds) from \citet{Pancino2017}, along with NGC~6121 (red open squares) from \citet{Marino2008}. The stars 544.1 and 2080.0 match trends seen in globular clusters and have much higher [Al/Fe] ratios compared to other stars in our sample with similar [Mg/Fe]. Star 2080.0 is unconfined and belongs to the disk dynamical group. However, its orbit is unusual for a disk star, with a pericenter of 0.19 kpc, $r_{apo}$=8.6 kpc and $z_{max}$=0.90 kpc. On the other hand, star 544.1 is on a typical inner bulge orbit with a pericenter of 0.77 kpc, $r_{apo}$=2.79 kpc and $z_{max}$=1.95 kpc. It is especially interesting to note that 544.1 specifically matches abundance trends from NGC~2808, one of the MW's most massive globular clusters, that is theorized to be part of the Gaia-Enceladus system \citep{Myeong2018}.

\citet{Schiavon2017} and \citet{Horta2021} estimate that $\sim$25\% of the stellar mass in the inner 2 kpc of the Galaxy are disrupted globular cluster stars, assuming nitrogen-rich stars are second-generation globular cluster stars. Given these results, it is expected that more than 2 out of our 241 stars with Al and Mg measurements would be second-generation globular cluster stars and would therefore show the Al-Mg anti-correlation. However, it is unclear whether all second-generation globular cluster stars can be detected using the Al-Mg anti-correlation. To perform an apples-to-apples comparison, further observations are required to determine the fraction of N-rich stars in our sample. Another possible explanation for the apparent lack of second-generation globular cluster stars in our sample could be the result of SkyMapper photometry for target selection. It is known that photometric selection of metal-poor stars using SkyMapper can be biased against selecting C-enhanced stars \citep{DaCosta2019}. As nitrogen-rich stars can frequently be C-rich \citep{Horta2020}, we may be biased against selecting nitrogen-rich stars and therefore second-generation globular cluster stars as well. Last, our results could be discrepant with estimates from \citet{Schiavon2017} and \citet{Horta2021} due to their assumption that all N-rich stars are disrupted globular cluster stars which would lead to an overestimate of the contribution of dissipated globular clusters to the stellar mass. Other origins for N-rich stars have been suggested \citep[e.g.,][]{Bekki2019}, which could explain the high fraction of N-rich stars in the inner Galaxy.

\section{Summary and Conclusions}
\label{sec:conclu} 
The Galactic bulge is a complex structure with overlapping stellar populations which can provide crucial information about the formation and evolution of the MW. The metal-poor population in the bulge is of special interest given that cosmological simulations predict they are some of the oldest stars in the Galaxy \citep{Salvadori2010,Tumlinson2010,Kobayashi2011a,Starkenburg2017a,El-Badry2018b}. However, this population has historically been difficult to study given that it compromises only $\sim$5\% of bulge stars. Now, with the recent advent of metallicity-sensitive photometric surveys \citep{Starkenburg2017b,Wolf2018,Casagrande2019}, we can target metal-poor stars in the Galactic bulge and study them in large numbers \citep{Arentsen2020b}. Recent work has determined that many metal-poor stars found in the bulge are actually halo interloping stars  \citep{Kunder2020,Lucey2021}. Therefore, it is necessary to combine dynamical and chemical information to disentangle the metal-poor bulge population and study its origins in detail.

In this work, we successfully target metal-poor stars in the Galactic bulge using Skymapper photometry and observe 555 stars with the VLT/GIRAFFE spectrograph. We report stellar parameters and abundances for 319 which have SNR > 20 $\rm{pixel^{-1}}$ and astrometry from \gaia\ DR2. The stellar parameters and abundances are determined using a $\chi^2$ fit to model spectra synthesized using SME with NLTE departure coefficients for Li, O, Na, Mg, Al, Si, Ca, Ti, Fe, and Ba \citep{Valenti1996,Piskunov2016}. We compare our stellar parameters to results for \gaia\ Benchmark stars \citep{Heiter2015,Jofre2014, Hawkins2016a}, the ARGOS survey \citep{Freeman2013}, and the HERBS survey \citep{Duong2019} and find they are generally consistent. We report elemental abundances for C, Na, Mg, Al, Si, Ca, Ti, Cr, Mn, Fe, Zn, Ba, and Ce.  Using results from \citet{Lucey2021}, which defines confined bulge stars as those with apocenters $<$3.5 kpc, we divide our sample into 5 groups based on their dynamics. We associate these groups with different Galactic structures. Specifically, we label them as the inner bulge population, the outer bulge, the halo, and the thick and thin disks.

Given these data we find evidence that:
\begin{enumerate}
    \item The halo stars which pass through the inner Galaxy have relatively low chemical complexity compared to the inner and outer bulge populations. Specifically the elemental abundances are highly correlated (mean correlation coefficient of 0.57) and have lower dimensionality (98.0\% of variance explained by 4 components) than the inner and outer bulge populations.   This may indicate that these halo stars formed in-situ in the disk and were heated to halo kinematics by a merger event \citep[e.g.,][]{Di_Matteo2019}. However, it is also possible these stars formed ex-situ in a single chemically simple progenitor or in many progenitors but the universe, in general, was less chemically complex. The abundances for this halo population are consistent with a longer star formation timescale when compared to the outer and inner bulge population.
    
    \item The outer bulge population is very similar to the inner bulge population, although more chemically correlated (mean correlation coefficient of 0.53). The outer bulge is significantly distinct from the inner bulge stars in only its Na and Mn abundances, which both have metallicity-dependent yields in SNe II. Given that Na and Mn yields are higher in more metal-rich stars, it is likely the outer bulge stars were enriched by a more metal-rich population than the inner stars. This result is consistent with predictions from cosmological simulations that the outer bulge is younger than the inner bulge population \citep{Tumlinson2010}. 
    
    \item The confined (or inner) bulge population is more chemically complex (mean correlation coefficient of 0.38 and 96.6\% of the variance explained by 4 components) than the unconfined stars in the bulge. Additionally, results from simulations predict that confined, or tightly bound stars are generally older than unconfined, loosely bound stars \citep{Tumlinson2010}. Combined, these results indicate that older populations in the inner Galaxy are generally more chemically complex than younger populations of similar metallicity. Furthermore, this suggests that the universe is more chemically complex early on indicating either more diversity in chemical enrichment events or inhomogeneous mixing in the ISM. 
    
    \item We also find evidence that the Ba in low-metallicity ([Fe/H]$\lesssim$-1 dex) outer and inner bulge populations have similar origins to the $\alpha$-elements. Specifically, we find the Ba and $\alpha$-element abundances are positively correlated in these populations, but negatively correlated in the halo population.
    
    \item In our inner bulge population, we find one star that may show a signature of PISNe. Explicitly, this star has [Fe/H]=-1.07 dex and [Ca/Mg]=0.83 dex, but its [Al/Mg] ratio (-0.30 dex) is higher and its [Si/(Mg,Ca,Fe)] ratio is lower than expected for PISNe yields. Further observations are needed to measure more chemical abundances for this star and compare them to model PISNe yields. 
    
    \item  We detect 2 stars whose chemistry is consistent with second-generation globular cluster stars in that they show the signature Mg-Al anti-correlation. One of these stars belongs to the inner bulge population while the other belongs to the outer bulge population. It is especially interesting that one star has a [Al/Mg] ratio that is similar to what is observed for NGC~2808 which is theorized to be the core of the Gaia-Enceladus system \citep{Myeong2018}. 
\end{enumerate}

In total, this work demonstrates the power and necessity of combining chemistry with dynamics to disentangle and separately study the metal-poor stellar populations in the bulge region. In future work,  we hope to achieve more precise dynamical parameters with improved astrometry from further \gaia\ data releases and the use of spectro-photometric distances.  In addition, our work signifies the need for more measurements of neutron-capture elements in metal-poor bulge stars. However, further work on nucleosynthetic yields of Ba from Population II and III stars is needed to understand the correlation between Ba and $\alpha$-elements observed in this work. Furthermore, more precise photometry in the bulge is required to continue to target the most metal-poor bulge stars in large numbers with lower contamination rates from metal-rich stars.

\appendix
\section{Online Table}
We show a section of the available online table in Table \ref{tab:table}. The online table includes all 319 stars for which we measure stellar parameters and elemental abundances in this work and have derived orbital properties from COMBS II. In the table, we include all measured properties of the stars, including the SNRs, positions, velocities, dynamical properties, stellar parameters, and elemental abundances for every atomic line used in this work, along with the associated uncertainties. 

\begin{table*}
\caption{Stellar Parameters, Elemental Abundances, Dynamical and Observed Properties.}
\label{tab:table}
\begin{tabular}{cccccccccccccc}
\hline\hline
Object & l & b & $\rm{SNR_{06}}$ & $\rm{T_{eff}}$ & \logg\ & [M/H] & [Fe/H] & $\rm{Fe_{scatter}}$ & [Mg/H] & $\rm{Mg_{scatter}}$ & ... \\
 &  (deg) & (deg) & $\rm{pixel^{-1}}$  & (K) &  &  &  &  &  &  &  \\
 \hline
10078.0 & 15.45 & -9.67 & 74.81 & 4774$\pm$84 & 1.92$\pm$0.3 & -1.61$\pm$0.17 & -1.29$\pm$0.04 & 0.06 & -1.36$\pm$0.12 & 0.0 & ...\\
10123.1 & 355.93 & -10.15 & 20.27 & 6137$\pm$127 & 4.0$\pm$0.36 & -0.07$\pm$0.2 & 0.16$\pm$0.18 & 0.0 & -0.15$\pm$0.17 & 0.0 & ...\\
1017.0 & 0.1 & -9.5 & 100.19 & 4921$\pm$83 & 2.0$\pm$0.3 & -1.29$\pm$0.17 & -1.09$\pm$0.03 & 0.07 & -0.81$\pm$0.06 & 0.01 & ...\\
10205.0 & 15.45 & -9.64 & 39.35 & 4773$\pm$90 & 2.75$\pm$0.31 & -1.12$\pm$0.18 & -1.31$\pm$0.09 & 0.04 & -0.74$\pm$0.07 & 0.02 & ...\\
1023.0 & 0.03 & -9.54 & 60.15 & 4946$\pm$84 & 2.5$\pm$0.3 & -0.65$\pm$0.17 & -0.43$\pm$0.04 & 0.03 & -0.27$\pm$0.04 & 0.05& ... \\
10272.0 & 15.44 & -9.63 & 35.09 & 5422$\pm$94 & 3.81$\pm$0.31 & -0.44$\pm$0.18 & -0.56$\pm$0.1 & 0.01 & -0.21$\pm$0.07 & 0.0 & ...\\
10331.0 & 15.45 & -9.65 & 53.55 & 4769$\pm$85 & 2.33$\pm$0.3 & -0.63$\pm$0.17 & -0.52$\pm$0.05 & 0.08 & -0.3$\pm$0.04 & 0.08 & ...\\
1036.0 & 0.06 & -9.53 & 61.26 & 5142$\pm$84 & 2.25$\pm$0.3 & -0.96$\pm$0.17 & -0.5$\pm$0.06 & 0.03 & -0.52$\pm$0.1 & 0.0& ... \\
10397.0 & 15.47 & -9.66 & 23.7 & 4699$\pm$114 & 3.72$\pm$0.34 & -0.53$\pm$0.19 & -0.82$\pm$0.08 & 0.05 & -0.23$\pm$0.14 & 0.11& ... \\
1070.0 & 359.93 & -9.61 & 54.5 & 4773$\pm$85 & 2.18$\pm$0.3 & -1.43$\pm$0.17 & -1.43$\pm$0.07 & 0.1 & -1.12$\pm$0.07 & 0.09 & ...\\
1080.0 & 0.11 & -9.53 & 53.06 & 4660$\pm$85 & 2.25$\pm$0.3 & -0.88$\pm$0.17 & -0.87$\pm$0.05 & 0.01 & -0.49$\pm$0.04 & 0.05 & ...\\
10859.0 & 15.44 & -9.58 & 23.68 & 5176$\pm$114 & 4.5$\pm$0.34 & -0.16$\pm$0.19 & -0.36$\pm$0.14 & 0.0 &  & & ... \\
10896.0 & 15.51 & -9.72 & 37.13 & 4767$\pm$92 & 2.5$\pm$0.31 & -1.22$\pm$0.18 & -1.33$\pm$0.1 & 0.09 & -0.82$\pm$0.09 & 0.05& ... \\
1097.1 & 14.36 & -9.55 & 68.68 & 4908$\pm$84 & 2.25$\pm$0.3 & -1.0$\pm$0.17 & -0.74$\pm$0.04 & 0.02 & -0.55$\pm$0.04 & 0.06& ... \\
10985.0 & 15.48 & -9.64 & 22.03 & 6499$\pm$120 & 3.85$\pm$0.35 & -0.28$\pm$0.19 &  && & & ... \\
11004.0 & 15.47 & -9.64 & 55.42 & 4347$\pm$85 & 1.75$\pm$0.3 & -0.97$\pm$0.17 & -1.13$\pm$0.05 & 0.06 & -0.5$\pm$0.03 & 0.01& ... \\
1106.0 & 359.83 & -9.67 & 85.36 & 4660$\pm$83 & 1.71$\pm$0.3 & -1.9$\pm$0.17 & -1.85$\pm$0.05 & 0.13 &  &  & ...\\
1109.3 & 14.4 & -9.64 & 48.73 & 5192$\pm$86 & 3.44$\pm$0.3 & 0.33$\pm$0.17 & 0.11$\pm$0.12 & 0.15 &  &  & ...\\
11151.0 & 15.55 & -9.77 & 26.0 & 5150$\pm$108 & 3.19$\pm$0.33 & -0.13$\pm$0.19 & -0.16$\pm$0.1 & 0.0 & -0.04$\pm$0.07 & 0.0 & ...\\
1118.0 & 0.1 & -9.54 & 84.33 & 4449$\pm$83 & 1.93$\pm$0.3 & -1.02$\pm$0.17 & -1.1$\pm$0.03 & 0.08 & -0.54$\pm$0.02 & 0.02& ... \\
1129.0 & 359.84 & -9.67 & 38.71 & 4953$\pm$91 & 2.75$\pm$0.31 & -1.02$\pm$0.18 & -1.0$\pm$0.12 & 0.12 & -0.6$\pm$0.13 & 0.04 & ...\\
1178.0 & 359.86 & -9.68 & 54.79 & 4462$\pm$85 & 1.5$\pm$0.3 & -1.39$\pm$0.17 & -1.43$\pm$0.05 & 0.09 & -0.91$\pm$0.05 & 0.02 & ...\\
1189.0 & 0.18 & -9.52 & 32.26 & 5165$\pm$97 & 2.25$\pm$0.32 & -1.08$\pm$0.18 & -0.69$\pm$0.09 & 0.02 & -0.78$\pm$0.12 & 0.01 & ...\\
1194.0 & 359.87 & -9.67 & 60.37 & 5282$\pm$84 & 3.19$\pm$0.3 & -0.32$\pm$0.17 & -0.09$\pm$0.06 & 0.09 & -0.08$\pm$0.03 & 0.0 & ...\\
1196.2 & 355.73 & -10.23 & 33.14 & 5680$\pm$96 & 3.57$\pm$0.31 & -0.36$\pm$0.18 & -0.34$\pm$0.22 & 0.0 & -0.32$\pm$0.21 & 0.0 & ...\\
11976.0 & 355.91 & -10.02 & 24.6 & 4350$\pm$111 & 1.25$\pm$0.34 & -1.88$\pm$0.19 & -2.28$\pm$0.14 & 0.1 & -1.44$\pm$0.13 & 0.0 & ...\\
11985.1 & 14.33 & -9.7 & 24.29 & 4400$\pm$112 & 1.64$\pm$0.34 & -1.36$\pm$0.19 & -1.41$\pm$0.24 & 0.0 & -1.06$\pm$0.16 & 0.0& ... \\
1200.0 & 359.84 & -9.69 & 49.11 & 5909$\pm$86 & 3.5$\pm$0.3 & -0.41$\pm$0.17 & -0.06$\pm$0.09 & 0.05 & -0.18$\pm$0.09 & 0.0 & ...\\
12063.3 & 14.28 & -9.59 & 33.32 & 4340$\pm$95 & 1.25$\pm$0.31 & -1.42$\pm$0.18 & -1.5$\pm$0.09 & 0.12 & -0.94$\pm$0.07 & 0.06& ... \\
1212.0 & 0.0 & -9.61 & 242.84 & 5040$\pm$83 & 2.0$\pm$0.3 & -1.6$\pm$0.17 & -1.13$\pm$0.02 & 0.03 & -1.11$\pm$0.06 & 0.03 & ...\\
1214.0 & 360.0 & -9.62 & 77.14 & 4733$\pm$84 & 2.42$\pm$0.3 & -0.58$\pm$0.17 & -0.67$\pm$0.04 & 0.02 & -0.28$\pm$0.04 & 0.05 & ...\\
12311.3 & 14.28 & -9.63 & 90.9 & 4994$\pm$83 & 3.25$\pm$0.3 & -0.85$\pm$0.17 & -0.95$\pm$0.04 & 0.01 & -0.46$\pm$0.03 & 0.05 & ... \\
... & ... & ... & ... & ...& ... & ... & ... & ... &... & ... & ...\\
\hline
\end{tabular}

\flushleft{A section of the online table with the object names, Galactic longitudes (l) and latitudes (b), the SNR of the HR06 spectra ($\rm{SNR_{06}}$), the effective temperature (\teff), surface gravity (\logg), metallicity ([M/H]), Fe abundance ([Fe/H]), line-by-line scatter in the Fe abundance ($\rm{Fe_{scatter}}$), Mg abundance ([Mg/H]) and the line-by-line scatter in the Mg abundance ($\rm{Mg_{scatter}}$).  Also included in the online table is the probability of confinement, distance, Galactic positions (X,Y,Z), Galactic velocities (U,V,W), eccentricity, apocenter, pericenter, $z_{max}$, the z-component of the angular momentum ($L_z$) along with the associated asymmetric uncertainties for all of these quantities.  In addition, we include the \vsini, right ascension, declination, \gaia\ source ID, renormalized unit weight error (ruwe), $G$ band magnitude, the radial velocity (along with the associated uncertainty and scatter), the number of combined spectra and the SNR for both the HR06 and HR21 spectra. In addition to the final mean [X/H] abundance and associated uncertainty, we include all of the measured elemental abundances for each individual atomic line used, along with the associated uncertainty and the line-by-line scatter.    }
\end{table*}

\section*{Acknowledgements}
{\small 
This material is based upon work supported by the National Science Foundation Graduate Research Fellowship under Grant No. 000392968.
VPD is supported by STFC Consolidated grant \# ST/R000786/1.
T.B. acknowledges financial support by grant No. 2018-04857 from the Swedish Research Council. KH has been partially supported by a TDA/Scialog (2018-2020) grant funded by the Research Corporation and a TDA/Scialog grant (2019-2021) funded by the Heising-Simons Foundation. ML \& KH acknowledge support from
the National Science Foundation grant AST-1907417. KH is partially supported through the Wootton Center for Astrophysical Plasma Properties funded under the United States Department of Energy collaborative agreement DE-NA0003843.
CK acknowledges funding from the UK Science and Technology Facility Council (STFC) through grant ST/ R000905/1.}

This work has made use of data from the European Space Agency (ESA) mission {\it Gaia} (https://www.cosmos.esa.int/
gaia), processed by the {\it Gaia} Data Processing and Analysis Consortium (DPAC, https://www.cosmos.esa.int/web/gaia/dpac/
consortium). Funding for the DPAC has been provided by national institutions, in particular the institutions participating in
the Gaia Multilateral Agreement.

\section*{Data Availability}
The data underlying this article are available in the ESO Science Archive Facility at \url{http://archive.eso.org/}, and can be accessed with ESO programme ID 089.B-069.

\bibliography{bibliography}

\begin{thebibliography}{}
\makeatletter
\relax
\def\mn@urlcharsother{\let\do\@makeother \do\$\do\&\do\#\do\^\do\_\do\%\do\~}
\def\mn@doi{\begingroup\mn@urlcharsother \@ifnextchar [ {\mn@doi@}
  {\mn@doi@[]}}
\def\mn@doi@[#1]#2{\def\@tempa{#1}\ifx\@tempa\@empty \href
  {http://dx.doi.org/#2} {doi:#2}\else \href {http://dx.doi.org/#2} {#1}\fi
  \endgroup}
\def\mn@eprint#1#2{\mn@eprint@#1:#2::\@nil}
\def\mn@eprint@arXiv#1{\href {http://arxiv.org/abs/#1} {{\tt arXiv:#1}}}
\def\mn@eprint@dblp#1{\href {http://dblp.uni-trier.de/rec/bibtex/#1.xml}
  {dblp:#1}}
\def\mn@eprint@#1:#2:#3:#4\@nil{\def\@tempa {#1}\def\@tempb {#2}\def\@tempc
  {#3}\ifx \@tempc \@empty \let \@tempc \@tempb \let \@tempb \@tempa \fi \ifx
  \@tempb \@empty \def\@tempb {arXiv}\fi \@ifundefined
  {mn@eprint@\@tempb}{\@tempb:\@tempc}{\expandafter \expandafter \csname
  mn@eprint@\@tempb\endcsname \expandafter{\@tempc}}}

\bibitem[\protect\citeauthoryear{{Adibekyan}, {Sousa}, {Santos}, {Delgado
  Mena}, {Gonz{\'a}lez Hern{\'a}ndez}, {Israelian}, {Mayor}  \&
  {Khachatryan}}{{Adibekyan} et~al.}{2012}]{Adibekyan2012}
{Adibekyan} V.~Z.,  {Sousa} S.~G.,  {Santos} N.~C.,  {Delgado Mena} E.,
  {Gonz{\'a}lez Hern{\'a}ndez} J.~I.,  {Israelian} G.,  {Mayor} M.,
  {Khachatryan} G.,  2012, \mn@doi [\aap] {10.1051/0004-6361/201219401}, \href
  {http://adsabs.harvard.edu/abs/2012A%26A...545A..32A} {545, A32}

\bibitem[\protect\citeauthoryear{{Ahumada} et~al.,}{{Ahumada}
  et~al.}{2020}]{Ahumada2020}
{Ahumada} R.,  et~al., 2020, \mn@doi [\apjs] {10.3847/1538-4365/ab929e}, \href
  {https://ui.adsabs.harvard.edu/abs/2020ApJS..249....3A} {249, 3}

\bibitem[\protect\citeauthoryear{{Allende Prieto}}{{Allende
  Prieto}}{2004}]{Allende-Prieto2004}
{Allende Prieto} C.,  2004, \mn@doi [Astronomische Nachrichten]
  {10.1002/asna.200410291}, \href
  {https://ui.adsabs.harvard.edu/abs/2004AN....325..604A} {325, 604}

\bibitem[\protect\citeauthoryear{{Allende Prieto}, {Beers}, {Wilhelm},
  {Newberg}, {Rockosi}, {Yanny}  \& {Lee}}{{Allende Prieto}
  et~al.}{2006}]{AllendePrieto2006}
{Allende Prieto} C.,  {Beers} T.~C.,  {Wilhelm} R.,  {Newberg} H.~J.,
  {Rockosi} C.~M.,  {Yanny} B.,   {Lee} Y.~S.,  2006, \mn@doi [\apj]
  {10.1086/498131}, \href {http://adsabs.harvard.edu/abs/2006ApJ...636..804A}
  {636, 804}

\bibitem[\protect\citeauthoryear{{Allende Prieto} et~al.,}{{Allende Prieto}
  et~al.}{2008}]{AllendePrieto2008}
{Allende Prieto} C.,  et~al., 2008, \mn@doi [\aj]
  {10.1088/0004-6256/136/5/2070}, \href
  {http://adsabs.harvard.edu/abs/2008AJ....136.2070A} {136, 2070}

\bibitem[\protect\citeauthoryear{{Allende Prieto}, {Hubeny}  \&
  {Smith}}{{Allende Prieto} et~al.}{2009}]{Allende-Prieto2016}
{Allende Prieto} C.,  {Hubeny} I.,   {Smith} J.~A.,  2009, \mn@doi [\mnras]
  {10.1111/j.1365-2966.2009.14775.x}, \href
  {https://ui.adsabs.harvard.edu/abs/2009MNRAS.396..759A} {396, 759}

\bibitem[\protect\citeauthoryear{{Alonso}, {Arribas}  \&
  {Mart{\'\i}nez-Roger}}{{Alonso} et~al.}{1999}]{Alonso1999}
{Alonso} A.,  {Arribas} S.,   {Mart{\'\i}nez-Roger} C.,  1999, \mn@doi [\aaps]
  {10.1051/aas:1999521}, \href
  {https://ui.adsabs.harvard.edu/abs/1999A&AS..140..261A} {140, 261}

\bibitem[\protect\citeauthoryear{{Amarsi} \& {Asplund}}{{Amarsi} \&
  {Asplund}}{2017}]{Amarsi2017}
{Amarsi} A.~M.,  {Asplund} M.,  2017, \mn@doi [\mnras] {10.1093/mnras/stw2445},
  \href {https://ui.adsabs.harvard.edu/abs/2017MNRAS.464..264A} {464, 264}

\bibitem[\protect\citeauthoryear{{Amarsi}, {Asplund}, {Collet}  \&
  {Leenaarts}}{{Amarsi} et~al.}{2016}]{Amarsi2016}
{Amarsi} A.~M.,  {Asplund} M.,  {Collet} R.,   {Leenaarts} J.,  2016, \mn@doi
  [\mnras] {10.1093/mnras/stv2608}, \href
  {https://ui.adsabs.harvard.edu/abs/2016MNRAS.455.3735A} {455, 3735}

\bibitem[\protect\citeauthoryear{{Arentsen} et~al.,}{{Arentsen}
  et~al.}{2020a}]{Arentsen2020}
{Arentsen} A.,  et~al., 2020a, \mn@doi [\mnras] {10.1093/mnrasl/slz156}, \href
  {https://ui.adsabs.harvard.edu/abs/2020MNRAS.491L..11A} {491, L11}

\bibitem[\protect\citeauthoryear{{Arentsen} et~al.,}{{Arentsen}
  et~al.}{2020b}]{Arentsen2020b}
{Arentsen} A.,  et~al., 2020b, \mn@doi [\mnras] {10.1093/mnras/staa1661}, \href
  {https://ui.adsabs.harvard.edu/abs/2020MNRAS.496.4964A} {496, 4964}

\bibitem[\protect\citeauthoryear{{Arentsen} et~al.,}{{Arentsen}
  et~al.}{2021}]{Arentsen2021}
{Arentsen} A.,  et~al., 2021, arXiv e-prints, \href
  {https://ui.adsabs.harvard.edu/abs/2021arXiv210503441A} {p. arXiv:2105.03441}

\bibitem[\protect\citeauthoryear{{Armandroff} \& {Da Costa}}{{Armandroff} \&
  {Da Costa}}{1991}]{Armandroff1991}
{Armandroff} T.~E.,  {Da Costa} G.~S.,  1991, \mn@doi [\aj] {10.1086/115769},
  \href {https://ui.adsabs.harvard.edu/abs/1991AJ....101.1329A} {101, 1329}

\bibitem[\protect\citeauthoryear{{Armandroff} \& {Zinn}}{{Armandroff} \&
  {Zinn}}{1988}]{Armandroff1988}
{Armandroff} T.~E.,  {Zinn} R.,  1988, \mn@doi [\aj] {10.1086/114792}, \href
  {https://ui.adsabs.harvard.edu/abs/1988AJ.....96...92A} {96, 92}

\bibitem[\protect\citeauthoryear{{Babusiaux} et~al.,}{{Babusiaux}
  et~al.}{2010}]{Babusiaux2010}
{Babusiaux} C.,  et~al., 2010, \mn@doi [\aap] {10.1051/0004-6361/201014353},
  \href {https://ui.adsabs.harvard.edu/abs/2010A&A...519A..77B} {519, A77}

\bibitem[\protect\citeauthoryear{{Barkat}, {Rakavy}  \& {Sack}}{{Barkat}
  et~al.}{1967}]{Barkar1967}
{Barkat} Z.,  {Rakavy} G.,   {Sack} N.,  1967, \mn@doi [\prl]
  {10.1103/PhysRevLett.18.379}, \href
  {https://ui.adsabs.harvard.edu/abs/1967PhRvL..18..379B} {18, 379}

\bibitem[\protect\citeauthoryear{{Battaglia}, {Irwin}, {Tolstoy}, {Hill},
  {Helmi}, {Letarte}  \& {Jablonka}}{{Battaglia} et~al.}{2008}]{Battaglia2008}
{Battaglia} G.,  {Irwin} M.,  {Tolstoy} E.,  {Hill} V.,  {Helmi} A.,  {Letarte}
  B.,   {Jablonka} P.,  2008, \mn@doi [\mnras]
  {10.1111/j.1365-2966.2007.12532.x}, \href
  {https://ui.adsabs.harvard.edu/abs/2008MNRAS.383..183B} {383, 183}

\bibitem[\protect\citeauthoryear{{Battistini} \& {Bensby}}{{Battistini} \&
  {Bensby}}{2015}]{Battistini2015}
{Battistini} C.,  {Bensby} T.,  2015, \mn@doi [\aap]
  {10.1051/0004-6361/201425327}, \href
  {http://adsabs.harvard.edu/abs/2015A%26A...577A...9B} {577, A9}

\bibitem[\protect\citeauthoryear{{Battistini} \& {Bensby}}{{Battistini} \&
  {Bensby}}{2016}]{Battistini2016}
{Battistini} C.,  {Bensby} T.,  2016, \mn@doi [\aap]
  {10.1051/0004-6361/201527385}, \href
  {http://adsabs.harvard.edu/abs/2016A%26A...586A..49B} {586, A49}

\bibitem[\protect\citeauthoryear{{Beers}, {Norris}, {Placco}, {Lee}, {Rossi},
  {Carollo}  \& {Masseron}}{{Beers} et~al.}{2014}]{Beers2014}
{Beers} T.~C.,  {Norris} J.~E.,  {Placco} V.~M.,  {Lee} Y.~S.,  {Rossi} S.,
  {Carollo} D.,   {Masseron} T.,  2014, \mn@doi [\apj]
  {10.1088/0004-637X/794/1/58}, \href
  {http://adsabs.harvard.edu/abs/2014ApJ...794...58B} {794, 58}

\bibitem[\protect\citeauthoryear{{Bekki}}{{Bekki}}{2019}]{Bekki2019}
{Bekki} K.,  2019, \mn@doi [\mnras] {10.1093/mnras/stz2732}, \href
  {https://ui.adsabs.harvard.edu/abs/2019MNRAS.490.4007B} {490, 4007}

\bibitem[\protect\citeauthoryear{{Bensby} et~al.,}{{Bensby}
  et~al.}{2013}]{Bensby2013}
{Bensby} T.,  et~al., 2013, \mn@doi [\aap] {10.1051/0004-6361/201220678}, \href
  {http://adsabs.harvard.edu/abs/2013A%26A...549A.147B} {549, A147}

\bibitem[\protect\citeauthoryear{{Bensby}, {Feltzing}  \& {Oey}}{{Bensby}
  et~al.}{2014}]{Bensby2014}
{Bensby} T.,  {Feltzing} S.,   {Oey} M.~S.,  2014, \mn@doi [\aap]
  {10.1051/0004-6361/201322631}, \href
  {http://adsabs.harvard.edu/abs/2014A%26A...562A..71B} {562, A71}

\bibitem[\protect\citeauthoryear{{Bensby} et~al.,}{{Bensby}
  et~al.}{2017}]{Bensby2017}
{Bensby} T.,  et~al., 2017, \mn@doi [\aap] {10.1051/0004-6361/201730560}, \href
  {https://ui.adsabs.harvard.edu/abs/2017A&A...605A..89B} {605, A89}

\bibitem[\protect\citeauthoryear{{Bessell}, {Castelli}  \& {Plez}}{{Bessell}
  et~al.}{1998}]{Bessell1998}
{Bessell} M.~S.,  {Castelli} F.,   {Plez} B.,  1998, \aap, \href
  {https://ui.adsabs.harvard.edu/abs/1998A&A...333..231B} {333, 231}

\bibitem[\protect\citeauthoryear{{Blanco-Cuaresma}, {Soubiran}, {Heiter}  \&
  {Jofr{\'e}}}{{Blanco-Cuaresma} et~al.}{2014}]{Blanco-Cuaresma2014}
{Blanco-Cuaresma} S.,  {Soubiran} C.,  {Heiter} U.,   {Jofr{\'e}} P.,  2014,
  \mn@doi [\aap] {10.1051/0004-6361/201423945}, \href
  {http://adsabs.harvard.edu/abs/2014A%26A...569A.111B} {569, A111}

\bibitem[\protect\citeauthoryear{{Bournaud}}{{Bournaud}}{2016}]{Bournaud2016}
{Bournaud} F.,  2016, in {Laurikainen} E.,  {Peletier} R.,   {Gadotti} D.,
  eds,  Astrophysics and Space Science Library Vol. 418, Galactic Bulges.
  p.~355 (\mn@eprint {arXiv} {1503.07660}),
  \mn@doi{10.1007/978-3-319-19378-6_13}

\bibitem[\protect\citeauthoryear{{Bovy}}{{Bovy}}{2015}]{Bovy2015}
{Bovy} J.,  2015, \mn@doi [\apjs] {10.1088/0067-0049/216/2/29}, \href
  {https://ui.adsabs.harvard.edu/abs/2015ApJS..216...29B} {216, 29}

\bibitem[\protect\citeauthoryear{{Bromm}}{{Bromm}}{2013}]{Bromm2013}
{Bromm} V.,  2013, \mn@doi [Reports on Progress in Physics]
  {10.1088/0034-4885/76/11/112901}, \href
  {https://ui.adsabs.harvard.edu/\#abs/2013RPPh...76k2901B} {76, 112901}

\bibitem[\protect\citeauthoryear{{Brook}, {Kawata}, {Scannapieco}, {Martel}  \&
  {Gibson}}{{Brook} et~al.}{2007}]{Brook2007}
{Brook} C.~B.,  {Kawata} D.,  {Scannapieco} E.,  {Martel} H.,   {Gibson} B.~K.,
   2007, \mn@doi [\apj] {10.1086/511514}, \href
  {https://ui.adsabs.harvard.edu/\#abs/2007ApJ...661...10B} {661, 10}

\bibitem[\protect\citeauthoryear{{Brown} et~al.,}{{Brown}
  et~al.}{2010}]{Brown2010}
{Brown} T.~M.,  et~al., 2010, \mn@doi [\apjl] {10.1088/2041-8205/725/1/L19},
  \href {https://ui.adsabs.harvard.edu/abs/2010ApJ...725L..19B} {725, L19}

\bibitem[\protect\citeauthoryear{{Buder} et~al.,}{{Buder}
  et~al.}{2018}]{Buder2018}
{Buder} S.,  et~al., 2018, preprint, \href
  {http://adsabs.harvard.edu/abs/2018arXiv180406041B} {} (\mn@eprint {arXiv}
  {1804.06041})

\bibitem[\protect\citeauthoryear{{Bureau} \& {Athanassoula}}{{Bureau} \&
  {Athanassoula}}{2005}]{Bureau2005}
{Bureau} M.,  {Athanassoula} E.,  2005, \mn@doi [\apj] {10.1086/430056}, \href
  {https://ui.adsabs.harvard.edu/abs/2005ApJ...626..159B} {626, 159}

\bibitem[\protect\citeauthoryear{{Calamida} et~al.,}{{Calamida}
  et~al.}{2014}]{Calamida2014}
{Calamida} A.,  et~al., 2014, \mn@doi [\apj] {10.1088/0004-637X/790/2/164},
  \href {https://ui.adsabs.harvard.edu/abs/2014ApJ...790..164C} {790, 164}

\bibitem[\protect\citeauthoryear{{Carollo} et~al.,}{{Carollo}
  et~al.}{2019}]{Carollo2019}
{Carollo} D.,  et~al., 2019, \mn@doi [\apj] {10.3847/1538-4357/ab517c}, \href
  {https://ui.adsabs.harvard.edu/abs/2019ApJ...887...22C} {887, 22}

\bibitem[\protect\citeauthoryear{{Carrillo}, {Hawkins}, {Bowler}, {Cochran}  \&
  {Vanderburg}}{{Carrillo} et~al.}{2020}]{Carrillo2020}
{Carrillo} A.,  {Hawkins} K.,  {Bowler} B.~P.,  {Cochran} W.,   {Vanderburg}
  A.,  2020, \mn@doi [\mnras] {10.1093/mnras/stz3255}, \href
  {https://ui.adsabs.harvard.edu/abs/2020MNRAS.491.4365C} {491, 4365}

\bibitem[\protect\citeauthoryear{{Carroll}}{{Carroll}}{1933a}]{Carroll1933a}
{Carroll} J.~A.,  1933a, \mn@doi [\mnras] {10.1093/mnras/93.7.478}, \href
  {https://ui.adsabs.harvard.edu/abs/1933MNRAS..93..478C} {93, 478}

\bibitem[\protect\citeauthoryear{{Carroll}}{{Carroll}}{1933b}]{Carroll1933b}
{Carroll} J.~A.,  1933b, \mn@doi [\mnras] {10.1093/mnras/93.9.680}, \href
  {https://ui.adsabs.harvard.edu/abs/1933MNRAS..93..680C} {93, 680}

\bibitem[\protect\citeauthoryear{{Casagrande}, {Wolf}, {Mackey}, {Nordland er},
  {Yong}  \& {Bessell}}{{Casagrande} et~al.}{2019}]{Casagrande2019}
{Casagrande} L.,  {Wolf} C.,  {Mackey} A.~D.,  {Nordland er} T.,  {Yong} D.,
  {Bessell} M.,  2019, \mn@doi [\mnras] {10.1093/mnras/sty2878}, \href
  {https://ui.adsabs.harvard.edu/\#abs/2019MNRAS.482.2770C} {482, 2770}

\bibitem[\protect\citeauthoryear{{Cescutti} \& {Chiappini}}{{Cescutti} \&
  {Chiappini}}{2014}]{Cescutti2014}
{Cescutti} G.,  {Chiappini} C.,  2014, \mn@doi [\aap]
  {10.1051/0004-6361/201423432}, \href
  {https://ui.adsabs.harvard.edu/abs/2014A&A...565A..51C} {565, A51}

\bibitem[\protect\citeauthoryear{{Cescutti}, {Chiappini}, {Hirschi}, {Meynet}
  \& {Frischknecht}}{{Cescutti} et~al.}{2013}]{Cescutti2013}
{Cescutti} G.,  {Chiappini} C.,  {Hirschi} R.,  {Meynet} G.,   {Frischknecht}
  U.,  2013, \mn@doi [\aap] {10.1051/0004-6361/201220809}, \href
  {https://ui.adsabs.harvard.edu/abs/2013A&A...553A..51C} {553, A51}

\bibitem[\protect\citeauthoryear{{Cescutti}, {Chiappini}  \&
  {Hirschi}}{{Cescutti} et~al.}{2018}]{Cescutti2018}
{Cescutti} G.,  {Chiappini} C.,   {Hirschi} R.,  2018, in {Chiappini} C.,
  {Minchev} I.,  {Starkenburg} E.,   {Valentini} M.,  eds,  IAU Symposium Vol.
  334, Rediscovering Our Galaxy. pp 94--97 (\mn@eprint {arXiv} {1710.11014}),
  \mn@doi{10.1017/S1743921317008183}

\bibitem[\protect\citeauthoryear{{Choi}, {Dotter}, {Conroy}, {Cantiello},
  {Paxton}  \& {Johnson}}{{Choi} et~al.}{2016}]{Choi2016}
{Choi} J.,  {Dotter} A.,  {Conroy} C.,  {Cantiello} M.,  {Paxton} B.,
  {Johnson} B.~D.,  2016, \mn@doi [\apj] {10.3847/0004-637X/823/2/102}, \href
  {https://ui.adsabs.harvard.edu/abs/2016ApJ...823..102C} {823, 102}

\bibitem[\protect\citeauthoryear{{Christlieb}, {Sch{\"o}rck}, {Frebel},
  {Beers}, {Wisotzki}  \& {Reimers}}{{Christlieb}
  et~al.}{2008}]{Christlieb2008}
{Christlieb} N.,  {Sch{\"o}rck} T.,  {Frebel} A.,  {Beers} T.~C.,  {Wisotzki}
  L.,   {Reimers} D.,  2008, \mn@doi [\aap] {10.1051/0004-6361:20078748}, \href
  {https://ui.adsabs.harvard.edu/\#abs/2008A&A...484..721C} {484, 721}

\bibitem[\protect\citeauthoryear{{Clarkson} et~al.,}{{Clarkson}
  et~al.}{2011}]{Clarkson2011}
{Clarkson} W.~I.,  et~al., 2011, \mn@doi [\apj] {10.1088/0004-637X/735/1/37},
  \href {https://ui.adsabs.harvard.edu/abs/2011ApJ...735...37C} {735, 37}

\bibitem[\protect\citeauthoryear{{Cole}, {Smecker-Hane}, {Tolstoy}, {Bosler}
  \& {Gallagher}}{{Cole} et~al.}{2004}]{Cole2004}
{Cole} A.~A.,  {Smecker-Hane} T.~A.,  {Tolstoy} E.,  {Bosler} T.~L.,
  {Gallagher} J.~S.,  2004, \mn@doi [\mnras]
  {10.1111/j.1365-2966.2004.07223.x}, \href
  {https://ui.adsabs.harvard.edu/abs/2004MNRAS.347..367C} {347, 367}

\bibitem[\protect\citeauthoryear{{Combes} \& {Sanders}}{{Combes} \&
  {Sanders}}{1981}]{Combes1981}
{Combes} F.,  {Sanders} R.~H.,  1981, \aap, \href
  {https://ui.adsabs.harvard.edu/abs/1981A&A....96..164C} {96, 164}

\bibitem[\protect\citeauthoryear{{Combes}, {Debbasch}, {Friedli}  \&
  {Pfenniger}}{{Combes} et~al.}{1990}]{Combes1990}
{Combes} F.,  {Debbasch} F.,  {Friedli} D.,   {Pfenniger} D.,  1990, \aap,
  \href {https://ui.adsabs.harvard.edu/abs/1990A&A...233...82C} {233, 82}

\bibitem[\protect\citeauthoryear{{Cowan}, {Thielemann}  \& {Truran}}{{Cowan}
  et~al.}{1991}]{Cowan1991}
{Cowan} J.~J.,  {Thielemann} F.-K.,   {Truran} J.~W.,  1991, \mn@doi [\physrep]
  {10.1016/0370-1573(91)90070-3}, \href
  {https://ui.adsabs.harvard.edu/abs/1991PhR...208..267C} {208, 267}

\bibitem[\protect\citeauthoryear{{Da Costa} et~al.,}{{Da Costa}
  et~al.}{2019}]{DaCosta2019}
{Da Costa} G.~S.,  et~al., 2019, \mn@doi [\mnras] {10.1093/mnras/stz2550},
  \href {https://ui.adsabs.harvard.edu/abs/2019MNRAS.489.5900D} {489, 5900}

\bibitem[\protect\citeauthoryear{{Debattista}, {Mayer}, {Carollo}, {Moore},
  {Wadsley}  \& {Quinn}}{{Debattista} et~al.}{2006}]{Debattista2006}
{Debattista} V.~P.,  {Mayer} L.,  {Carollo} C.~M.,  {Moore} B.,  {Wadsley} J.,
   {Quinn} T.,  2006, \mn@doi [\apj] {10.1086/504147}, \href
  {https://ui.adsabs.harvard.edu/abs/2006ApJ...645..209D} {645, 209}

\bibitem[\protect\citeauthoryear{{Debattista}, {Ness}, {Gonzalez}, {Freeman},
  {Zoccali}  \& {Minniti}}{{Debattista} et~al.}{2017}]{Debattista2017}
{Debattista} V.~P.,  {Ness} M.,  {Gonzalez} O.~A.,  {Freeman} K.,  {Zoccali}
  M.,   {Minniti} D.,  2017, \mn@doi [\mnras] {10.1093/mnras/stx947}, \href
  {https://ui.adsabs.harvard.edu/abs/2017MNRAS.469.1587D} {469, 1587}

\bibitem[\protect\citeauthoryear{{Dehnen}}{{Dehnen}}{2000}]{Dehnen2000}
{Dehnen} W.,  2000, \mn@doi [\aj] {10.1086/301226}, \href
  {https://ui.adsabs.harvard.edu/abs/2000AJ....119..800D} {119, 800}

\bibitem[\protect\citeauthoryear{{Di Matteo}, {Haywood}, {Lehnert}, {Katz},
  {Khoperskov}, {Snaith}, {G{\'o}mez}  \& {Robichon}}{{Di Matteo}
  et~al.}{2019}]{Di_Matteo2019}
{Di Matteo} P.,  {Haywood} M.,  {Lehnert} M.~D.,  {Katz} D.,  {Khoperskov} S.,
  {Snaith} O.~N.,  {G{\'o}mez} A.,   {Robichon} N.,  2019, \mn@doi [\aap]
  {10.1051/0004-6361/201834929}, \href
  {https://ui.adsabs.harvard.edu/abs/2019A&A...632A...4D} {632, A4}

\bibitem[\protect\citeauthoryear{{Diemand}, {Kuhlen}, {Madau}, {Zemp}, {Moore},
  {Potter}  \& {Stadel}}{{Diemand} et~al.}{2008}]{Diemand2008}
{Diemand} J.,  {Kuhlen} M.,  {Madau} P.,  {Zemp} M.,  {Moore} B.,  {Potter} D.,
    {Stadel} J.,  2008, \mn@doi [\nat] {10.1038/nature07153}, \href
  {https://ui.adsabs.harvard.edu/\#abs/2008Natur.454..735D} {454, 735}

\bibitem[\protect\citeauthoryear{{Dotter}}{{Dotter}}{2016}]{Dotter2016}
{Dotter} A.,  2016, \mn@doi [\apjs] {10.3847/0067-0049/222/1/8}, \href
  {https://ui.adsabs.harvard.edu/abs/2016ApJS..222....8D} {222, 8}

\bibitem[\protect\citeauthoryear{{Duong}, {Asplund}, {Nataf}, {Freeman}, {Ness}
   \& {Howes}}{{Duong} et~al.}{2019a}]{Duong2019}
{Duong} L.,  {Asplund} M.,  {Nataf} D.~M.,  {Freeman} K.~C.,  {Ness} M.,
  {Howes} L.~M.,  2019a, arXiv e-prints, \href
  {https://ui.adsabs.harvard.edu/\#abs/2019arXiv190309706D} {p.
  arXiv:1903.09706}

\bibitem[\protect\citeauthoryear{{Duong}, {Asplund}, {Nataf}, {Freeman}  \&
  {Ness}}{{Duong} et~al.}{2019b}]{Duong2019b}
{Duong} L.,  {Asplund} M.,  {Nataf} D.~M.,  {Freeman} K.~C.,   {Ness} M.,
  2019b, \mn@doi [\mnras] {10.1093/mnras/stz1183}, \href
  {https://ui.adsabs.harvard.edu/abs/2019MNRAS.486.5349D} {486, 5349}

\bibitem[\protect\citeauthoryear{{El-Badry}, {Rix}, {Ting}, {Weisz},
  {Bergemann}, {Cargile}, {Conroy}  \& {Eilers}}{{El-Badry}
  et~al.}{2018a}]{El-Badry2018a}
{El-Badry} K.,  {Rix} H.-W.,  {Ting} Y.-S.,  {Weisz} D.~R.,  {Bergemann} M.,
  {Cargile} P.,  {Conroy} C.,   {Eilers} A.-C.,  2018a, \mn@doi [\mnras]
  {10.1093/mnras/stx2758}, \href
  {https://ui.adsabs.harvard.edu/abs/2018MNRAS.473.5043E} {473, 5043}

\bibitem[\protect\citeauthoryear{{El-Badry} et~al.,}{{El-Badry}
  et~al.}{2018b}]{El-Badry2018b}
{El-Badry} K.,  et~al., 2018b, \mn@doi [\mnras] {10.1093/mnras/sty1864}, \href
  {https://ui.adsabs.harvard.edu/abs/2018MNRAS.480..652E} {480, 652}

\bibitem[\protect\citeauthoryear{{Fasano} \& {Franceschini}}{{Fasano} \&
  {Franceschini}}{1987}]{Fasano1987}
{Fasano} G.,  {Franceschini} A.,  1987, \mn@doi [\mnras]
  {10.1093/mnras/225.1.155}, \href
  {https://ui.adsabs.harvard.edu/abs/1987MNRAS.225..155F} {225, 155}

\bibitem[\protect\citeauthoryear{{Fern{\'a}ndez-Trincado}
  et~al.,}{{Fern{\'a}ndez-Trincado} et~al.}{2017}]{Fernandez2017}
{Fern{\'a}ndez-Trincado} J.~G.,  et~al., 2017, \mn@doi [\apj]
  {10.3847/2041-8213/aa8032}, \href
  {https://ui.adsabs.harvard.edu/\#abs/2017ApJ...846L...2F} {846, L2}

\bibitem[\protect\citeauthoryear{{Frebel} et~al.,}{{Frebel}
  et~al.}{2006}]{Frebel2006}
{Frebel} A.,  et~al., 2006, \mn@doi [\apj] {10.1086/508506}, \href
  {https://ui.adsabs.harvard.edu/\#abs/2006ApJ...652.1585F} {652, 1585}

\bibitem[\protect\citeauthoryear{{Freeman} et~al.,}{{Freeman}
  et~al.}{2013}]{Freeman2013}
{Freeman} K.,  et~al., 2013, \mn@doi [\mnras] {10.1093/mnras/sts305}, \href
  {https://ui.adsabs.harvard.edu/\#abs/2013MNRAS.428.3660F} {428, 3660}

\bibitem[\protect\citeauthoryear{{Fulbright}, {McWilliam}  \&
  {Rich}}{{Fulbright} et~al.}{2007}]{Fulbright2007}
{Fulbright} J.~P.,  {McWilliam} A.,   {Rich} R.~M.,  2007, \mn@doi [\apj]
  {10.1086/513710}, \href
  {https://ui.adsabs.harvard.edu/abs/2007ApJ...661.1152F} {661, 1152}

\bibitem[\protect\citeauthoryear{{Garc{\'\i}a P{\'e}rez} et~al.,}{{Garc{\'\i}a
  P{\'e}rez} et~al.}{2018}]{GarciaPerez2019}
{Garc{\'\i}a P{\'e}rez} A.~E.,  et~al., 2018, \mn@doi [\apj]
  {10.3847/1538-4357/aa9d88}, \href
  {https://ui.adsabs.harvard.edu/abs/2018ApJ...852...91G} {852, 91}

\bibitem[\protect\citeauthoryear{{Gilmore} et~al.,}{{Gilmore}
  et~al.}{2012}]{Gilmore2012}
{Gilmore} G.,  et~al., 2012, The Messenger, \href
  {http://adsabs.harvard.edu/abs/2012Msngr.147...25G} {147, 25}

\bibitem[\protect\citeauthoryear{{Gonzalez} et~al.,}{{Gonzalez}
  et~al.}{2015}]{Gonzalez2015}
{Gonzalez} O.~A.,  et~al., 2015, \mn@doi [\aap] {10.1051/0004-6361/201526737},
  \href {http://adsabs.harvard.edu/abs/2015A%26A...584A..46G} {584, A46}

\bibitem[\protect\citeauthoryear{{Gratton}, {Sneden}  \& {Carretta}}{{Gratton}
  et~al.}{2004}]{Gratton2004}
{Gratton} R.,  {Sneden} C.,   {Carretta} E.,  2004, \mn@doi [Annual Review of
  Astronomy and Astrophysics] {10.1146/annurev.astro.42.053102.133945}, \href
  {https://ui.adsabs.harvard.edu/\#abs/2004ARA&A..42..385G} {42, 385}

\bibitem[\protect\citeauthoryear{{Grevesse}, {Asplund}  \& {Sauval}}{{Grevesse}
  et~al.}{2007}]{Grevesse2007}
{Grevesse} N.,  {Asplund} M.,   {Sauval} A.~J.,  2007, \mn@doi [\ssr]
  {10.1007/s11214-007-9173-7}, \href
  {https://ui.adsabs.harvard.edu/abs/2007SSRv..130..105G} {130, 105}

\bibitem[\protect\citeauthoryear{{Guedes}, {Mayer}, {Carollo}  \&
  {Madau}}{{Guedes} et~al.}{2013}]{Guedes2013}
{Guedes} J.,  {Mayer} L.,  {Carollo} M.,   {Madau} P.,  2013, \mn@doi [\apj]
  {10.1088/0004-637X/772/1/36}, \href
  {https://ui.adsabs.harvard.edu/abs/2013ApJ...772...36G} {772, 36}

\bibitem[\protect\citeauthoryear{{Gustafsson}, {Edvardsson}, {Eriksson},
  {J{\o}rgensen}, {Nordlund}  \& {Plez}}{{Gustafsson}
  et~al.}{2008}]{Gustafsson2008}
{Gustafsson} B.,  {Edvardsson} B.,  {Eriksson} K.,  {J{\o}rgensen} U.~G.,
  {Nordlund} {\AA}.,   {Plez} B.,  2008, \mn@doi [\aap]
  {10.1051/0004-6361:200809724}, \href
  {http://adsabs.harvard.edu/abs/2008A%26A...486..951G} {486, 951}

\bibitem[\protect\citeauthoryear{{Hawkins} et~al.,}{{Hawkins}
  et~al.}{2015}]{Hawkins2015}
{Hawkins} K.,  et~al., 2015, \mn@doi [\mnras] {10.1093/mnras/stu2574}, \href
  {http://adsabs.harvard.edu/abs/2015MNRAS.447.2046H} {447, 2046}

\bibitem[\protect\citeauthoryear{{Hawkins} et~al.,}{{Hawkins}
  et~al.}{2016}]{Hawkins2016a}
{Hawkins} K.,  et~al., 2016, \mn@doi [\aap] {10.1051/0004-6361/201628268},
  \href {http://adsabs.harvard.edu/abs/2016A%26A...592A..70H} {592, A70}

\bibitem[\protect\citeauthoryear{{Heger} \& {Woosley}}{{Heger} \&
  {Woosley}}{2010}]{Heger2010}
{Heger} A.,  {Woosley} S.~E.,  2010, \mn@doi [\apj]
  {10.1088/0004-637X/724/1/341}, \href
  {https://ui.adsabs.harvard.edu/\#abs/2010ApJ...724..341H} {724, 341}

\bibitem[\protect\citeauthoryear{{Heiter}, {Jofr{\'e}}, {Gustafsson}, {Korn},
  {Soubiran}  \& {Th{\'e}venin}}{{Heiter} et~al.}{2015}]{Heiter2015}
{Heiter} U.,  {Jofr{\'e}} P.,  {Gustafsson} B.,  {Korn} A.~J.,  {Soubiran} C.,
   {Th{\'e}venin} F.,  2015, \mn@doi [\aap] {10.1051/0004-6361/201526319},
  \href {http://adsabs.harvard.edu/abs/2015A%26A...582A..49H} {582, A49, Paper
  I}

\bibitem[\protect\citeauthoryear{{Heiter} et~al.,}{{Heiter}
  et~al.}{2020}]{Heiter2020}
{Heiter} U.,  et~al., 2020, arXiv e-prints, \href
  {https://ui.adsabs.harvard.edu/abs/2020arXiv201102049H} {p. arXiv:2011.02049}

\bibitem[\protect\citeauthoryear{{Hill} et~al.,}{{Hill}
  et~al.}{2011}]{Hill2011}
{Hill} V.,  et~al., 2011, \mn@doi [\aap] {10.1051/0004-6361/200913757}, \href
  {http://adsabs.harvard.edu/abs/2011A%26A...534A..80H} {534, A80}

\bibitem[\protect\citeauthoryear{{Hirano}, {Hosokawa}, {Yoshida}, {Omukai}  \&
  {Yorke}}{{Hirano} et~al.}{2015}]{Hirano2015}
{Hirano} S.,  {Hosokawa} T.,  {Yoshida} N.,  {Omukai} K.,   {Yorke} H.~W.,
  2015, \mn@doi [\mnras] {10.1093/mnras/stv044}, \href
  {https://ui.adsabs.harvard.edu/abs/2015MNRAS.448..568H} {448, 568}

\bibitem[\protect\citeauthoryear{{Horta} et~al.,}{{Horta}
  et~al.}{2020}]{Horta2020}
{Horta} D.,  et~al., 2020, arXiv e-prints, \href
  {https://ui.adsabs.harvard.edu/abs/2020arXiv200710374H} {p. arXiv:2007.10374}

\bibitem[\protect\citeauthoryear{{Horta} et~al.,}{{Horta}
  et~al.}{2021}]{Horta2021}
{Horta} D.,  et~al., 2021, \mn@doi [\mnras] {10.1093/mnras/staa3598}, \href
  {https://ui.adsabs.harvard.edu/abs/2021MNRAS.500.5462H} {500, 5462}

\bibitem[\protect\citeauthoryear{{Howard} et~al.,}{{Howard}
  et~al.}{2009}]{Howard2009}
{Howard} C.~D.,  et~al., 2009, \mn@doi [\apjl] {10.1088/0004-637X/702/2/L153},
  \href {https://ui.adsabs.harvard.edu/abs/2009ApJ...702L.153H} {702, L153}

\bibitem[\protect\citeauthoryear{{Howes} et~al.,}{{Howes}
  et~al.}{2014}]{Howes2014}
{Howes} L.~M.,  et~al., 2014, \mn@doi [\mnras] {10.1093/mnras/stu1991}, \href
  {https://ui.adsabs.harvard.edu/\#abs/2014MNRAS.445.4241H} {445, 4241}

\bibitem[\protect\citeauthoryear{{Howes} et~al.,}{{Howes}
  et~al.}{2015}]{Howes2015}
{Howes} L.~M.,  et~al., 2015, \mn@doi [\nat] {10.1038/nature15747}, \href
  {https://ui.adsabs.harvard.edu/\#abs/2015Natur.527..484H} {527, 484}

\bibitem[\protect\citeauthoryear{{Howes} et~al.,}{{Howes}
  et~al.}{2016}]{Howes2016}
{Howes} L.~M.,  et~al., 2016, \mn@doi [\mnras] {10.1093/mnras/stw1004}, \href
  {https://ui.adsabs.harvard.edu/abs/2016MNRAS.460..884H} {460, 884}

\bibitem[\protect\citeauthoryear{{Iwamoto}, {Brachwitz}, {Nomoto}, {Kishimoto},
  {Umeda}, {Hix}  \& {Thielemann}}{{Iwamoto} et~al.}{1999}]{Iwamoto1999}
{Iwamoto} K.,  {Brachwitz} F.,  {Nomoto} K.,  {Kishimoto} N.,  {Umeda} H.,
  {Hix} W.~R.,   {Thielemann} F.-K.,  1999, \mn@doi [\apjs] {10.1086/313278},
  \href {http://adsabs.harvard.edu/abs/1999ApJS..125..439I} {125, 439}

\bibitem[\protect\citeauthoryear{{Jofr{\'e}} et~al.,}{{Jofr{\'e}}
  et~al.}{2014}]{Jofre2014}
{Jofr{\'e}} P.,  et~al., 2014, \mn@doi [\aap] {10.1051/0004-6361/201322440},
  \href {http://adsabs.harvard.edu/abs/2014A%26A...564A.133J} {564, A133, Paper
  III}

\bibitem[\protect\citeauthoryear{{Johnson}, {Rich}, {Kobayashi}  \&
  {Fulbright}}{{Johnson} et~al.}{2012}]{Johnson2012}
{Johnson} C.~I.,  {Rich} R.~M.,  {Kobayashi} C.,   {Fulbright} J.~P.,  2012,
  \mn@doi [\apj] {10.1088/0004-637X/749/2/175}, \href
  {https://ui.adsabs.harvard.edu/\#abs/2012ApJ...749..175J} {749, 175}

\bibitem[\protect\citeauthoryear{{Johnson}, {Rich}, {Kobayashi}, {Kunder},
  {Pilachowski}, {Koch}  \& {de Propris}}{{Johnson}
  et~al.}{2013}]{Johnson2013a}
{Johnson} C.~I.,  {Rich} R.~M.,  {Kobayashi} C.,  {Kunder} A.,  {Pilachowski}
  C.~A.,  {Koch} A.,   {de Propris} R.,  2013, \mn@doi [\apj]
  {10.1088/0004-637X/765/2/157}, \href
  {https://ui.adsabs.harvard.edu/\#abs/2013ApJ...765..157J} {765, 157}

\bibitem[\protect\citeauthoryear{{Johnson}, {Rich}, {Kobayashi}, {Kunder}  \&
  {Koch}}{{Johnson} et~al.}{2014}]{Johnson2014}
{Johnson} C.~I.,  {Rich} R.~M.,  {Kobayashi} C.,  {Kunder} A.,   {Koch} A.,
  2014, \mn@doi [\aj] {10.1088/0004-6256/148/4/67}, \href
  {https://ui.adsabs.harvard.edu/abs/2014AJ....148...67J} {148, 67}

\bibitem[\protect\citeauthoryear{{Johnson} et~al.,}{{Johnson}
  et~al.}{2020}]{Johnson2020}
{Johnson} C.~I.,  et~al., 2020, \mn@doi [\mnras] {10.1093/mnras/staa2393},
  \href {https://ui.adsabs.harvard.edu/abs/2020MNRAS.tmp.2083J} {}

\bibitem[\protect\citeauthoryear{{Karlsson}, {Johnson}  \& {Bromm}}{{Karlsson}
  et~al.}{2008}]{Karlsson2008}
{Karlsson} T.,  {Johnson} J.~L.,   {Bromm} V.,  2008, \mn@doi [\apj]
  {10.1086/533520}, \href
  {https://ui.adsabs.harvard.edu/\#abs/2008ApJ...679....6K} {679, 6}

\bibitem[\protect\citeauthoryear{{Kauffmann}, {White}  \&
  {Guiderdoni}}{{Kauffmann} et~al.}{1993}]{Kauffmann1993}
{Kauffmann} G.,  {White} S.~D.~M.,   {Guiderdoni} B.,  1993, \mn@doi [\mnras]
  {10.1093/mnras/264.1.201}, \href
  {https://ui.adsabs.harvard.edu/abs/1993MNRAS.264..201K} {264, 201}

\bibitem[\protect\citeauthoryear{{Keller} et~al.,}{{Keller}
  et~al.}{2014}]{Keller2014}
{Keller} S.~C.,  et~al., 2014, \mn@doi [\nat] {10.1038/nature12990}, \href
  {https://ui.adsabs.harvard.edu/\#abs/2014Natur.506..463K} {506, 463}

\bibitem[\protect\citeauthoryear{{Kobayashi} \& {Nakasato}}{{Kobayashi} \&
  {Nakasato}}{2011}]{Kobayashi2011a}
{Kobayashi} C.,  {Nakasato} N.,  2011, \mn@doi [\apj]
  {10.1088/0004-637X/729/1/16}, \href
  {http://adsabs.harvard.edu/abs/2011ApJ...729...16K} {729, 16}

\bibitem[\protect\citeauthoryear{{Kobayashi}, {Umeda}, {Nomoto}, {Tominaga}  \&
  {Ohkubo}}{{Kobayashi} et~al.}{2006}]{Kobayashi2006}
{Kobayashi} C.,  {Umeda} H.,  {Nomoto} K.,  {Tominaga} N.,   {Ohkubo} T.,
  2006, \mn@doi [\apj] {10.1086/508914}, \href
  {http://adsabs.harvard.edu/abs/2006ApJ...653.1145K} {653, 1145}

\bibitem[\protect\citeauthoryear{{Kobayashi}, {Karakas}  \&
  {Umeda}}{{Kobayashi} et~al.}{2011a}]{Kobayashi2011c}
{Kobayashi} C.,  {Karakas} A.~I.,   {Umeda} H.,  2011a, \mn@doi [\mnras]
  {10.1111/j.1365-2966.2011.18621.x}, \href
  {https://ui.adsabs.harvard.edu/\#abs/2011MNRAS.414.3231K} {414, 3231}

\bibitem[\protect\citeauthoryear{{Kobayashi}, {Tominaga}  \&
  {Nomoto}}{{Kobayashi} et~al.}{2011b}]{Kobayashi2011b}
{Kobayashi} C.,  {Tominaga} N.,   {Nomoto} K.,  2011b, \mn@doi [\apj]
  {10.1088/2041-8205/730/2/L14}, \href
  {https://ui.adsabs.harvard.edu/\#abs/2011ApJ...730L..14K} {730, L14}

\bibitem[\protect\citeauthoryear{{Kobayashi}, {Ishigaki}, {Tominaga}  \&
  {Nomoto}}{{Kobayashi} et~al.}{2014}]{Kobayashi2014}
{Kobayashi} C.,  {Ishigaki} M.~N.,  {Tominaga} N.,   {Nomoto} K.,  2014,
  \mn@doi [\apjl] {10.1088/2041-8205/785/1/L5}, \href
  {https://ui.adsabs.harvard.edu/abs/2014ApJ...785L...5K} {785, L5}

\bibitem[\protect\citeauthoryear{{Kobayashi}, {Karakas}  \&
  {Lugaro}}{{Kobayashi} et~al.}{2020}]{Kobayashi2020}
{Kobayashi} C.,  {Karakas} A.~I.,   {Lugaro} M.,  2020, \mn@doi [\apj]
  {10.3847/1538-4357/abae65}, \href
  {https://ui.adsabs.harvard.edu/abs/2020ApJ...900..179K} {900, 179}

\bibitem[\protect\citeauthoryear{{Koch}, {Reichert}, {Hansen}, {Hampel},
  {Stancliffe}, {Karakas}  \& {Arcones}}{{Koch} et~al.}{2019}]{Koch2019}
{Koch} A.,  {Reichert} M.,  {Hansen} C.~J.,  {Hampel} M.,  {Stancliffe} R.~J.,
  {Karakas} A.,   {Arcones} A.,  2019, \mn@doi [\aap]
  {10.1051/0004-6361/201834241}, \href
  {https://ui.adsabs.harvard.edu/abs/2019A&A...622A.159K} {622, A159}

\bibitem[\protect\citeauthoryear{{Kruijssen}}{{Kruijssen}}{2015}]{Kruijssen2015}
{Kruijssen} J.~M.~D.,  2015, \mn@doi [\mnras] {10.1093/mnras/stv2026}, \href
  {https://ui.adsabs.harvard.edu/\#abs/2015MNRAS.454.1658K} {454, 1658}

\bibitem[\protect\citeauthoryear{{Kuijken} \& {Rich}}{{Kuijken} \&
  {Rich}}{2002}]{Kuijken2002}
{Kuijken} K.,  {Rich} R.~M.,  2002, \mn@doi [\aj] {10.1086/342540}, \href
  {https://ui.adsabs.harvard.edu/abs/2002AJ....124.2054K} {124, 2054}

\bibitem[\protect\citeauthoryear{{Kunder} et~al.,}{{Kunder}
  et~al.}{2016}]{Kunder2016}
{Kunder} A.,  et~al., 2016, \mn@doi [\apjl] {10.3847/2041-8205/821/2/L25},
  \href {https://ui.adsabs.harvard.edu/abs/2016ApJ...821L..25K} {821, L25}

\bibitem[\protect\citeauthoryear{{Kunder} et~al.,}{{Kunder}
  et~al.}{2020}]{Kunder2020}
{Kunder} A.,  et~al., 2020, \mn@doi [\aj] {10.3847/1538-3881/ab8d35}, \href
  {https://ui.adsabs.harvard.edu/abs/2020AJ....159..270K} {159, 270}

\bibitem[\protect\citeauthoryear{{Li} et~al.,}{{Li} et~al.}{2017}]{Li2017}
{Li} T.~S.,  et~al., 2017, \mn@doi [\apj] {10.3847/1538-4357/aa6113}, \href
  {https://ui.adsabs.harvard.edu/abs/2017ApJ...838....8L} {838, 8}

\bibitem[\protect\citeauthoryear{{Lind}, {Asplund}  \& {Barklem}}{{Lind}
  et~al.}{2009}]{Lind2009}
{Lind} K.,  {Asplund} M.,   {Barklem} P.~S.,  2009, \mn@doi [\aap]
  {10.1051/0004-6361/200912221}, \href
  {https://ui.adsabs.harvard.edu/abs/2009A&A...503..541L} {503, 541}

\bibitem[\protect\citeauthoryear{{Lind}, {Asplund}, {Barklem}  \&
  {Belyaev}}{{Lind} et~al.}{2011}]{Lind2011}
{Lind} K.,  {Asplund} M.,  {Barklem} P.~S.,   {Belyaev} A.~K.,  2011, \mn@doi
  [\aap] {10.1051/0004-6361/201016095}, \href
  {https://ui.adsabs.harvard.edu/abs/2011A&A...528A.103L} {528, A103}

\bibitem[\protect\citeauthoryear{{Lindegren}}{{Lindegren}}{2018}]{Lindegren2018b}
{Lindegren} L.,  2018, Technical report, {Re-normalising the astrometric
  chi-square in Gaia DR2}

\bibitem[\protect\citeauthoryear{{Lucey} et~al.,}{{Lucey}
  et~al.}{2019}]{Lucey2019}
{Lucey} M.,  et~al., 2019, \mn@doi [\mnras] {10.1093/mnras/stz1847}, \href
  {https://ui.adsabs.harvard.edu/abs/2019MNRAS.488.2283L} {488, 2283}

\bibitem[\protect\citeauthoryear{{Lucey} et~al.,}{{Lucey}
  et~al.}{2020}]{Lucey2021}
{Lucey} M.,  et~al., 2020, arXiv e-prints, \href
  {https://ui.adsabs.harvard.edu/abs/2020arXiv200903886L} {p. arXiv:2009.03886}

\bibitem[\protect\citeauthoryear{{Marino}, {Villanova}, {Piotto}, {Milone},
  {Momany}, {Bedin}  \& {Medling}}{{Marino} et~al.}{2008}]{Marino2008}
{Marino} A.~F.,  {Villanova} S.,  {Piotto} G.,  {Milone} A.~P.,  {Momany} Y.,
  {Bedin} L.~R.,   {Medling} A.~M.,  2008, \mn@doi [\aap]
  {10.1051/0004-6361:200810389}, \href
  {https://ui.adsabs.harvard.edu/\#abs/2008A&A...490..625M} {490, 625}

\bibitem[\protect\citeauthoryear{{Mashonkina} et~al.,}{{Mashonkina}
  et~al.}{2008}]{Mashonkina2008}
{Mashonkina} L.,  et~al., 2008, \mn@doi [\aap] {10.1051/0004-6361:20078060},
  \href {https://ui.adsabs.harvard.edu/abs/2008A&A...478..529M} {478, 529}

\bibitem[\protect\citeauthoryear{{Masseron}, {Merle}  \& {Hawkins}}{{Masseron}
  et~al.}{2016}]{Masseron2016}
{Masseron} T.,  {Merle} T.,   {Hawkins} K.,  2016, {BACCHUS: Brussels Automatic
  Code for Characterizing High accUracy Spectra}, Astrophysics Source Code
  Library (\mn@eprint {ascl} {1605.004}), \mn@doi{10.20356/C4TG6R}

\bibitem[\protect\citeauthoryear{{Mayor} et~al.,}{{Mayor}
  et~al.}{2003}]{Mayor2003}
{Mayor} M.,  et~al., 2003, The Messenger, \href
  {https://ui.adsabs.harvard.edu/abs/2003Msngr.114...20M} {114, 20}

\bibitem[\protect\citeauthoryear{{McWilliam} \& {Rich}}{{McWilliam} \&
  {Rich}}{1994}]{McWilliam1994}
{McWilliam} A.,  {Rich} R.~M.,  1994, \mn@doi [\apjs] {10.1086/191954}, \href
  {http://adsabs.harvard.edu/abs/1994ApJS...91..749M} {91, 749}

\bibitem[\protect\citeauthoryear{{McWilliam} \& {Rich}}{{McWilliam} \&
  {Rich}}{2004}]{McWilliam2004}
{McWilliam} A.,  {Rich} R.~M.,  2004, in {McWilliam} A.,  {Rauch} M.,  eds,
  Origin and Evolution of the Elements. p.~38 (\mn@eprint {arXiv}
  {astro-ph/0312628})

\bibitem[\protect\citeauthoryear{{Merritt} \& {Sellwood}}{{Merritt} \&
  {Sellwood}}{1994}]{Merritt1994}
{Merritt} D.,  {Sellwood} J.~A.,  1994, \mn@doi [\apj] {10.1086/174005}, \href
  {https://ui.adsabs.harvard.edu/abs/1994ApJ...425..551M} {425, 551}

\bibitem[\protect\citeauthoryear{{Monari}, {Famaey}, {Siebert}, {Grand },
  {Kawata}  \& {Boily}}{{Monari} et~al.}{2016}]{Monari2016}
{Monari} G.,  {Famaey} B.,  {Siebert} A.,  {Grand } R. J.~J.,  {Kawata} D.,
  {Boily} C.,  2016, \mn@doi [\mnras] {10.1093/mnras/stw1564}, \href
  {https://ui.adsabs.harvard.edu/abs/2016MNRAS.461.3835M} {461, 3835}

\bibitem[\protect\citeauthoryear{{Myeong}, {Evans}, {Belokurov}, {Sanders}  \&
  {Koposov}}{{Myeong} et~al.}{2018}]{Myeong2018}
{Myeong} G.~C.,  {Evans} N.~W.,  {Belokurov} V.,  {Sanders} J.~L.,   {Koposov}
  S.~E.,  2018, \mn@doi [\apjl] {10.3847/2041-8213/aad7f7}, \href
  {https://ui.adsabs.harvard.edu/abs/2018ApJ...863L..28M} {863, L28}

\bibitem[\protect\citeauthoryear{{Ness} \& {Freeman}}{{Ness} \&
  {Freeman}}{2016}]{Ness2016}
{Ness} M.,  {Freeman} K.,  2016, \mn@doi [Publications of the Astronomical
  Society of Australia] {10.1017/pasa.2015.51}, \href
  {https://ui.adsabs.harvard.edu/abs/2016PASA...33...22N} {33, e022}

\bibitem[\protect\citeauthoryear{{Ness} et~al.,}{{Ness}
  et~al.}{2013a}]{Ness2013a}
{Ness} M.,  et~al., 2013a, \mn@doi [\mnras] {10.1093/mnras/sts629}, \href
  {https://ui.adsabs.harvard.edu/abs/2013MNRAS.430..836N} {430, 836}

\bibitem[\protect\citeauthoryear{{Ness} et~al.,}{{Ness}
  et~al.}{2013b}]{Ness2013b}
{Ness} M.,  et~al., 2013b, \mn@doi [\mnras] {10.1093/mnras/stt533}, \href
  {https://ui.adsabs.harvard.edu/abs/2013MNRAS.432.2092N} {432, 2092}

\bibitem[\protect\citeauthoryear{{Nishimura}, {Takiwaki}  \&
  {Thielemann}}{{Nishimura} et~al.}{2015}]{Nishimura2015}
{Nishimura} N.,  {Takiwaki} T.,   {Thielemann} F.-K.,  2015, \mn@doi [\apj]
  {10.1088/0004-637X/810/2/109}, \href
  {https://ui.adsabs.harvard.edu/abs/2015ApJ...810..109N} {810, 109}

\bibitem[\protect\citeauthoryear{{Nomoto}, {Kobayashi}  \& {Tominaga}}{{Nomoto}
  et~al.}{2013}]{Nomoto2013}
{Nomoto} K.,  {Kobayashi} C.,   {Tominaga} N.,  2013, \mn@doi [\araa]
  {10.1146/annurev-astro-082812-140956}, \href
  {http://adsabs.harvard.edu/abs/2013ARA%26A..51..457N} {51, 457}

\bibitem[\protect\citeauthoryear{{Nordlander} \& {Lind}}{{Nordlander} \&
  {Lind}}{2017}]{Nordlander2017}
{Nordlander} T.,  {Lind} K.,  2017, \mn@doi [\aap]
  {10.1051/0004-6361/201730427}, \href
  {https://ui.adsabs.harvard.edu/abs/2017A&A...607A..75N} {607, A75}

\bibitem[\protect\citeauthoryear{{Norris}, {Christlieb}, {Korn}, {Eriksson},
  {Bessell}, {Beers}, {Wisotzki}  \& {Reimers}}{{Norris}
  et~al.}{2007}]{Norris2007}
{Norris} J.~E.,  {Christlieb} N.,  {Korn} A.~J.,  {Eriksson} K.,  {Bessell}
  M.~S.,  {Beers} T.~C.,  {Wisotzki} L.,   {Reimers} D.,  2007, \mn@doi [\apj]
  {10.1086/521919}, \href
  {https://ui.adsabs.harvard.edu/\#abs/2007ApJ...670..774N} {670, 774}

\bibitem[\protect\citeauthoryear{{Olszewski}, {Schommer}, {Suntzeff}  \&
  {Harris}}{{Olszewski} et~al.}{1991}]{Olszewski1991}
{Olszewski} E.~W.,  {Schommer} R.~A.,  {Suntzeff} N.~B.,   {Harris} H.~C.,
  1991, \mn@doi [\aj] {10.1086/115701}, \href
  {https://ui.adsabs.harvard.edu/abs/1991AJ....101..515O} {101, 515}

\bibitem[\protect\citeauthoryear{{Ortolani}, {Renzini}, {Gilmozzi}, {Marconi},
  {Barbuy}, {Bica}  \& {Rich}}{{Ortolani} et~al.}{1995}]{Ortolani1995}
{Ortolani} S.,  {Renzini} A.,  {Gilmozzi} R.,  {Marconi} G.,  {Barbuy} B.,
  {Bica} E.,   {Rich} R.~M.,  1995, \mn@doi [\nat] {10.1038/377701a0}, \href
  {https://ui.adsabs.harvard.edu/abs/1995Natur.377..701O} {377, 701}

\bibitem[\protect\citeauthoryear{{Osorio}, {Barklem}, {Lind}, {Belyaev},
  {Spielfiedel}, {Guitou}  \& {Feautrier}}{{Osorio} et~al.}{2015}]{Osorio2015}
{Osorio} Y.,  {Barklem} P.~S.,  {Lind} K.,  {Belyaev} A.~K.,  {Spielfiedel} A.,
   {Guitou} M.,   {Feautrier} N.,  2015, \mn@doi [\aap]
  {10.1051/0004-6361/201525846}, \href
  {https://ui.adsabs.harvard.edu/abs/2015A&A...579A..53O} {579, A53}

\bibitem[\protect\citeauthoryear{{Pancino} et~al.,}{{Pancino}
  et~al.}{2017}]{Pancino2017}
{Pancino} E.,  et~al., 2017, \mn@doi [\aap] {10.1051/0004-6361/201730474},
  \href {https://ui.adsabs.harvard.edu/\#abs/2017A&A...601A.112P} {601, A112}

\bibitem[\protect\citeauthoryear{{Pasquini} et~al.,}{{Pasquini}
  et~al.}{2002}]{Pasquini2002}
{Pasquini} L.,  et~al., 2002, The Messenger, \href
  {https://ui.adsabs.harvard.edu/abs/2002Msngr.110....1P} {110, 1}

\bibitem[\protect\citeauthoryear{{Paxton}, {Bildsten}, {Dotter}, {Herwig},
  {Lesaffre}  \& {Timmes}}{{Paxton} et~al.}{2011}]{Paxton2011}
{Paxton} B.,  {Bildsten} L.,  {Dotter} A.,  {Herwig} F.,  {Lesaffre} P.,
  {Timmes} F.,  2011, \mn@doi [\apjs] {10.1088/0067-0049/192/1/3}, \href
  {https://ui.adsabs.harvard.edu/abs/2011ApJS..192....3P} {192, 3}

\bibitem[\protect\citeauthoryear{{Paxton} et~al.,}{{Paxton}
  et~al.}{2013}]{Paxton2013}
{Paxton} B.,  et~al., 2013, \mn@doi [\apjs] {10.1088/0067-0049/208/1/4}, \href
  {https://ui.adsabs.harvard.edu/abs/2013ApJS..208....4P} {208, 4}

\bibitem[\protect\citeauthoryear{{Paxton} et~al.,}{{Paxton}
  et~al.}{2015}]{Paxton2015}
{Paxton} B.,  et~al., 2015, \mn@doi [\apjs] {10.1088/0067-0049/220/1/15}, \href
  {https://ui.adsabs.harvard.edu/abs/2015ApJS..220...15P} {220, 15}

\bibitem[\protect\citeauthoryear{{Peacock}}{{Peacock}}{1983}]{Peacock1983}
{Peacock} J.~A.,  1983, \mn@doi [\mnras] {10.1093/mnras/202.3.615}, \href
  {https://ui.adsabs.harvard.edu/abs/1983MNRAS.202..615P} {202, 615}

\bibitem[\protect\citeauthoryear{{Piskunov} \& {Valenti}}{{Piskunov} \&
  {Valenti}}{2017}]{Piskunov2016}
{Piskunov} N.,  {Valenti} J.~A.,  2017, \mn@doi [\aap]
  {10.1051/0004-6361/201629124}, \href
  {https://ui.adsabs.harvard.edu/abs/2017A&A...597A..16P} {597, A16}

\bibitem[\protect\citeauthoryear{{Placco}, {Frebel}, {Beers}  \&
  {Stancliffe}}{{Placco} et~al.}{2014}]{Placco2014}
{Placco} V.~M.,  {Frebel} A.,  {Beers} T.~C.,   {Stancliffe} R.~J.,  2014,
  \mn@doi [\apj] {10.1088/0004-637X/797/1/21}, \href
  {https://ui.adsabs.harvard.edu/abs/2014ApJ...797...21P} {797, 21}

\bibitem[\protect\citeauthoryear{{Portail}, {Gerhard}, {Wegg}  \&
  {Ness}}{{Portail} et~al.}{2017}]{Portail2017}
{Portail} M.,  {Gerhard} O.,  {Wegg} C.,   {Ness} M.,  2017, \mn@doi [\mnras]
  {10.1093/mnras/stw2819}, \href
  {https://ui.adsabs.harvard.edu/abs/2017MNRAS.465.1621P} {465, 1621}

\bibitem[\protect\citeauthoryear{{Quillen}}{{Quillen}}{2002}]{Quillen2002}
{Quillen} A.~C.,  2002, \mn@doi [\aj] {10.1086/341753}, \href
  {https://ui.adsabs.harvard.edu/abs/2002AJ....124..722Q} {124, 722}

\bibitem[\protect\citeauthoryear{{Quillen}, {Minchev}, {Sharma}, {Qin}  \& {Di
  Matteo}}{{Quillen} et~al.}{2014}]{Quillen2014}
{Quillen} A.~C.,  {Minchev} I.,  {Sharma} S.,  {Qin} Y.-J.,   {Di Matteo} P.,
  2014, \mn@doi [\mnras] {10.1093/mnras/stt1972}, \href
  {https://ui.adsabs.harvard.edu/abs/2014MNRAS.437.1284Q} {437, 1284}

\bibitem[\protect\citeauthoryear{{Raha}, {Sellwood}, {James}  \& {Kahn}}{{Raha}
  et~al.}{1991}]{Raha1991}
{Raha} N.,  {Sellwood} J.~A.,  {James} R.~A.,   {Kahn} F.~D.,  1991, \mn@doi
  [\nat] {10.1038/352411a0}, \href
  {https://ui.adsabs.harvard.edu/abs/1991Natur.352..411R} {352, 411}

\bibitem[\protect\citeauthoryear{{Rakavy}, {Shaviv}  \& {Zinamon}}{{Rakavy}
  et~al.}{1967}]{Rakavy1967}
{Rakavy} G.,  {Shaviv} G.,   {Zinamon} Z.,  1967, \mn@doi [\apj]
  {10.1086/149318}, \href
  {https://ui.adsabs.harvard.edu/abs/1967ApJ...150..131R} {150, 131}

\bibitem[\protect\citeauthoryear{{Rich} \& {McWilliam}}{{Rich} \&
  {McWilliam}}{2000}]{Rich2000}
{Rich} R.~M.,  {McWilliam} A.,  2000, in {Bergeron} J.,  ed.,  Society of
  Photo-Optical Instrumentation Engineers (SPIE) Conference Series Vol. 4005,
  Discoveries and Research Prospects from 8- to 10-Meter-Class Telescopes. pp
  150--161 (\mn@eprint {arXiv} {astro-ph/0005113}), \mn@doi{10.1117/12.390138}

\bibitem[\protect\citeauthoryear{{Roederer}, {Preston}, {Thompson}, {Shectman},
  {Sneden}, {Burley}  \& {Kelson}}{{Roederer} et~al.}{2014}]{Roederer2014}
{Roederer} I.~U.,  {Preston} G.~W.,  {Thompson} I.~B.,  {Shectman} S.~A.,
  {Sneden} C.,  {Burley} G.~S.,   {Kelson} D.~D.,  2014, \mn@doi [\aj]
  {10.1088/0004-6256/147/6/136}, \href
  {http://adsabs.harvard.edu/abs/2014AJ....147..136R} {147, 136}

\bibitem[\protect\citeauthoryear{{Rojas-Arriagada} et~al.,}{{Rojas-Arriagada}
  et~al.}{2014}]{Rojas-Arriagada2014}
{Rojas-Arriagada} A.,  et~al., 2014, \mn@doi [\aap]
  {10.1051/0004-6361/201424121}, \href
  {http://adsabs.harvard.edu/abs/2014A%26A...569A.103R} {569, A103}

\bibitem[\protect\citeauthoryear{{Rojas-Arriagada} et~al.,}{{Rojas-Arriagada}
  et~al.}{2017}]{Rojas-Arriagada2017}
{Rojas-Arriagada} A.,  et~al., 2017, \mn@doi [\aap]
  {10.1051/0004-6361/201629160}, \href
  {https://ui.adsabs.harvard.edu/abs/2017A&A...601A.140R} {601, A140}

\bibitem[\protect\citeauthoryear{{Rojas-Arriagada} et~al.,}{{Rojas-Arriagada}
  et~al.}{2020}]{Rojas-Arriagada2020}
{Rojas-Arriagada} A.,  et~al., 2020, arXiv e-prints, \href
  {https://ui.adsabs.harvard.edu/abs/2020arXiv200713967R} {p. arXiv:2007.13967}

\bibitem[\protect\citeauthoryear{{Rosswog}, {Liebend{\"o}rfer}, {Thielemann},
  {Davies}, {Benz}  \& {Piran}}{{Rosswog} et~al.}{1999}]{Rosswog1999}
{Rosswog} S.,  {Liebend{\"o}rfer} M.,  {Thielemann} F.~K.,  {Davies} M.~B.,
  {Benz} W.,   {Piran} T.,  1999, \aap, \href
  {https://ui.adsabs.harvard.edu/abs/1999A&A...341..499R} {341, 499}

\bibitem[\protect\citeauthoryear{{Salvadori}, {Ferrara}, {Schneider},
  {Scannapieco}  \& {Kawata}}{{Salvadori} et~al.}{2010}]{Salvadori2010}
{Salvadori} S.,  {Ferrara} A.,  {Schneider} R.,  {Scannapieco} E.,   {Kawata}
  D.,  2010, \mn@doi [\mnras] {10.1111/j.1745-3933.2009.00772.x}, \href
  {https://ui.adsabs.harvard.edu/\#abs/2010MNRAS.401L...5S} {401, L5}

\bibitem[\protect\citeauthoryear{{Santistevan}, {Wetzel}, {El-Badry},
  {Bland-Hawthorn}, {Boylan-Kolchin}, {Bailin}, {Faucher-Gigu{\`e}re}  \&
  {Benincasa}}{{Santistevan} et~al.}{2020}]{Santistevan2020}
{Santistevan} I.~B.,  {Wetzel} A.,  {El-Badry} K.,  {Bland-Hawthorn} J.,
  {Boylan-Kolchin} M.,  {Bailin} J.,  {Faucher-Gigu{\`e}re} C.-A.,
  {Benincasa} S.,  2020, \mn@doi [\mnras] {10.1093/mnras/staa1923}, \href
  {https://ui.adsabs.harvard.edu/abs/2020MNRAS.497..747S} {497, 747}

\bibitem[\protect\citeauthoryear{{Schiavon} et~al.,}{{Schiavon}
  et~al.}{2017}]{Schiavon2017}
{Schiavon} R.~P.,  et~al., 2017, \mn@doi [\mnras] {10.1093/mnras/stw3093},
  \href {https://ui.adsabs.harvard.edu/\#abs/2017MNRAS.466.1010S} {466, 1010}

\bibitem[\protect\citeauthoryear{{Sellwood} \& {Gerhard}}{{Sellwood} \&
  {Gerhard}}{2020}]{Sellwood2020}
{Sellwood} J.~A.,  {Gerhard} O.,  2020, \mn@doi [\mnras]
  {10.1093/mnras/staa1336}, \href
  {https://ui.adsabs.harvard.edu/abs/2020MNRAS.495.3175S} {495, 3175}

\bibitem[\protect\citeauthoryear{{Shapiro}, {Genzel}  \& {F{\"o}rster
  Schreiber}}{{Shapiro} et~al.}{2010}]{Shapiro2010}
{Shapiro} K.~L.,  {Genzel} R.,   {F{\"o}rster Schreiber} N.~M.,  2010, \mn@doi
  [\mnras] {10.1111/j.1745-3933.2010.00810.x}, \href
  {https://ui.adsabs.harvard.edu/\#abs/2010MNRAS.403L..36S} {403, L36}

\bibitem[\protect\citeauthoryear{{Shen}, {Rich}, {Kormendy}, {Howard}, {De
  Propris}  \& {Kunder}}{{Shen} et~al.}{2010}]{Shen2010}
{Shen} J.,  {Rich} R.~M.,  {Kormendy} J.,  {Howard} C.~D.,  {De Propris} R.,
  {Kunder} A.,  2010, \mn@doi [\apjl] {10.1088/2041-8205/720/1/L72}, \href
  {https://ui.adsabs.harvard.edu/abs/2010ApJ...720L..72S} {720, L72}

\bibitem[\protect\citeauthoryear{{Simmerer}, {Sneden}, {Cowan}, {Collier},
  {Woolf}  \& {Lawler}}{{Simmerer} et~al.}{2004}]{Simmerer2004}
{Simmerer} J.,  {Sneden} C.,  {Cowan} J.~J.,  {Collier} J.,  {Woolf} V.~M.,
  {Lawler} J.~E.,  2004, \mn@doi [\apj] {10.1086/424504}, \href
  {https://ui.adsabs.harvard.edu/abs/2004ApJ...617.1091S} {617, 1091}

\bibitem[\protect\citeauthoryear{{Sitnova}, {Yakovleva}, {Belyaev}  \&
  {Mashonkina}}{{Sitnova} et~al.}{2020}]{Sitnova2020}
{Sitnova} T.~M.,  {Yakovleva} S.~A.,  {Belyaev} A.~K.,   {Mashonkina} L.~I.,
  2020, \mn@doi [Astronomy Letters] {10.1134/S1063773720010041}, \href
  {https://ui.adsabs.harvard.edu/abs/2020AstL...46..120S} {46, 120}

\bibitem[\protect\citeauthoryear{{Smiljanic} et~al.,}{{Smiljanic}
  et~al.}{2014}]{Smiljanic2014}
{Smiljanic} R.,  et~al., 2014, \mn@doi [\aap] {10.1051/0004-6361/201423937},
  \href {http://adsabs.harvard.edu/abs/2014A%26A...570A.122S} {570, A122}

\bibitem[\protect\citeauthoryear{{Starkenburg} et~al.,}{{Starkenburg}
  et~al.}{2010}]{Starkenburg2010}
{Starkenburg} E.,  et~al., 2010, \mn@doi [\aap] {10.1051/0004-6361/200913759},
  \href {https://ui.adsabs.harvard.edu/abs/2010A&A...513A..34S} {513, A34}

\bibitem[\protect\citeauthoryear{{Starkenburg}, {Oman}, {Navarro}, {Crain},
  {Fattahi}, {Frenk}, {Sawala}  \& {Schaye}}{{Starkenburg}
  et~al.}{2017a}]{Starkenburg2017a}
{Starkenburg} E.,  {Oman} K.~A.,  {Navarro} J.~F.,  {Crain} R.~A.,  {Fattahi}
  A.,  {Frenk} C.~S.,  {Sawala} T.,   {Schaye} J.,  2017a, \mn@doi [\mnras]
  {10.1093/mnras/stw2873}, \href
  {https://ui.adsabs.harvard.edu/abs/2017MNRAS.465.2212S} {465, 2212}

\bibitem[\protect\citeauthoryear{{Starkenburg} et~al.,}{{Starkenburg}
  et~al.}{2017b}]{Starkenburg2017b}
{Starkenburg} E.,  et~al., 2017b, \mn@doi [\mnras] {10.1093/mnras/stx1068},
  \href {https://ui.adsabs.harvard.edu/\#abs/2017MNRAS.471.2587S} {471, 2587}

\bibitem[\protect\citeauthoryear{{Takahashi}, {Yoshida}  \&
  {Umeda}}{{Takahashi} et~al.}{2018}]{Takahashi2018}
{Takahashi} K.,  {Yoshida} T.,   {Umeda} H.,  2018, \mn@doi [\apj]
  {10.3847/1538-4357/aab95f}, \href
  {https://ui.adsabs.harvard.edu/\#abs/2018ApJ...857..111T} {857, 111}

\bibitem[\protect\citeauthoryear{{Truran}}{{Truran}}{1981}]{Truran1981}
{Truran} J.~W.,  1981, \aap, \href
  {https://ui.adsabs.harvard.edu/abs/1981A&A....97..391T} {97, 391}

\bibitem[\protect\citeauthoryear{{Tumlinson}}{{Tumlinson}}{2006}]{Tumlinson2006}
{Tumlinson} J.,  2006, \mn@doi [\apj] {10.1086/500383}, \href
  {https://ui.adsabs.harvard.edu/\#abs/2006ApJ...641....1T} {641, 1}

\bibitem[\protect\citeauthoryear{{Tumlinson}}{{Tumlinson}}{2010}]{Tumlinson2010}
{Tumlinson} J.,  2010, \mn@doi [\apj] {10.1088/0004-637X/708/2/1398}, \href
  {https://ui.adsabs.harvard.edu/\#abs/2010ApJ...708.1398T} {708, 1398}

\bibitem[\protect\citeauthoryear{{Valenti} \& {Piskunov}}{{Valenti} \&
  {Piskunov}}{1996}]{Valenti1996}
{Valenti} J.~A.,  {Piskunov} N.,  1996, \aaps, \href
  {https://ui.adsabs.harvard.edu/abs/1996A&AS..118..595V} {118, 595}

\bibitem[\protect\citeauthoryear{{Valenti}, {Zoccali}, {Renzini}, {Brown},
  {Gonzalez}, {Minniti}, {Debattista}  \& {Mayer}}{{Valenti}
  et~al.}{2013}]{Valenti2013}
{Valenti} E.,  {Zoccali} M.,  {Renzini} A.,  {Brown} T.~M.,  {Gonzalez} O.~A.,
  {Minniti} D.,  {Debattista} V.~P.,   {Mayer} L.,  2013, \mn@doi [\aap]
  {10.1051/0004-6361/201321962}, \href
  {https://ui.adsabs.harvard.edu/abs/2013A&A...559A..98V} {559, A98}

\bibitem[\protect\citeauthoryear{{Van der Swaelmen}, {Hill}, {Primas}  \&
  {Cole}}{{Van der Swaelmen} et~al.}{2013}]{VanderSwaelmen2013}
{Van der Swaelmen} M.,  {Hill} V.,  {Primas} F.,   {Cole} A.~A.,  2013, \mn@doi
  [\aap] {10.1051/0004-6361/201321109}, \href
  {http://adsabs.harvard.edu/abs/2013A%26A...560A..44V} {560, A44}

\bibitem[\protect\citeauthoryear{{Vincenzo} \& {Kobayashi}}{{Vincenzo} \&
  {Kobayashi}}{2018}]{Vincenzo2018a}
{Vincenzo} F.,  {Kobayashi} C.,  2018, \mn@doi [\aap]
  {10.1051/0004-6361/201732395}, \href
  {https://ui.adsabs.harvard.edu/abs/2018A&A...610L..16V} {610, L16}

\bibitem[\protect\citeauthoryear{{Wegg}, {Gerhard}  \& {Portail}}{{Wegg}
  et~al.}{2015}]{Wegg2015}
{Wegg} C.,  {Gerhard} O.,   {Portail} M.,  2015, \mn@doi [\mnras]
  {10.1093/mnras/stv745}, \href
  {https://ui.adsabs.harvard.edu/abs/2015MNRAS.450.4050W} {450, 4050}

\bibitem[\protect\citeauthoryear{{White} \& {Springel}}{{White} \&
  {Springel}}{2000}]{White2000}
{White} S. D.~M.,  {Springel} V.,  2000, in {Weiss} A.,  {Abel} T.~G.,   {Hill}
  V.,  eds, The First Stars. p.~327 (\mn@eprint {arXiv} {astro-ph/9911378}),
  \mn@doi{10.1007/10719504_62}

\bibitem[\protect\citeauthoryear{{Winteler}, {K{\"a}ppeli}, {Perego},
  {Arcones}, {Vasset}, {Nishimura}, {Liebend{\"o}rfer}  \&
  {Thielemann}}{{Winteler} et~al.}{2012}]{Winteler2012}
{Winteler} C.,  {K{\"a}ppeli} R.,  {Perego} A.,  {Arcones} A.,  {Vasset} N.,
  {Nishimura} N.,  {Liebend{\"o}rfer} M.,   {Thielemann} F.~K.,  2012, \mn@doi
  [\apjl] {10.1088/2041-8205/750/1/L22}, \href
  {https://ui.adsabs.harvard.edu/abs/2012ApJ...750L..22W} {750, L22}

\bibitem[\protect\citeauthoryear{{Wolf} et~al.,}{{Wolf}
  et~al.}{2018}]{Wolf2018}
{Wolf} C.,  et~al., 2018, \mn@doi [Publications of the Astronomical Society of
  Australia] {10.1017/pasa.2018.5}, \href
  {https://ui.adsabs.harvard.edu/\#abs/2018PASA...35...10W} {35, e010}

\bibitem[\protect\citeauthoryear{{Woosley} \& {Heger}}{{Woosley} \&
  {Heger}}{2006}]{Woosley2006}
{Woosley} S.~E.,  {Heger} A.,  2006, \mn@doi [\apj] {10.1086/498500}, \href
  {https://ui.adsabs.harvard.edu/abs/2006ApJ...637..914W} {637, 914}

\bibitem[\protect\citeauthoryear{{Woosley} \& {Weaver}}{{Woosley} \&
  {Weaver}}{1995}]{Woosley1995}
{Woosley} S.~E.,  {Weaver} T.~A.,  1995, \mn@doi [\apjs] {10.1086/192237},
  \href {http://adsabs.harvard.edu/abs/1995ApJS..101..181W} {101, 181}

\bibitem[\protect\citeauthoryear{{Woosley}, {Heger}  \& {Weaver}}{{Woosley}
  et~al.}{2002}]{Woosley2002}
{Woosley} S.~E.,  {Heger} A.,   {Weaver} T.~A.,  2002, \mn@doi [Reviews of
  Modern Physics] {10.1103/RevModPhys.74.1015}, \href
  {https://ui.adsabs.harvard.edu/abs/2002RvMP...74.1015W} {74, 1015}

\bibitem[\protect\citeauthoryear{{Wylie}, {Gerhard}, {Ness}, {Clarke},
  {Freeman}  \& {Bland-Hawthorn}}{{Wylie} et~al.}{2021}]{Wylie2021}
{Wylie} S.~M.,  {Gerhard} O.~E.,  {Ness} M.~K.,  {Clarke} J.~P.,  {Freeman}
  K.~C.,   {Bland-Hawthorn} J.,  2021, arXiv e-prints, \href
  {https://ui.adsabs.harvard.edu/abs/2021arXiv210614298W} {p. arXiv:2106.14298}

\bibitem[\protect\citeauthoryear{{Yong} et~al.,}{{Yong}
  et~al.}{2013}]{Yong2013}
{Yong} D.,  et~al., 2013, \mn@doi [\apj] {10.1088/0004-637X/762/1/26}, \href
  {http://adsabs.harvard.edu/abs/2013ApJ...762...26Y} {762, 26}

\bibitem[\protect\citeauthoryear{{Zoccali} et~al.,}{{Zoccali}
  et~al.}{2003}]{Zoccali2003}
{Zoccali} M.,  et~al., 2003, \mn@doi [\aap] {10.1051/0004-6361:20021604}, \href
  {https://ui.adsabs.harvard.edu/abs/2003A&A...399..931Z} {399, 931}

\bibitem[\protect\citeauthoryear{{Zoccali}, {Hill}, {Lecureur}, {Barbuy},
  {Renzini}, {Minniti}, {G{\'o}mez}  \& {Ortolani}}{{Zoccali}
  et~al.}{2008}]{Zoccali2008}
{Zoccali} M.,  {Hill} V.,  {Lecureur} A.,  {Barbuy} B.,  {Renzini} A.,
  {Minniti} D.,  {G{\'o}mez} A.,   {Ortolani} S.,  2008, \mn@doi [\aap]
  {10.1051/0004-6361:200809394}, \href
  {http://adsabs.harvard.edu/abs/2008A%26A...486..177Z} {486, 177}

\bibitem[\protect\citeauthoryear{{Zoccali} et~al.,}{{Zoccali}
  et~al.}{2014}]{Zoccali2014}
{Zoccali} M.,  et~al., 2014, \mn@doi [\aap] {10.1051/0004-6361/201323120},
  \href {https://ui.adsabs.harvard.edu/\#abs/2014A&A...562A..66Z} {562, A66}

\bibitem[\protect\citeauthoryear{{Zoccali} et~al.,}{{Zoccali}
  et~al.}{2017}]{Zoccali2017}
{Zoccali} M.,  et~al., 2017, \mn@doi [\aap] {10.1051/0004-6361/201629805},
  \href {https://ui.adsabs.harvard.edu/\#abs/2017A&A...599A..12Z} {599, A12}

\makeatother
\end{thebibliography}
\bsp	
\label{lastpage}
\end{document}